\documentclass[aps]{revtex4}
\usepackage{eurosym}
\usepackage{amsfonts}
\usepackage{amsmath}
\usepackage{amssymb,epsf}
\usepackage{color}
\usepackage{xcolor}
\usepackage{graphicx}
\usepackage{epstopdf}
\usepackage{float}
\usepackage{caption}
\usepackage{subfig}
\usepackage{hyperref}
\usepackage{mathtools}
\usepackage{hyperref}
\usepackage{booktabs}
\usepackage{orcidlink}

\begin{document}

\title{Topological Mod(A)Max AdS black holes}
\author{B. Eslam Panah\,\orcidlink{0000-0002-1447-3760}}
\email{eslampanah@umz.ac.ir}
\affiliation{Department of Physics, University of Mazandaran, Babolsar, Iran}
\author{B. Hamil\,\orcidlink{0000-0002-7043-6104}}
\email{hamilbilel@gmail.com/bilel.hamil@umc.edu.dz}
\affiliation{Laboratoire de Physique Math\'{e}matique et Physique Subatomique,LPMPS,
Facult\'{e} des Sciences Exactes, Universit\'{e} Constantine 1, Constantine,
Algeria}
\author{Manuel E. Rodrigues\,\orcidlink{0000-0001-8586-0285}}
\email{esialg@gmail.com}
\affiliation{Faculdade de F\'{\i}sica, Programa de P\'{o}s-Gradua\c{c}\~{a}o em F\'{\i}%
sica, Universidade Federal do Par\'{a}, 66075-110, Bel\'{e}m, Par\'{a},
Brazill}
\affiliation{Faculdade de Ci\^{e}ncias Exatas e Tecnologia, Universidade Federal do Par%
\'{a}, Campus Universit\'{a}rio de Abaetetuba, 68440-000, Abaetetuba, Par%
\'{a}, Brazil}

\begin{abstract}
In this work, we construct new classes of topological black hole solutions
in anti-de Sitter (AdS) spacetime using a novel model of nonlinear
electrodynamics called Modification Maxwell (ModMax) and Modification
phantom or Modification anti-Maxwell (ModAMax). We then evaluate the
thermodynamic quantities and verify the first law of thermodynamics. Our
study examines how the parameters of the ModMax and ModAMax fields, as well
as the topological constant, affect the black hole solutions, thermodynamic
quantities, and local and global thermal stabilities. Furthermore, within the framework of extended phase space thermodynamics, we
analyze the Joule-Thomson expansion process and determine the inversion
curves. This analysis reveals that the ModMax and ModAMax parameters
significantly alter the cooling and heating behavior of these AdS black
holes, depending on their topology. Finally, by treating these topological
Mod(A)Max AdS black holes as heat engines, we assess their efficiencies,
demonstrating that the parameters of nonlinear electrodynamics and horizon
topology play crucial roles in enhancing or suppressing the system's
thermodynamic performance.
\end{abstract}

\maketitle

\section{Introduction}

Within the realm of physics, one occasionally encounters systems whose
behavior defies ordinary expectations, displaying negative energy densities.
Such configurations appear in several distinct contexts \cite{Visser}. Among
the best-known are: (a) the traditional Casimir effect, observed between two
uncharged parallel plates as well as in its topological extension; (b) the
squeezed vacuum state, which emerges in quantum optics; (c) the
Hartle--Hawking vacuum, a quantum state relevant to black hole
thermodynamics; and (d) the so-called phantom field, a hypothetical entity
invoked to explain the accelerated expansion of the cosmos \cite{Caldwell}.

Comparable scenarios arise within the framework of gravitation. A
historically important instance is the quasi-charged bridge introduced by
Einstein and Rosen in 1935 \cite{Einstein}. In that construction, the system
acquires an effective electric charge with an inverted sign relative to the
Reissner--Nordstrom geometry, expressed formally as ($q^{2}\rightarrow-q^{2} 
$). This feature effectively introduces into the action the first
manifestation of a spin--$1$ phantom field, characterized by its negative
energy contribution.

Phantom fields are not limited to vector forms. The anti-Fisher solution 
\cite{Bronnikov} provides a notable example of a scalar, or spin-0, phantom
configuration. When scalar and electromagnetic phantom components are
coupled, the resulting models are known as
Einstein-anti-Maxwell-anti-Dilaton systems \cite{Gerard}. Various extensions
of these models have been developed, including those incorporating a
cosmological constant \cite{Jardim2012}, as well as regular configurations
free from singularities \cite{Fabris}. The Sigma-model approach has also
been employed to construct rotating phantom solutions \cite{Gerard2}.

The study of phantom and related systems has expanded considerably,
encompassing analyses of photon trajectories \cite{Azreg}, gravitational
collapse \cite{Na}, lensing phenomena \cite{Gy}, global monopoles \cite{Chen}%
, quasi-black holes \cite{Bronnikov2}, dyonic black holes \cite{Abishev},
wormhole geometries \cite{Huang}, spherical accretion processes \cite{Azreg2}%
, topological black holes within $f(R)$ gravity \cite{Rodrigues}, and, more
recently, the intriguing class of black--bounce spacetimes \cite{Silva}.

Linear Maxwell electrodynamics possesses a remarkably elegant structure. It
is invariant under the $U(1)$ gauge group and is governed by second--order,
linear differential equations in the electromagnetic potentials. These
features ensure both mathematical simplicity and physical consistency.
However, when such assumptions are relaxed, a variety of richer and more
exotic behaviors emerge that are absent in the linear theory.

If the $U(1)$ gauge symmetry is broken, the resulting field equations
generally become of higher order in the potentials, as exemplified in the
formulations of Podolsky and Proca electrodynamics \cite%
{Podolsky1942,Bopp1940,Proca1936}. On the other hand, allowing for nonlinear
dependencies in the electromagnetic field invariants gives rise to entire
families of modified theories collectively known as nonlinear
electrodynamics (NED). Such frameworks predict novel physical effects,
including magnetic birefringence, in which light propagates
anisotropically--its velocity depending on both the orientation of the
external magnetic field and the direction of the light's polarization \cite%
{Cadene2013}.

The origins of NED trace back to Born and Infeld in 1934 \cite{Born1934},
who sought to eliminate the divergent self-energy of point charges inherent
in Maxwell's formulation. Their theory successfully regularized the field
energy near charged particles, thereby removing the singularity at the
charge's position. A few years later, in 1936, Euler and Heisenberg
developed a quantum-field-based form of NED to describe photon--photon
scattering \cite{Heisenberg1936}. Soon after, Hoffmann extended the
Born-Infeld Lagrangian to include gravitational coupling \cite{Hoffmann1935}%
, establishing the foundation for subsequent studies in NED within curved
spacetime.

Since then, NEDs have been invoked across a wide range of physical
scenarios, including the ionisation of the hydrogen atom \cite{hidrogenio},
baryogenesis \cite{bariogenese}, cosmic microwave background polarisation 
\cite{CMB}, multiple refraction phenomena \cite{multi}, neutrino
astrophysics \cite{neutrino}, unidirectional light propagation \cite{luz},
pulsar emission models \cite{pulsar}, cosmological inflation \cite{inflacao}%
, photon gas thermodynamics \cite{termodinamica}, and the accelerated
expansion of the Universe \cite{aceleracao}.

Experimental exploration of NED effects has been an active and rapidly
advancing field. Laboratory tests include the PVLAS (Polarizzazione del
Vuoto by LASERs) experiment \cite{PVLAS}, LSW (Light Shining through Walls) 
\cite{LSW}, BMV (Bir\'{e}fringence Magn\'{e}tique du Vide, Toulouse) \cite%
{BMV}, VH (Vacuum Hohlraum, photon collider) \cite{VH}, XFELS (X-ray Free
Electron LASERS) \cite{XFELS}, ELI (Extreme Light Infrastructure) \cite{ELI}%
, and SULF (Shanghai Ultra-Laser Facilities) \cite{SULF}. Forthcoming
facilities--the Station of Extreme Light (SEL) in Shanghai, expected to
operate in 2025, and the ExaWatt Centre for Extreme Light Studies (XCELS) in
Russia, anticipated for 2026--may provide decisive experimental confirmation
of NED effects.

In 2020, a new model of NED was proposed, retaining the same symmetries as
Maxwell's theory \cite{ModMaxI,ModMaxII}. This model is known as the
Modification of Maxwell (ModMax) theory. The ModMax theory is characterized
by a dimensionless parameter $\gamma$, referred to as the ModMax parameter,
which reduces to Maxwell's theory when $\gamma=0$. Research has examined
magnetic \cite{ModMaxMagnetic} and electromagnetized black holes \cite%
{ModMaxElecMag} in the presence of ModMax theory. Additionally, various
black hole solutions, including accelerating \cite%
{acc1ModMax,acc2ModMax,acc3ModMax} and BTZ \cite{BTZModMax} black holes,
have been derived from ModMax theory. The impact of the ModMax parameter on
several black hole properties has also been investigated, including shadow 
\cite{ShadowModMax}, quasinormal modes \cite{ShadowModMax,quasiModMax},
greybody radiation \cite{ShadowModMax}, light propagation \cite{LightModMax}%
, thermodynamic topology \cite{TheTopoModMax}, emission rates \cite%
{quasiModMax}, and gravitational lensing \cite{GraLensModMax}. Moreover, the
combination of ModMax with other modified theories of gravity, such as $F(R)$
\cite{FRModMax}, massive \cite{massiveModMax}, and Kalb-Ramond \cite%
{KR1ModMax,KR2ModMax,KR3ModMax} theories, has led to the exploration of new
black hole solutions and their properties. In this regard, further studies
on the intriguing aspects of ModMax theory are evaluated in Refs. \cite%
{MM1,MM2,MM3,MM4,MM5,MM6,MM7,MM8,MM9,MM10,MM11,MM12,MM13,MM14,MM15,MM16,MM17,MM18,MM19,MM20,MM21,MM22,MM23}%
.

On the other hand, the topology of the event horizon of a four-dimensional
asymptotically flat stationary black hole is uniquely characterized as a
two-sphere, denoted as $S^{2}$ \cite{S21,S22}. This determination is derived
from Hawking's theorem, which asserts that the integrated Ricci scalar
curvature, computed with respect to the induced metric on the event horizon,
must be positive \cite{S21}. However, in spacetimes that do not exhibit
asymptotic flatness, the spherical topology of the black hole's horizon is
not mandated; stationary black holes can manifest nontrivial topologies. It
has been discussed that in four-dimensional Einstein theory coupled with a
Maxwell field, asymptotically AdS spacetimes may possess black hole
solutions whose event horizons can exhibit zero or negative constant
curvature, resulting in topologies that deviate from the two-sphere $S^{2}$ 
\cite%
{Toplogy1,Toplogy2,Toplogy3,Toplogy4,Toplogy5,Toplogy6,Toplogy7,Toplogy9,Toplogy10}%
.

Bekenstein and Hawking identified an analogy between the geometric
properties of black holes and thermodynamic variables, enhancing our
understanding of the connection between gravity and classical thermodynamics 
\cite{BekHaw}. The thermodynamic properties of black holes, especially phase
transitions and thermal stability, are crucial as they provide insights into
the fundamental structure of spacetime geometry. Phase transitions are
significant across various fields, including elementary particles \cite%
{Kleinert}, classical thermodynamics \cite{Callen}, condensed matter \cite%
{Greer}, black holes \cite{Zou}, and cosmology \cite{Layzer}. An intriguing
aspect of black hole thermodynamics is the phase transition occurring in
anti-de Sitter (AdS) spacetime, which is informed by AdS/CFT duality \cite%
{AdSCFT1,AdSCFT2} and sheds light on the quantum nature of gravity.
Recently, the study of phase transitions in AdS black holes within an
extended phase space has garnered considerable attention. In this context,
the cosmological constant is treated as a variable, recognized as
thermodynamic pressure \cite{LambdaPI}, while the mass of the black hole is
interpreted as enthalpy \cite{LambdaPII}.

Recent advances in black hole thermodynamics include the Joule--Thomson (JT)
expansion \cite{Adine} and the development of holographic heat engines \cite%
{Johnson}. The JT expansion has been applied to black holes in AdS
spacetime, inspired by the established analogy between AdS black holes and
van der Waals fluids. In classical thermodynamics, JT process describes the
expansion of a gas from a region of high pressure to one of low pressure
through a porous medium, during which the enthalpy remains constant. This
process is inherently adiabatic and isoenthalpic, providing a useful
framework for analyzing the thermal behavior of thermodynamic systems. In
this context, the JT expansion has been utilized as an isoenthalpic tool to
explore the heating and cooling characteristics of black holes. The first
detailed investigation of the JT expansion in relation to black holes was
reported in \cite{Adine}. More recently, the JT expansion has been extended
to various black hole solutions, including quintessence AdS black holes \cite%
{Ghaffarnejad}, charged AdS black holes in arbitrary dimensions \cite{JXMo1}%
, Kerr-AdS black holes \cite{Okcu}, Kerr-Newman-AdS black holes \cite{ZWZhao}%
, Gauss-Bonnet AdS black holes \cite{Lan}, black holes in Lovelock gravity 
\cite{GQLi}, and black holes in NED \cite{Kuang}. Additionally, C. V.
Johnson introduced the concept of a conventional heat engine into black hole
thermodynamics. By defining a closed cycle in the pressure-volume ($P-V$)
plane for a charged AdS black hole, he demonstrated that useful mechanical
work can be extracted from its heat energy \cite{Johnson}. This approach
revealed that mechanical work can be derived from the heat energy of both
static and stationary AdS black holes, unlike the Penrose process, which is
limited to rotating black holes but can occur in both AdS and flat
spacetimes. Furthermore, Johnson proposed that black hole heat engines might
have intriguing holographic interpretations, as the thermodynamic cycle can
be seen as a trajectory through a family of holographically dual large-$N$
field theories \cite{Johnson}. This concept has since garnered significant
attention and has been extended to numerous other black hole configurations 
\cite%
{Setare,Belhaj,Pedraza,MZhang,FLiang,RBMann,Bhamidipati,JXMo,SHHendi,EslamJH}%
.

These studies prompt us to investigate topological AdS black holes with
modified Maxwell and phantom (anti-Maxwell) fields. Our goal is to examine
how the parameters of these modifications, along with the topological
constant, affect black hole solutions, thermodynamic quantities, and the
efficiency of heat engines when these black holes are treated as heat engine
machines.

\section{ModMax and ModAMax theory and Topological Black Hole Solutions}

The action describing the coupling of Einstein's gravity with the
cosmological constant in the presence of the Mod(A)Max electrodynamic fields
is expressed as 
\begin{equation}
\mathcal{I}=\frac{1}{16\pi }\int_{\partial \mathcal{M}}d^{4}x\sqrt{-g}\left[
R-2\Lambda -4\eta \mathcal{L}\right] ,  \label{Action}
\end{equation}%
where $g=\text{det}(g_{\mu \nu })$ is the determinant of the metric tensor $%
g_{\mu \nu }$. Also, $\Lambda $ and $R$ are related to the cosmological
constant and the Ricci scalar, respectively. In addition, $\eta =+1$, and $%
\eta =-1$ are related to Maxwell and anti-Maxwell (phantom) cases,
respectively. It is notable that, Mod(A)Max devotes to both ModMax and
ModAMax fields. In other words, Mod(A)Max is related to ModMax field when $%
\eta =+1$ or the ModAMax field when $\eta =-1$. In the action (\ref{Action}%
), $\mathcal{L}$ refers to ModMax's Lagrangian and is defined as \cite%
{ModMaxI,ModMaxII} 
\begin{equation}
\mathcal{L}=\mathcal{S}\cosh \gamma -\sqrt{\mathcal{S}^{2}+\mathcal{P}^{2}}%
\sinh \gamma ,  \label{ModMaxL}
\end{equation}%
where $\gamma $ is known as ModMax's parameter,. This parameter is a
dimensionless quantity. Furthermore, $\mathcal{S}$ and $\mathcal{P}$
represent a true scalar and a pseudoscalar, respectively, in the following
expressions 
\begin{equation}
\mathcal{S}=\frac{\mathcal{F}}{4},~~~\&~~~\mathcal{P}=\frac{\widetilde{%
\mathcal{F}}}{4},
\end{equation}%
in the equations above, the term $\mathcal{F}=F_{\mu \nu }F^{\mu \nu }$ is
referred to as the Maxwell invariant. Moreover, $F_{\mu \nu }$ represents
the electromagnetic tensor field and is defined as $F_{\mu \nu }=\partial
_{\mu }A_{\nu }-\partial _{\nu }A_{\mu }$, with $A_{\mu }$ being the gauge
potential. Also, $\widetilde{\mathcal{F}}=$ $F_{\mu \nu }\widetilde{F}^{\mu
\nu }$, where $\widetilde{F}^{\mu \nu }=\frac{1}{2}\epsilon _{\mu \nu
}^{~~~\rho \lambda }F_{\rho \lambda }$. By considering $\gamma =0$, the
Lagrangian of ModMax (Eq. (\ref{ModMaxL})) reduces to the Maxwell theory as $%
\mathcal{L}=\frac{\mathcal{F}}{4}$.

In order to get electrically charged black holes solutions, we set $\mathcal{%
P}=0$ in the Lagrangian of ModMax (Eq. (\ref{ModMaxL})). So, the generalized
Einstein-$\Lambda $ equations in the presence of the modification Maxwell
(ModMax) and the modification anti-Maxwell (ModAMax) fields write in the
following form \cite{MM1} 
\begin{eqnarray}
G_{\mu \nu }+\Lambda g_{\mu \nu } &=&8\pi \mathrm{T}_{\mu \nu },  \label{eq1}
\\
&&  \notag \\
\partial _{\mu }\left( \sqrt{-g}\widetilde{E}^{\mu \nu }\right) &=&0,
\label{eq2}
\end{eqnarray}%
where $\mathrm{T}_{\mu \nu }$ is the energy-momentum tensor in the presence
of Mod(A)Max field which is given by 
\begin{equation}
8\pi \mathrm{T}^{\mu \nu }=2\eta \left( F^{\mu \sigma }F_{~~\sigma }^{\nu
}e^{-\gamma }\right) -2\eta e^{-\gamma }\mathcal{S}g^{\mu \nu },  \label{eq3}
\end{equation}%
and $\widetilde{E}_{\mu \nu }$ is defined as 
\begin{equation}
\widetilde{E}_{\mu \nu }=\frac{\partial \mathcal{L}}{\partial F^{\mu \nu }}%
=2\left( \mathcal{L}_{\mathcal{S}}F_{\mu \nu }\right) ,  \label{eq4}
\end{equation}%
where $\mathcal{L}_{\mathcal{S}}=\frac{\partial \mathcal{L}}{\partial 
\mathcal{S}}$.

For charged case, the equation (\ref{eq2}), turns to 
\begin{equation}
\partial _{\mu }\left( \sqrt{-g}e^{-\gamma }F^{\mu \nu }\right) =0.
\label{Maxwell Equation}
\end{equation}

We consider a topological four-dimensional static metric as 
\begin{equation}
ds^{2}=-f\left( r\right) dt^{2}+\frac{dr^{2}}{f(r)}+r^{2}d\Omega _{k}^{2},
\label{Metric}
\end{equation}%
where $f(r)$ is the metric function, which we are going to extract it. In
addition, $d\Omega _{k}^{2}$ is given by 
\begin{equation}
d\Omega _{k}^{2}=\left\{ 
\begin{array}{ccc}
d\theta ^{2}+\sin ^{2}\theta d\varphi ^{2} &  & k=1 \\ 
d\theta ^{2}+d\varphi ^{2} &  & k=0 \\ 
d\theta ^{2}+\sinh ^{2}\theta d\varphi ^{2} &  & k=-1%
\end{array}%
\right. .
\end{equation}

Applying the following gauge potential, we can have a radial electric field 
\begin{equation}
A_{\mu }=h(r)\delta _{\mu }^{t},  \label{gauge
potential}
\end{equation}%
and using the metric (\ref{Metric}) and the equation (\ref{Maxwell Equation}%
), we find the following relation 
\begin{equation}
2h^{\prime }(r)+rh^{\prime \prime }(r)=0,  \label{heq}
\end{equation}%
where the prime and double prime are, respectively, the first and second
derivatives with respect to $r$. By solving the equation (\ref{heq}), we
obtain 
\begin{equation}
h(r)=-\frac{q}{r},  \label{h(r)}
\end{equation}%
where $q$ is an integration constant related to the electric charge.

To extract the metric function, $f(r)$, we investigate the equations (\ref%
{eq1}), (\ref{eq3}), (\ref{Metric}), and (\ref{h(r)}). We obtain the
following differential equations 
\begin{eqnarray}
&&eq_{tt}=eq_{rr}=rf^{\prime }(r)+f\left( r\right) +\Lambda r^{2}-k+\frac{%
\eta q^{2}e^{-\gamma }}{r^{2}}=0,  \label{eqENMax1} \\
&&  \notag \\
&&eq_{\theta \theta }=eq_{\varphi \varphi }=f^{\prime \prime }(r)+\frac{2}{r}%
f^{\prime }(r)+2\Lambda -\frac{2\eta q^{2}}{r^{4}}e^{-\gamma }=0,
\label{eqENMax2}
\end{eqnarray}%
which $eq_{tt}$, $eq_{rr}$, $eq_{\theta \theta }$, and $eq_{\varphi \varphi
} $ are representative $tt$, $rr$, $\theta \theta $, and $\varphi \varphi $
components of Eq. (\ref{eq1}), respectively. Applying the differential
equations (\ref{eqENMax1}) and (\ref{eqENMax2}), we can obtain an exact
solution for the metric function in the following form 
\begin{equation}
f(r)=k-\frac{m}{r}-\frac{\Lambda r^{2}}{3}+\frac{\eta q^{2}e^{-\gamma }}{%
r^{2}}=\left\{ 
\begin{array}{ccc}
k-\frac{m}{r}-\frac{\Lambda r^{2}}{3}+\frac{q^{2}e^{-\gamma }}{r^{2}}, &  & 
\text{ModMax} \\ 
&  &  \\ 
k-\frac{m}{r}-\frac{\Lambda r^{2}}{3}-\frac{q^{2}e^{-\gamma }}{r^{2}}, &  & 
\text{ModAMax}%
\end{array}%
\right. ,  \label{f(r)}
\end{equation}%
where $m$ is an integration constant that is related to the geometrical mass
of the black hole. Also, the metric function (\ref{f(r)}) satisfies all the
components of the field equation (\ref{eq1}), simultaneously. It is notable
that, $\eta =+1$ belongs to the ModMax solution and $\eta =-1$ is related to
the ModAMax solution. Moreover, the metric function Eq. (\ref{f(r)}) turns
to topological Reissner-Nordstrom AdS black hole when $\gamma =0$, and $\eta
=+1$, i.e. 
\begin{equation}
f(r)=k-\frac{m}{r}-\frac{\Lambda r^{2}}{3}+\frac{q^{2}}{r^{2}}.
\end{equation}%
and for $\gamma =0$, and $\eta =-1$, the metric function Eq. (\ref{f(r)})
reduces to topological anti-Maxwell (phantom) AdS black hole, i.e. 
\begin{equation}
f(r)=k-\frac{m}{r}-\frac{\Lambda r^{2}}{3}-\frac{q^{2}}{r^{2}}.
\end{equation}

We study the Kretschmann scalar to find the singularity of the metric
function (\ref{f(r)}). In this regard, we consider the metric (\ref{Metric})
and the metric function (\ref{f(r)}), and calculate the Kretschmann scalar
which leads to 
\begin{eqnarray}
R_{\alpha \beta \gamma \delta }R^{\alpha \beta \gamma \delta } &=&\frac{%
8\Lambda ^{2}}{3}+\frac{12m^{2}}{r^{6}}-\frac{48m\eta q^{2}e^{-\gamma }}{%
r^{7}}+\frac{56\eta ^{2}q^{4}e^{-2\gamma }}{r^{8}}  \notag \\
&&  \notag \\
&=&\left\{ 
\begin{array}{ccc}
\frac{8\Lambda ^{2}}{3}+\frac{12m^{2}}{r^{6}}-\frac{48mq^{2}e^{-\gamma }}{%
r^{7}}+\frac{56q^{4}e^{-2\gamma }}{r^{8}}, &  & \text{ModMax} \\ 
&  &  \\ 
\frac{8\Lambda ^{2}}{3}+\frac{12m^{2}}{r^{6}}+\frac{48mq^{2}e^{-\gamma }}{%
r^{7}}+\frac{56q^{4}e^{-2\gamma }}{r^{8}}, &  & \text{ModAMax}%
\end{array}%
\right. ,
\end{eqnarray}%
which diverges at $r=0$ (i.e., $\lim_{r\longrightarrow 0}R_{\alpha \beta
\gamma \delta }R^{\alpha \beta \gamma \delta }\longrightarrow \infty $).
This reveals that, there is a curvature singularity located at $r=0$. In
addition, the Kretschmann scalar is finite for $r\neq 0$.

The asymptotic behavior of spacetime is determined by 
\begin{eqnarray}
\lim_{r\longrightarrow \infty }R_{\alpha \beta \gamma \delta }R^{\alpha
\beta \gamma \delta } &\longrightarrow &\frac{8\Lambda ^{2}}{3},  \notag \\
&&  \notag \\
\lim_{r\longrightarrow \infty }f\left( r\right) &\longrightarrow &-\frac{%
\Lambda r^{2}}{3},
\end{eqnarray}%
where indicates that the spacetime will be asymptotically de sitter (dS)
when $\Lambda >0$, or anti-de Sitter (AdS) when $\Lambda <0$.

\begin{figure}[tbph]
\centering
\includegraphics[width=0.35\linewidth]{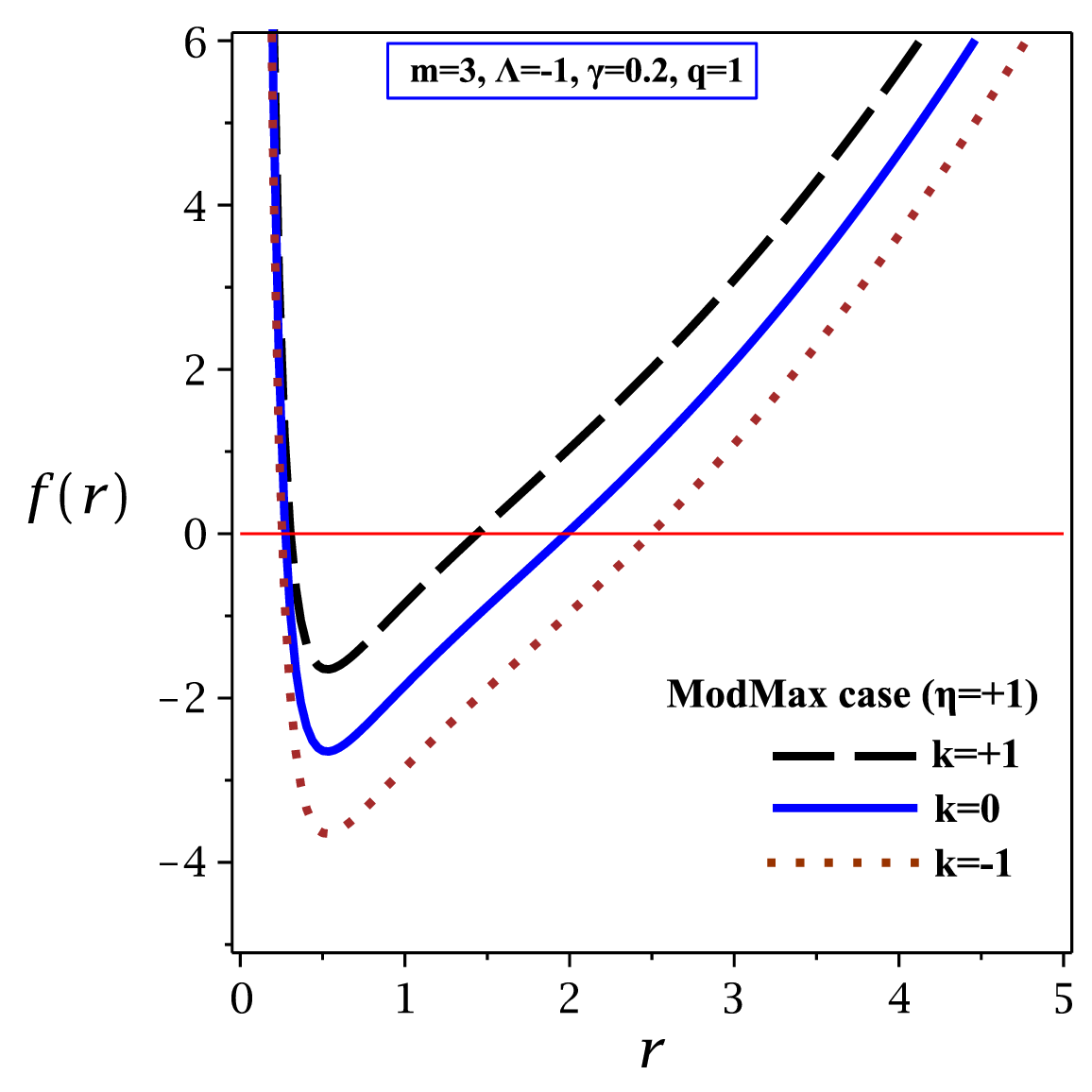} \includegraphics[width=0.35%
\linewidth]{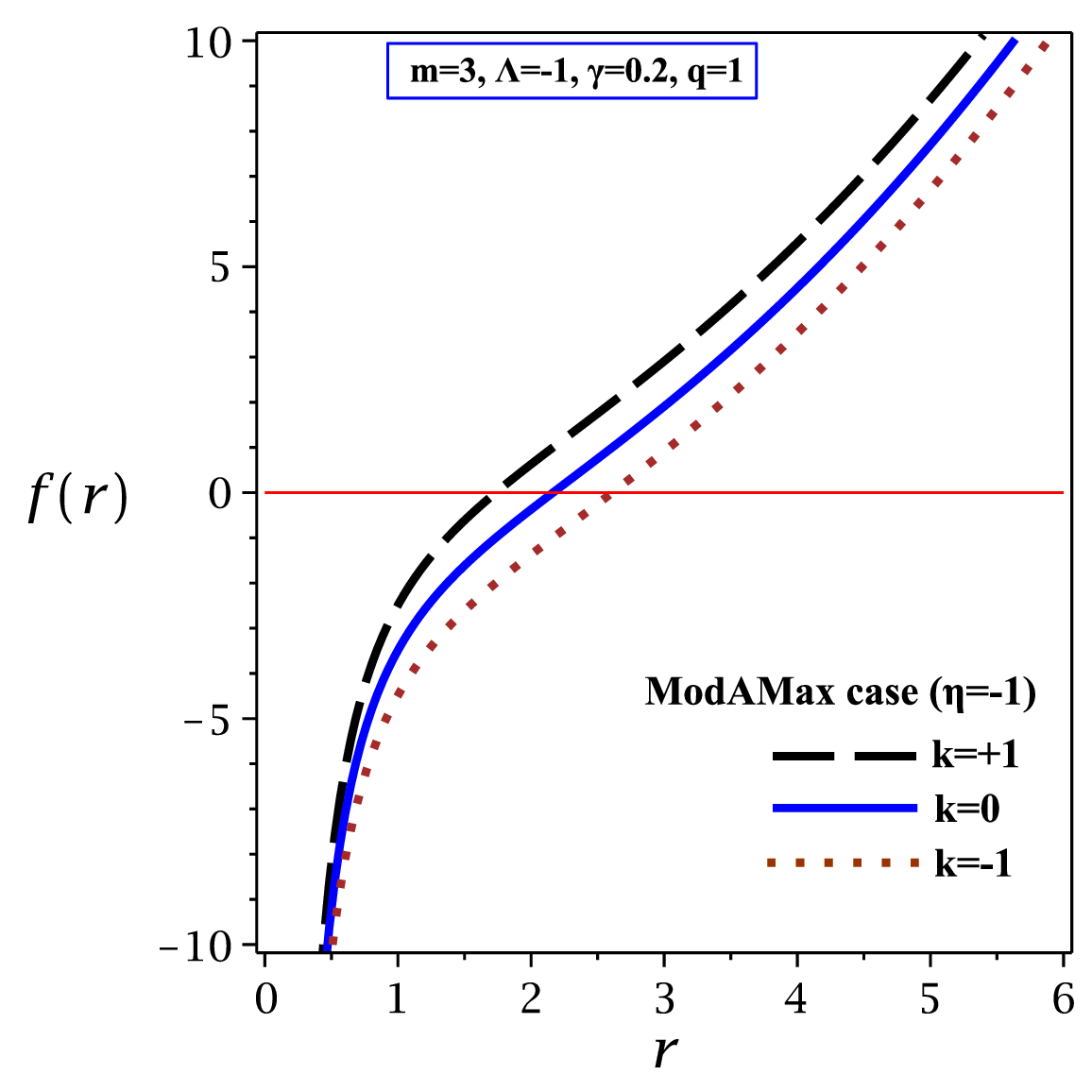} \newline
\caption{The metric function $f(r)$ versus $r$ for different values of the
topological constant ($k$). Left panel for ModMax case ($\protect\eta=+1$).
Right panel for ModAMax or phantom case ($\protect\eta=-1$).}
\label{Fig1}
\end{figure}

\begin{figure}[tbph]
\centering
\includegraphics[width=0.35\linewidth]{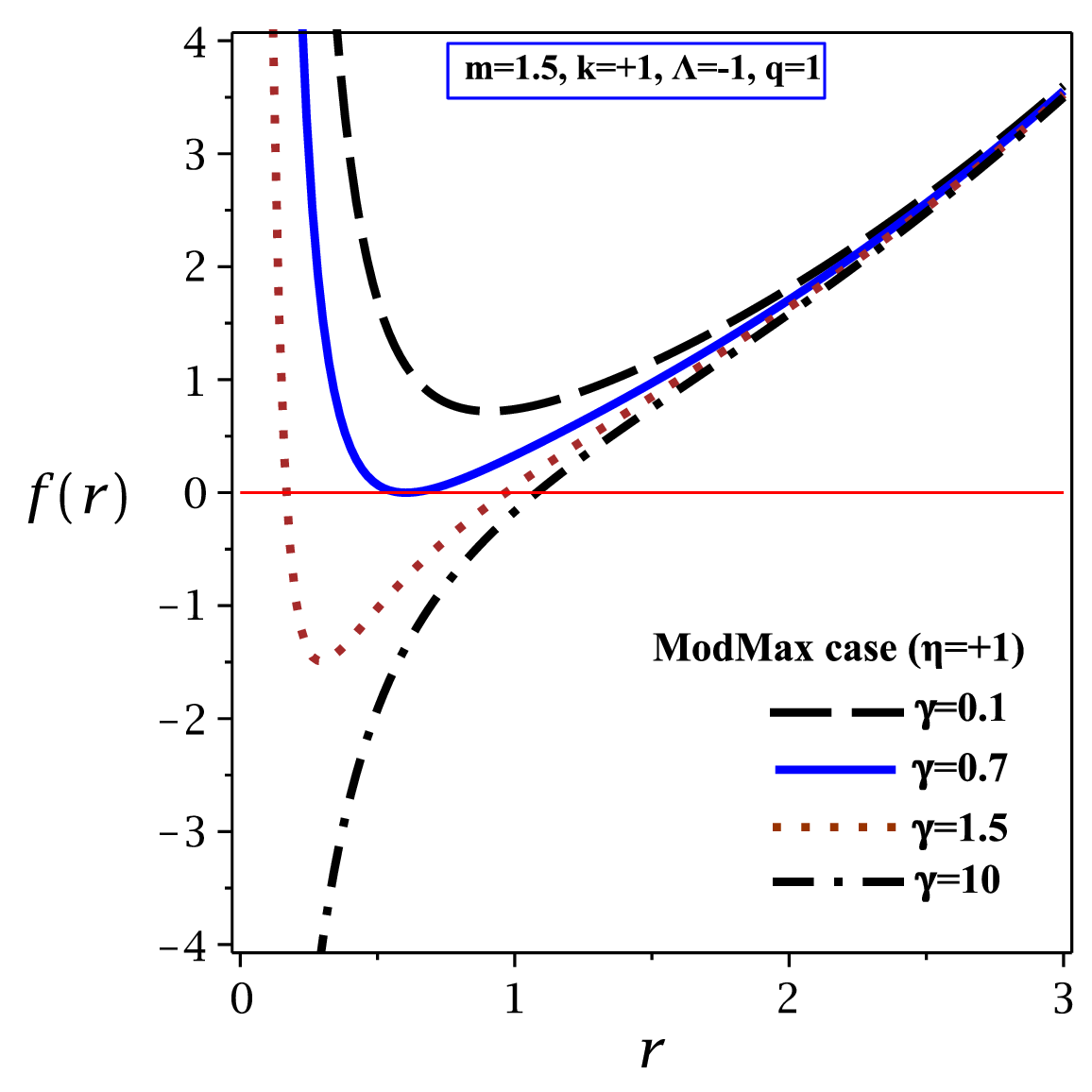} \includegraphics[width=0.35%
\linewidth]{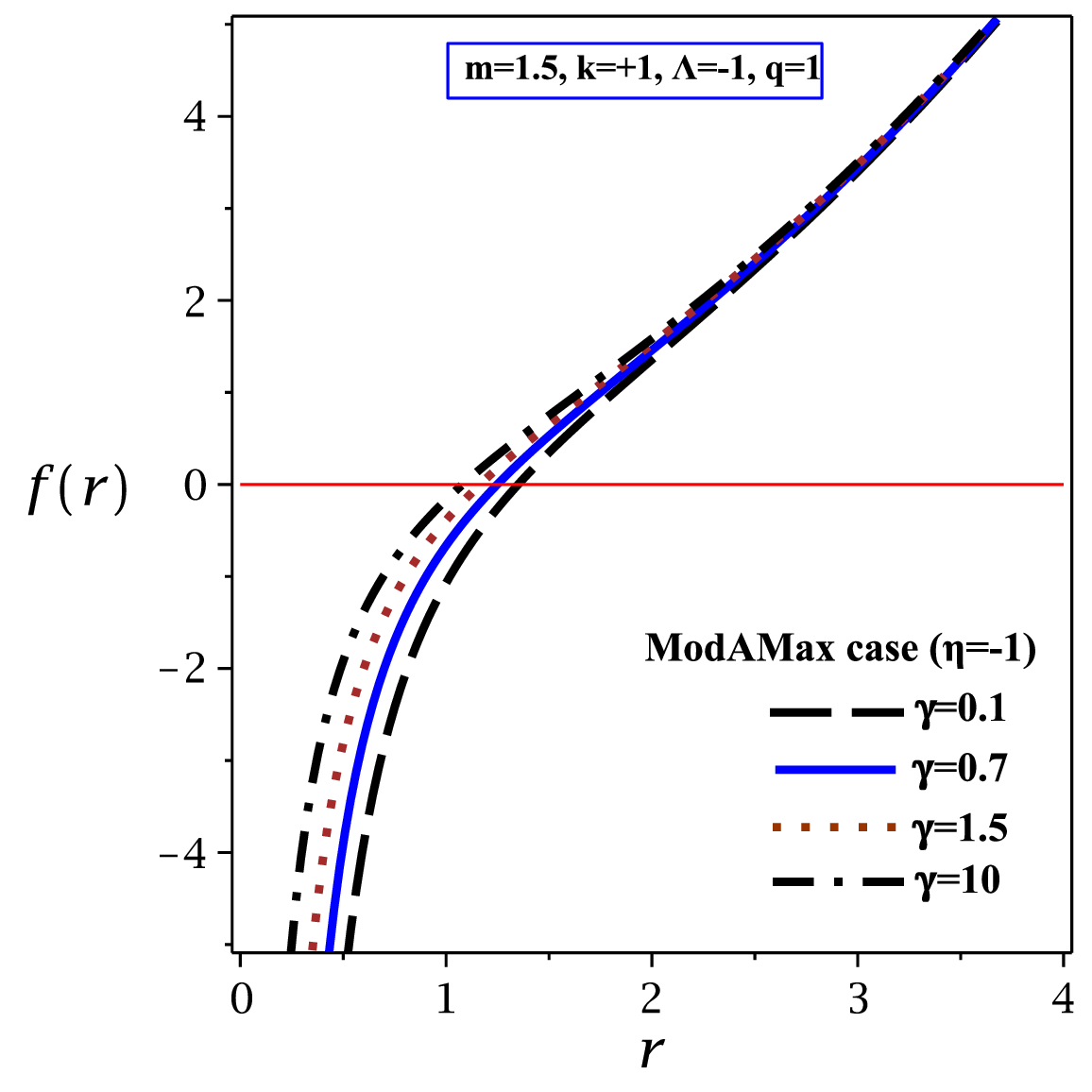} \newline
\caption{The metric function $f(r)$ versus $r$ for $k=+1$, and different
values of the ModMax's parameter. Left panel for ModMax case ($\protect\eta%
=+1$). Right panel for ModAMax or phantom case ($\protect\eta=-1$).}
\label{Fig2}
\end{figure}

\begin{figure}[tbph]
\centering
\includegraphics[width=0.35\linewidth]{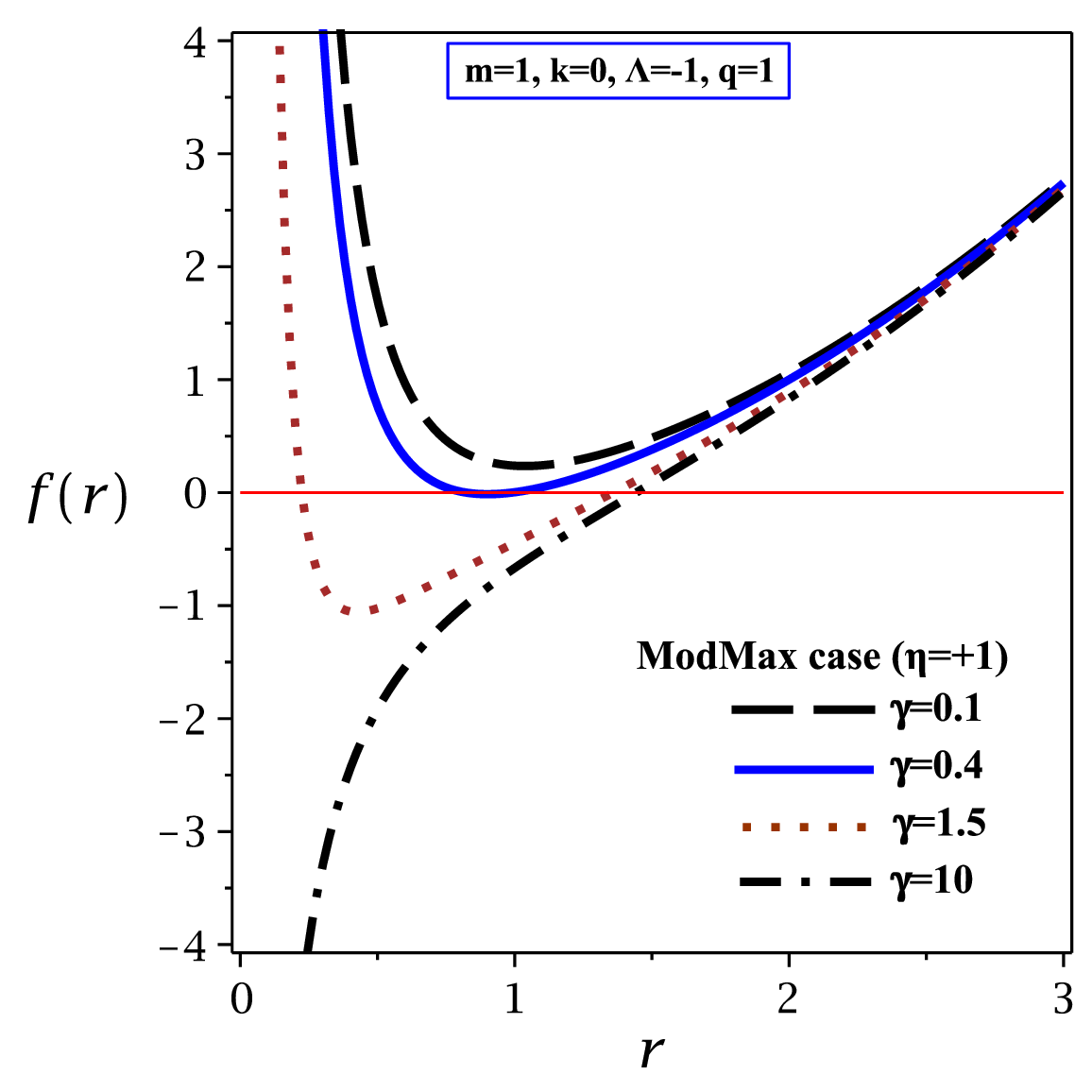} \includegraphics[width=0.35%
\linewidth]{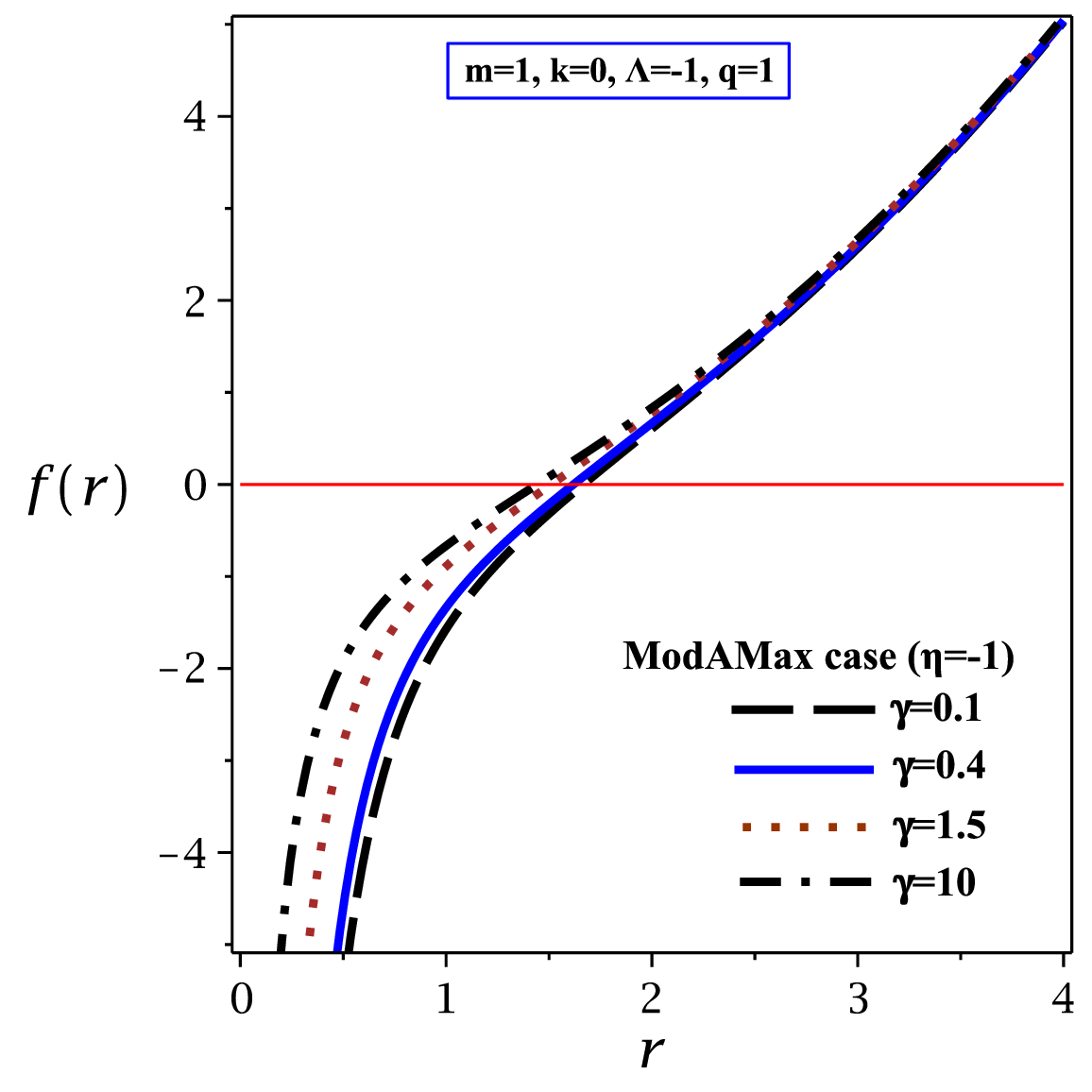} \newline
\caption{The metric function $f(r)$ versus $r$ for $k=0$, and different
values of the ModMax's parameter. Left panel for ModMax case ($\protect\eta%
=+1$). Right panel for ModAMax or phantom case ($\protect\eta=-1$).}
\label{Fig3}
\end{figure}

\begin{figure}[tbph]
\centering
\includegraphics[width=0.35\linewidth]{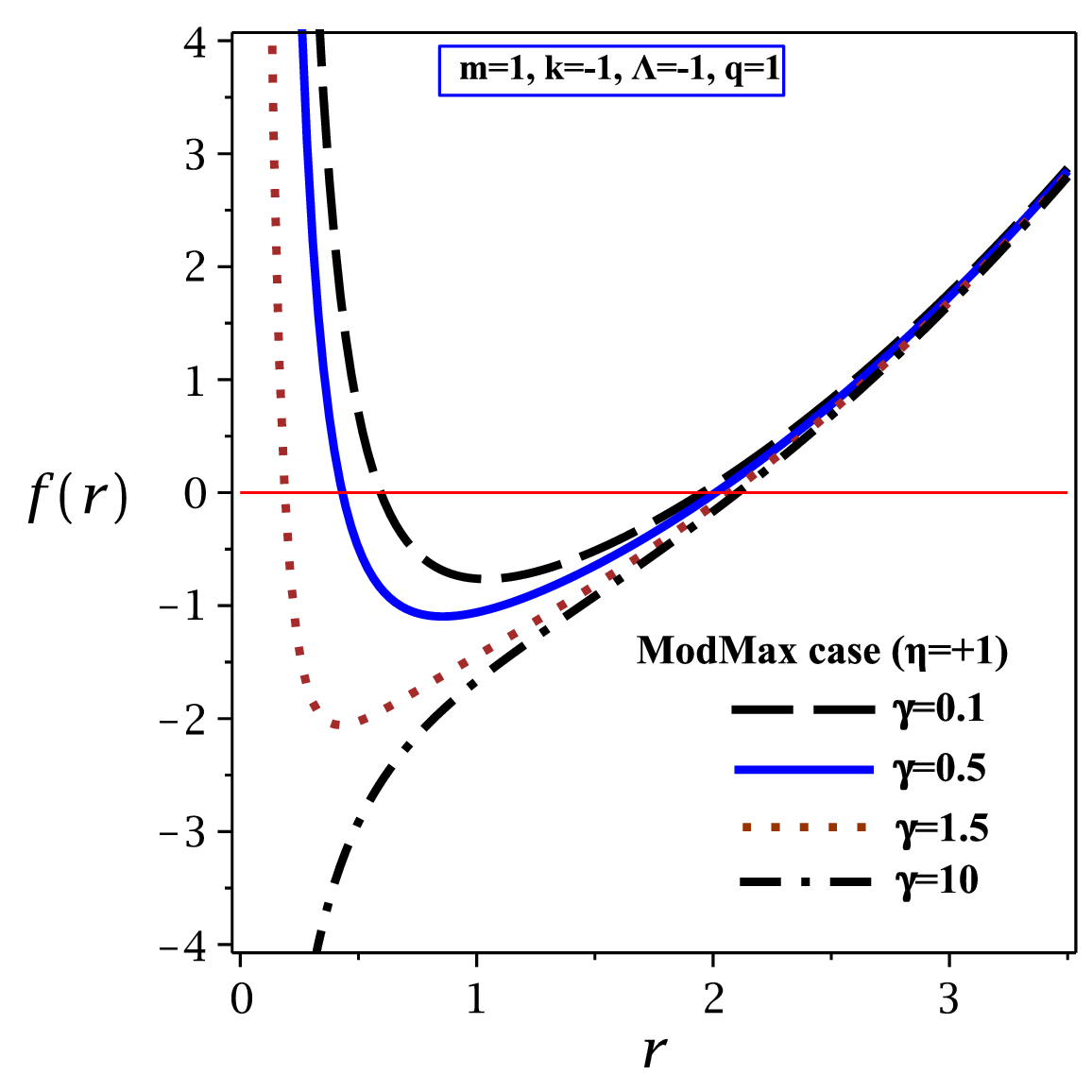} \includegraphics[width=0.35%
\linewidth]{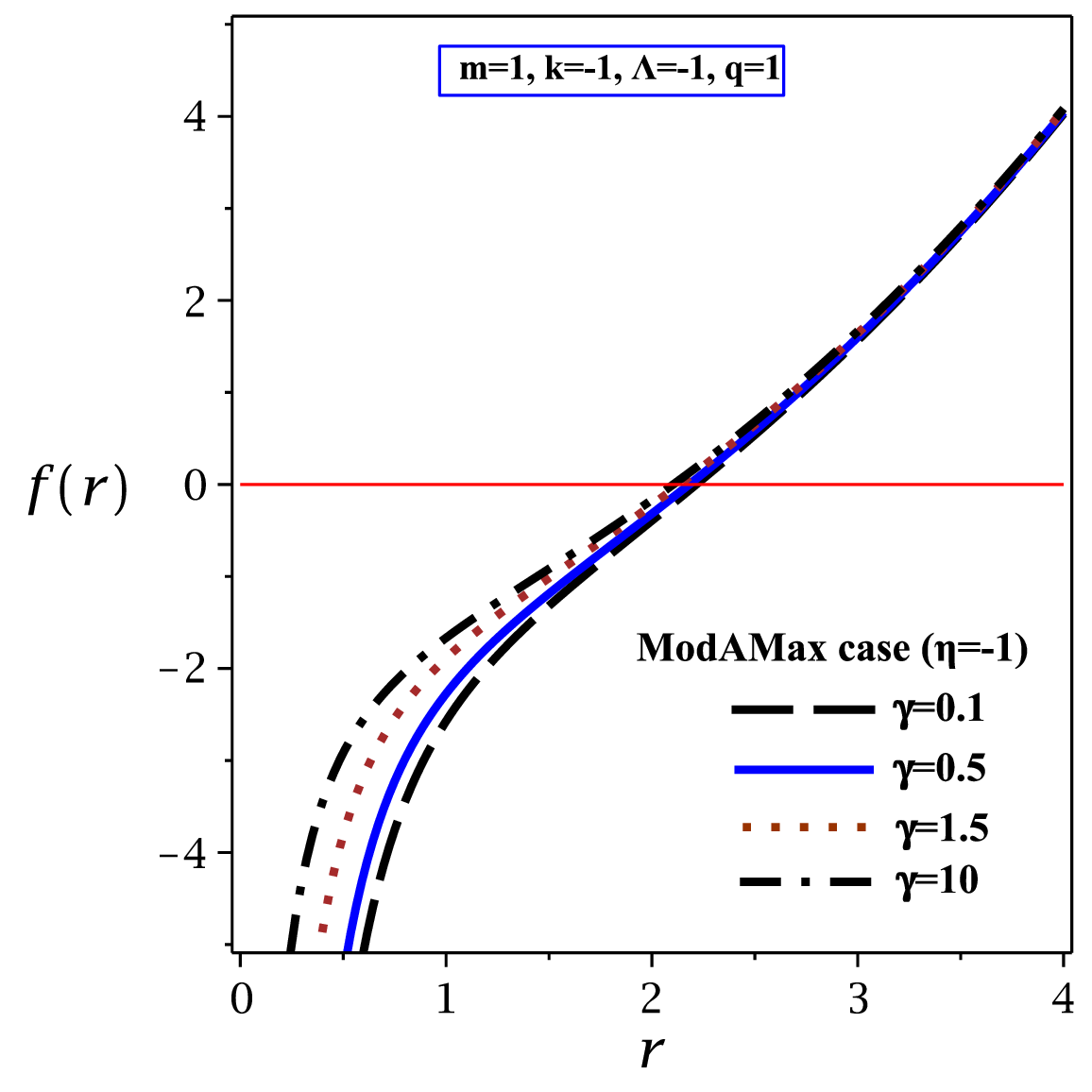} \newline
\caption{The metric function $f(r)$ versus $r$ for $k=-1$, and different
values of the ModMax's parameters. Left panel for ModMax case ($\protect\eta%
=+1$). Right panel for ModAMax or phantom case ($\protect\eta=-1$).}
\label{Fig4}
\end{figure}

We aim to study the effects of the Mod(A)Max parameters ($\gamma$ and $\eta$%
) and the topological constant $k$ on the derived metric function (Eq. \ref%
{f(r)}). To illustrate this, we plot $f\left( r\right) $ versus $r$ in Figs. %
\ref{Fig1} to \ref{Fig4}. Our findings are as follows:

\begin{description}
\item[\textbf{ModMax field ($\protect\eta=+1$):}] The effect of
topological constant reveals that smaller black holes are associated with a
positive topological constant ($k=+1$), while larger black holes correspond
to a negative topological constant ($k=-1$), see the left panel of Fig. \ref%
{Fig1}. We identify two roots, one corresponding to the inner horizon and
the other to the event horizon. In addition, the number of roots varies with
the parameter $\gamma$ (see the left panels in Figs. \ref{Fig2}-\ref{Fig4}).
Notably, for sufficiently large values of $\gamma$, only a single root
exists (see the dotted-dashed line in the left panels in Figs. \ref{Fig2} to %
\ref{Fig4}).

\item[\textbf{ModAMax field ($\protect\eta=-1$):}] The metric function
displays a single root related to the event horizon (see the right panels in
Figs. \ref{Fig1} to \ref{Fig4}). Results in the right panel of Fig. \ref%
{Fig1} indicate that the largest and smallest event horizons in the ModAMax
scenario correspond to $k=-1$ and $k=+1$, respectively, similar to the
ModMax case. Additionally, the event horizon of ModAMax black holes
decreases as $\gamma$ increases (see the right panels in Figs. \ref{Fig2} to %
\ref{Fig4}).
\end{description}

Our analysis reveals two distinct behaviors between ModMax and ModAMax AdS
black holes:

1) ModMax AdS black holes can have two roots, with the larger root
corresponding to the event horizon. In contrast, ModAMax AdS black holes
possess only one root, which is associated with the event horizon.

2) Increasing the parameter $\gamma$ results in an increase in the event
horizon of ModMax AdS black holes. Conversely, for ModAMax AdS black holes,
the event horizon decreases as $\gamma$ increases.


\section{Conserved and Thermodynamic Quantities: The First Law of
Thermodynamics}

To investigate the thermodynamic properties of these black holes, we first
calculate their Hawking temperature. This requires expressing the
geometrical mass $m$ in terms of the event horizon radius $r_{+}$, the
cosmological constant $\Lambda$, the electrical charge $q$, the ModMax
parameter $\gamma$, the phantom parameter $\eta$ (which includes the Maxwell
and anti-Maxwell parameters), and the topological constant $k$. By
considering the metric function given in Eq. (\ref{f(r)}), we solve the
equation $f(r)=0$ to determine the geometrical mass $m$, which leads to the following expression 
\begin{equation}
m=kr_{+}-\frac{\Lambda r_{+}^{3}}{3}+\frac{\eta q^{2}e^{-\gamma }}{r_{+}}%
=\left\{ 
\begin{array}{ccc}
kr_{+}-\frac{\Lambda r_{+}^{3}}{3}+\frac{q^{2}e^{-\gamma }}{r_{+}}, &  & 
\text{ModMax} \\ 
&  &  \\ 
kr_{+}-\frac{\Lambda r_{+}^{3}}{3}-\frac{q^{2}e^{-\gamma }}{r_{+}}, &  & 
\text{ModAMax}%
\end{array}%
\right. .  \label{mm}
\end{equation}

The Hawking temperature is defined as 
\begin{equation}
T=\frac{\kappa }{2\pi },  \label{TH1}
\end{equation}
where $\kappa $ is related to the surface gravity by 
\begin{equation}
\kappa =\sqrt{\frac{-1}{2}\left( \nabla _{\mu }\chi _{\nu }\right) \left(
\nabla ^{\mu }\chi ^{\nu}\right)}.  \label{kk}
\end{equation}

By using the metric (\ref{Metric}), the Killing vector $\chi =\partial _{t}$%
, and Eq. (\ref{kk}), we can extract the surface gravity which leads to 
\begin{equation}
\kappa =\frac{1}{2}\left. \frac{\partial f(r)}{\partial r}\right\vert
_{r=r_{+}},  \label{kk2}
\end{equation}%
so, by applying Eqs. (\ref{mm})-(\ref{kk2}), we can get the Hawking
temperature in the following form 
\begin{equation}
T=\frac{1}{4\pi }\left( \frac{k}{r_{+}}-\Lambda r_{+}-\frac{\eta
q^{2}e^{-\gamma }}{r_{+}^{3}}\right) =\left\{ 
\begin{array}{ccc}
\frac{1}{4\pi }\left( \frac{k}{r_{+}}-\Lambda r_{+}-\frac{q^{2}e^{-\gamma }}{%
r_{+}^{3}}\right) , &  & \text{ModMax} \\ 
&  &  \\ 
\frac{1}{4\pi }\left( \frac{k}{r_{+}}-\Lambda r_{+}+\frac{q^{2}e^{-\gamma }}{%
r_{+}^{3}}\right) , &  & \text{ModAMax}%
\end{array}%
\right. ,  \label{TemII}
\end{equation}%
where it depends on the $\Lambda $, $q$, $\gamma $, $\eta $, and $k$.

\begin{figure}[tbph]
\centering
\includegraphics[width=0.32\linewidth]{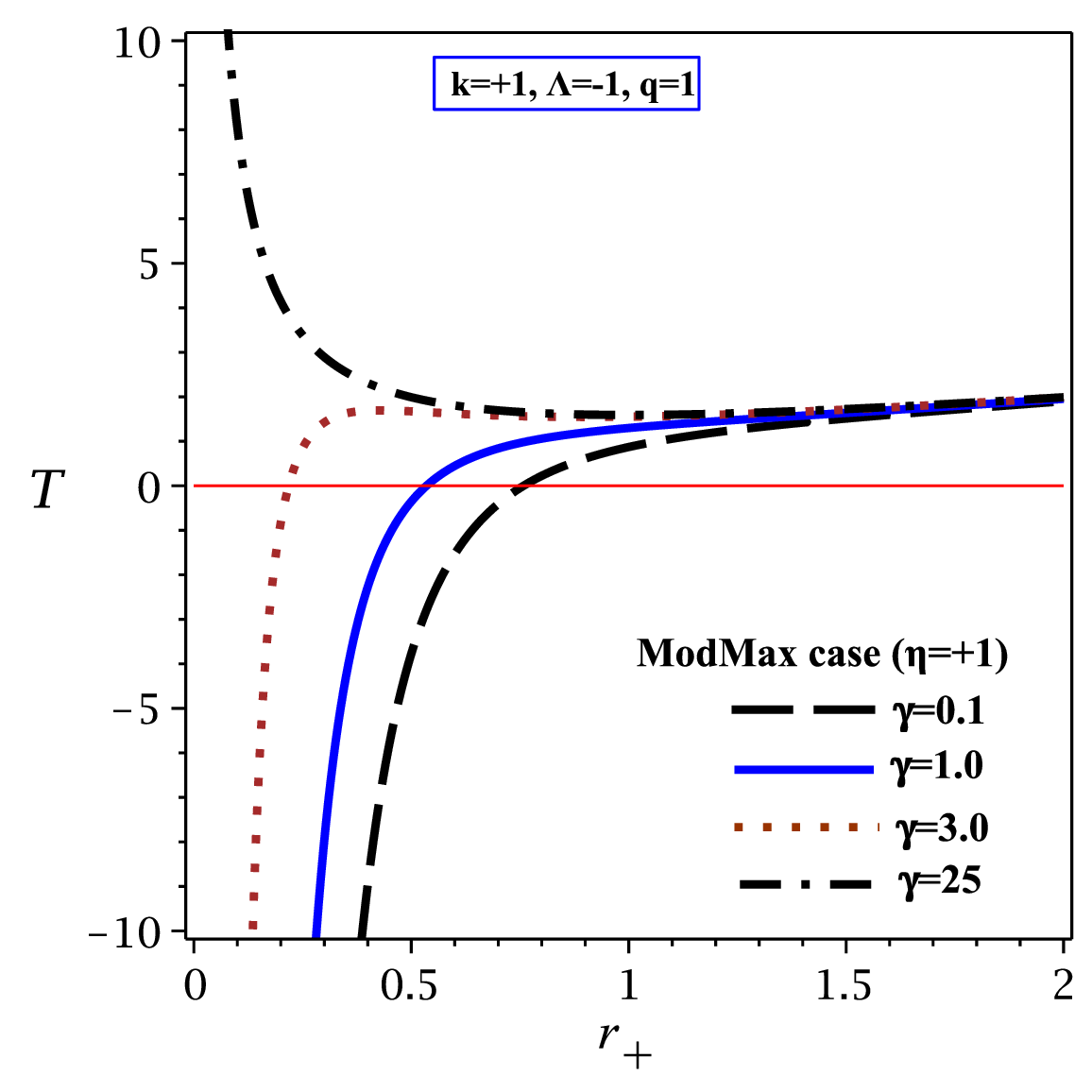} \includegraphics[width=0.32	%
\linewidth]{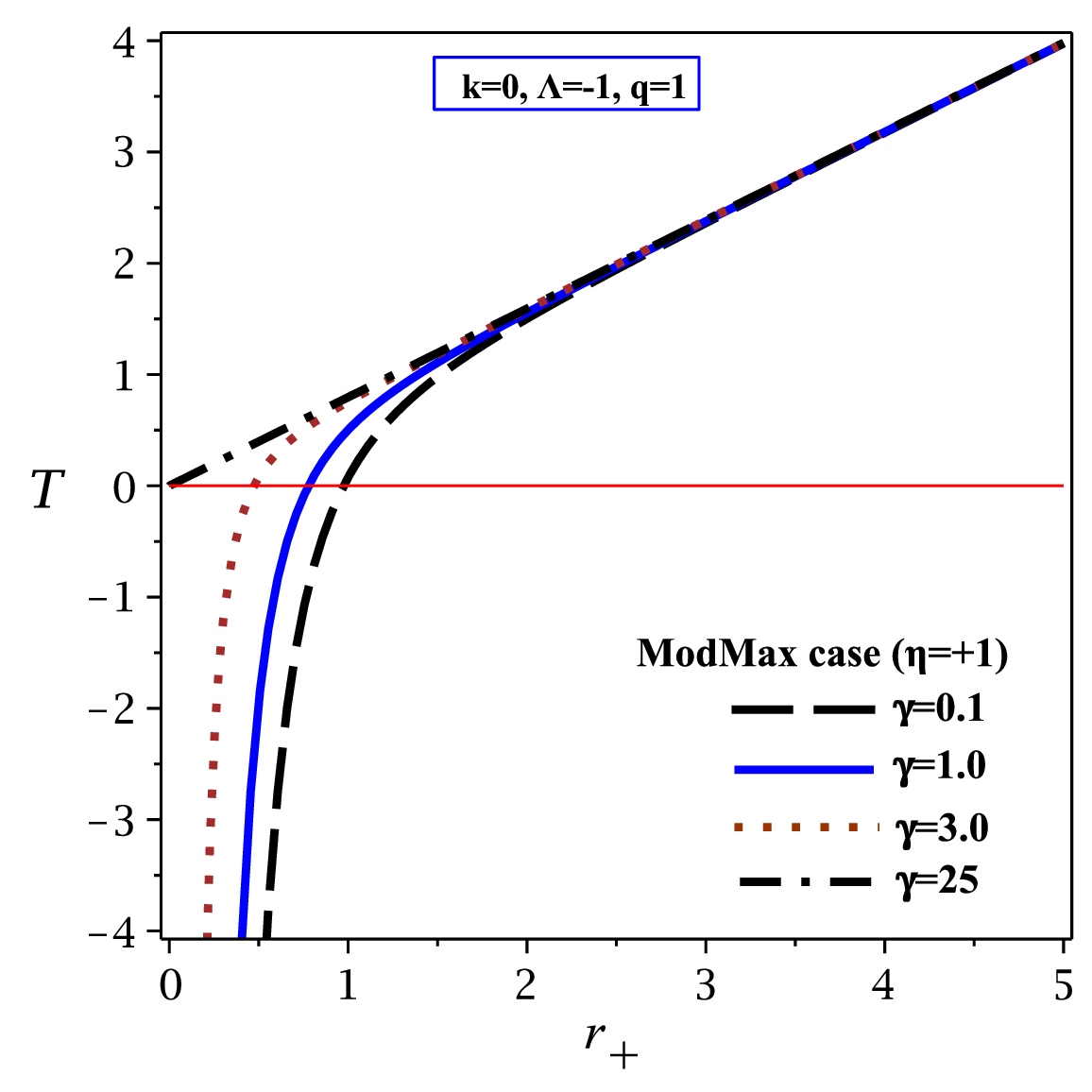} \includegraphics[width=0.32\linewidth]{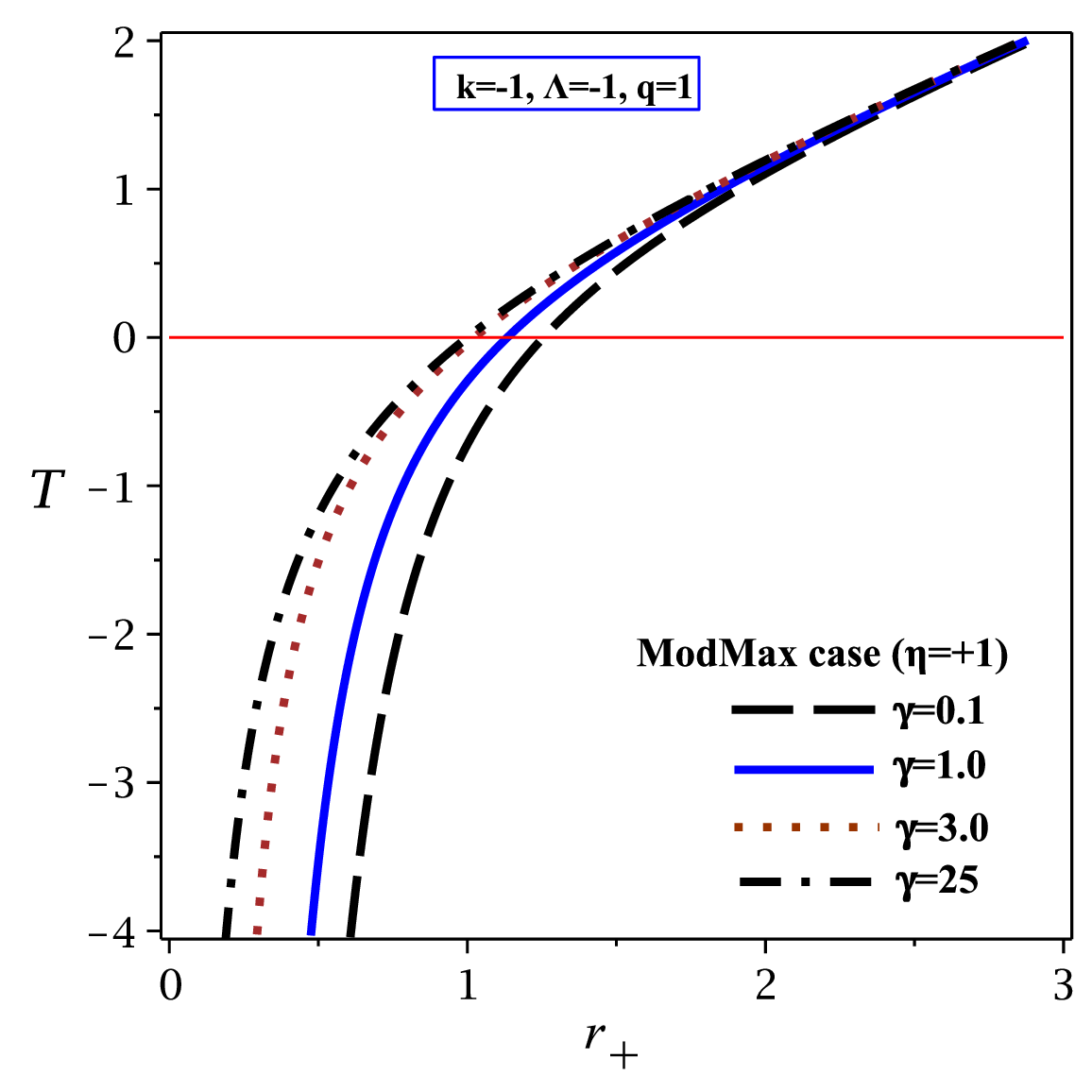} \newline
\caption{The Hawking temperature $T$ versus $r_{+}$ for $k=+1$ (left panel), 
$k=0$ (middle panel), and $k=-1$ (right panel) by considering the ModMax
field ($\protect\eta =+1$).}
\label{Fig5}
\end{figure}

For the Hawking temperature to be positive ($T>0$), the
system must obey the following condition 
\begin{equation}
kr_{+}^{2}-\Lambda r_{+}^{4}-\eta q^{2}e^{-\gamma }>0,
\end{equation}%
in the case of Mod(A)Max AdS black holes, the analysis gives rise to the
following two conditions 
\begin{equation}
\left\{ 
\begin{array}{ccc}
kr_{+}^{2}-\Lambda r_{+}^{4}-q^{2}e^{-\gamma }>0, &  & \text{ModMax} \\ 
&  &  \\ 
kr_{+}^{2}-\Lambda r_{+}^{4}+q^{2}e^{-\gamma }>0, &  & \text{ModAMax}%
\end{array}%
\right. ,
\end{equation}%
and for different values of the topological constant, the above conditions
in the ModMax case become 
\begin{equation}
T_{\text{ModMax}}>0~\Longrightarrow \left\{ 
\begin{array}{ccc}
r_{+}^{2}-\Lambda r_{+}^{4}-q^{2}e^{-\gamma }>0 &  & k=+1 \\ 
&  &  \\ 
-\Lambda r_{+}^{4}-q^{2}e^{-\gamma }>0 &  & k=0 \\ 
&  &  \\ 
-r_{+}^{2}-\Lambda r_{+}^{4}-q^{2}e^{-\gamma }>0 &  & k=-1%
\end{array}%
\right. ,
\end{equation}%
in which $T_{\text{ModMax}}$ represents the Hawking temperature associated
with ModMax black holes. Since the cosmological constant is negative ($%
\Lambda <0$) for AdS spacetimes, the Hawking temperature of ModMax AdS black
holes is determined by the topological constant ($k$), the electric charge ($%
q$), and the ModMax parameter ($\gamma $). The interplay among these
parameters results in three qualitatively different thermal behaviors: 

i) For $k=+1$, $T_{\text{ModMax}}$ is positive when $r_{+}^{2}\left(
1-\Lambda r_{+}^{2}\right) >q^{2}e^{-\gamma }$. It is notable that for the
large value of $\gamma $ (or $q=0$), $T_{\text{ModMax}}$ is always positive
(see dotted-dashed line in the left panel of Fig. \ref{Fig5}).

ii) For the case where $k=0$, the Hawking temperature of ModMax AdS black
holes remains positive under the condition $-\Lambda
r_{+}^{4}>q^{2}e^{-\gamma }$. Furthermore, $T_{\text{ModMax}}$ is always
positive when $q=0$ (or as $\gamma \rightarrow $ very large values); this is
illustrated by the dotted-dashed lines in the middle panel of Figure. \ref%
{Fig5}.

iii) For the case $k=-1$, maintaining a positive Hawking temperature ($T_{%
\text{ModMax}}>0$) requires the condition $-\Lambda r_{+}^{2}>1+\frac{%
3q^{2}e^{-\gamma }}{r_{+}^{2}}$.

We present Fig. \ref{Fig5} to further examine the effects of the ModMax
parameter and the topological constant on the temperature of ModMax AdS
black holes. Our findings are as follows:

1) The left panel of Fig. \ref{Fig5}: For $k=+1$, there is a critical value
for the ModMax parameter, denoted as $\gamma_{\text{critical}}$. When $%
\gamma > \gamma_{\text{critical}}$, the temperature of ModMax AdS black
holes remains positive, as illustrated by the dashed-dotted line in the left
panel of Fig. \ref{Fig5}. Conversely, for $\gamma < \gamma_{\text{critical}}$%
, there exists one root for the temperature in the following form 
\begin{equation}
r_{+_{T=0}}=\sqrt{\frac{-1+\sqrt{1-4\Lambda q^{2}e^{-\gamma }}}{-2\Lambda }},
\label{Tokk}
\end{equation}%
which the temperature is negative when $r_{+} < r_{+_{T=0}}$ and positive
when $r_{+} > r_{+_{T=0}}$, as shown by the dotted, continuous, and dashed
lines in the left panel of Fig. \ref{Fig5}. This indicates that large ModMax
AdS black holes possess positive temperatures, while small ModMax AdS black
holes have negative temperatures. Additionally, as the value of $\gamma$
increases, the range of positive temperatures also expands.

2) Middle panel in Fig. \ref{Fig5}: For $k=0$, there is one real root that
depends on the cosmological constant, the electric charge, and the ModMax
parameter, expressed as follows 
\begin{equation}
r_{+_{T=0}}=\sqrt{\frac{\sqrt{-4\Lambda q^{2}e^{-\gamma }}}{-2\Lambda }},
\label{T0k0}
\end{equation}%
where the root decreases as $\gamma $ increases and $q$ decreases. The
Hawking temperature is negative for $r_{+}<r_{+_{T=0}}$ (i.e., in this
range, $T$ is negative), while for $r_{+}>r_{+_{T=0}}$, the temperature of
ModMax AdS black holes becomes positive (see the middle panel in Fig. \ref%
{Fig5}). This indicates that small ModMax AdS black holes exhibit negative
temperatures, whereas larger ones have positive temperatures. According to
Eq. (\ref{T0k0}), the range of positive temperatures increases as $\gamma $
increases or $q$ decreases. In essence, the electric charge counteracts the
effects of the ModMax parameter. Notably, no real root exists when $\gamma $
takes on very large values or when $q=0$ (indicated by the dotted-dashed
line in the middle panel of Figure. \ref{Fig5}). An interesting
effect of the ModMax parameter on the Hawking temperature of these black
holes is observed: for very large values of $\gamma $ (or in the absence of $%
q$), the Hawking temperature vanishes as $r_{+}\rightarrow 0$ (i.e., $%
\lim_{r_{+}\longrightarrow 0}T\longrightarrow 0$).

3) Right panel in Fig. \ref{Fig5}: For $k=-1$, there is one real root, which
exists in the following form 
\begin{equation}
r_{+_{T=0}}=\sqrt{\frac{1+\sqrt{1-4\Lambda q^{2}e^{-\gamma }}}{-2\Lambda }},
\end{equation}%
and it depends on $\Lambda$, $q$, and $\gamma$. The real root decreases as $%
\gamma$ increases (or as $q$ decreases). The temperature is negative when $%
r_{+} < r_{+_{T=0}}$ and positive when $r_{+} > r_{+_{T=0}}$. Thus, the
temperatures of the small and large ModMax AdS black holes are negative and
positive, respectively.

\begin{figure}[tbph]
\centering
\includegraphics[width=0.32\linewidth]{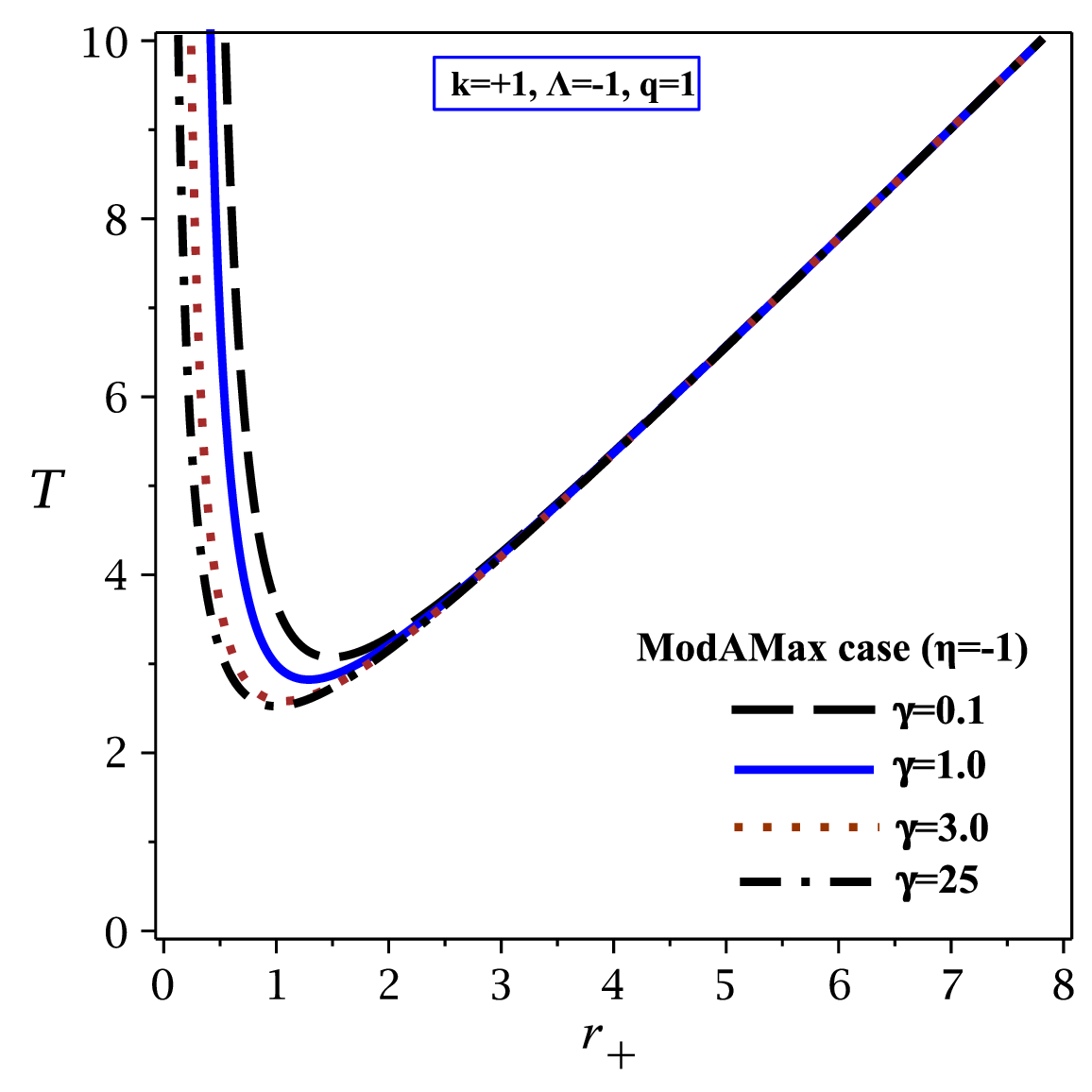} \includegraphics[width=0.32%
\linewidth]{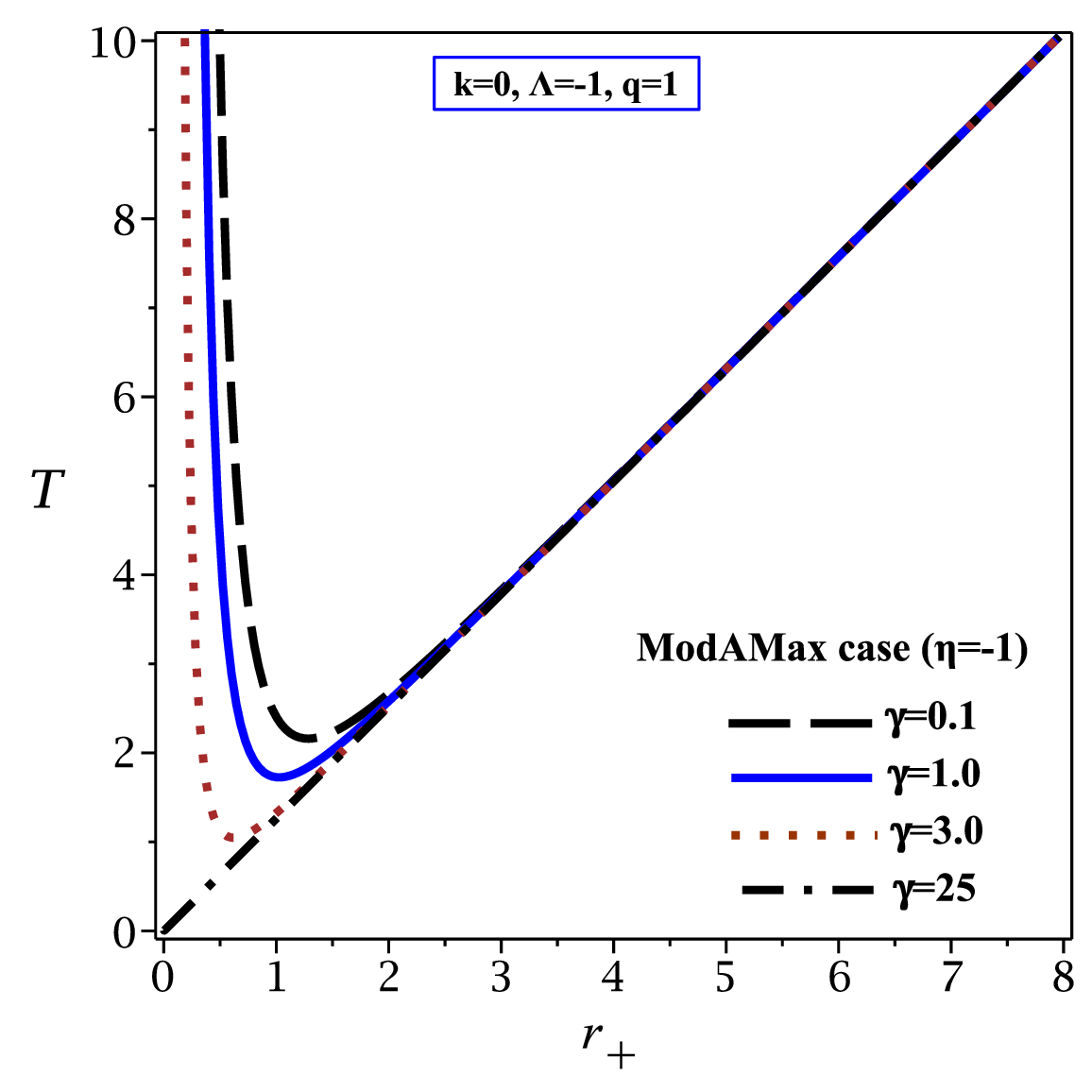} \includegraphics[width=0.32\linewidth]{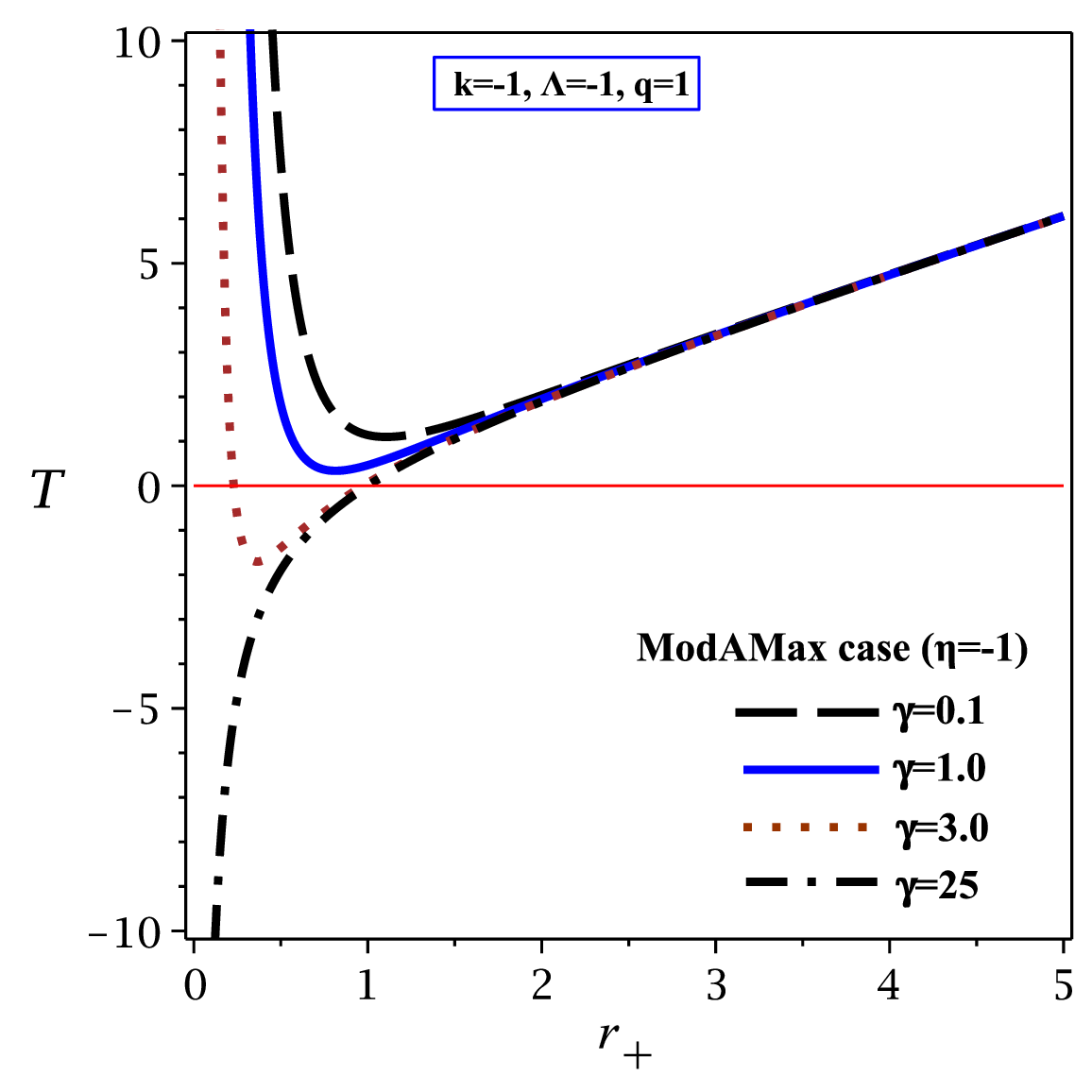} \newline
\caption{The Hawking temperature $T$ versus $r_{+}$ for $k=+1$ (left panel), 
$k=0$ (middle panel), and $k=-1$ (right panel) by considering the ModAMax field ($\protect\eta =-1$).}
\label{Fig6}
\end{figure}

For the Hawking temperature of ModAMax AdS black holes to be
positive, we find that 
\begin{equation}
T_{\text{ModAMax}}>0~\Longrightarrow \left\{ 
\begin{array}{ccc}
r_{+}^{2}-\Lambda r_{+}^{4}+q^{2}e^{-\gamma }>0 &  & k=+1 \\ 
&  &  \\ 
-\Lambda r_{+}^{4}+q^{2}e^{-\gamma }>0 &  & k=0 \\ 
&  &  \\ 
-r_{+}^{2}-\Lambda r_{+}^{4}+q^{2}e^{-\gamma }>0 &  & k=-1%
\end{array}%
\right. ,
\end{equation}%
where $T_{\text{ModAMax}}$ denotes the Hawking temperature of ModAMax black
holes. The Hawking temperature of ModAMax AdS black holes is a function of $%
k $, $q$, and $\gamma $, leading to three distinct thermodynamic behaviors: 

i) For $k=+1$: $T_{\text{ModAMax}}$ is always positive (as shown in the left
panel of Figure. \ref{Fig6}).

ii) For $k=0$: Similar to the $k=+1$ case, the Hawking temperature of
ModAMax AdS black holes is consistently positive (see the middle panel of
Figure. \ref{Fig6}).

iii) For $k=-1$: To ensure a positive Hawking temperature ($T_{\text{ModAMax}%
}>0$), the following condition must be met: $-\Lambda r_{+}^{2}+\frac{%
q^{2}e^{-\gamma }}{r_{+}^{2}}>1$. This condition explicitly reveals that the
Hawking temperature of ModAMax AdS black holes can become negative when $k=-1
$ (illustrated in the right panel of Figure. \ref{Fig6}).

We plot Fig. \ref{Fig6} to further examine the effects of the ModMax
parameter and the topological constant on the temperature of ModAMax AdS
black holes. Our results are as follows:

1) The left panel in Fig. \ref{Fig6} shows that for $k=+1$, the
Hawking temperature of ModAMax AdS black holes is consistently positive.
Varying the ModMax parameter has a minimal effect on the temperature of
small black holes.

2) The middle panel in Fig. \ref{Fig6} illustrates that for $k=0$,
no roots exist. As the radius decreases, the Hawking temperature initially
decreases and reaches a minimum value that depends on $\gamma$. Beyond this
point, the temperature rises sharply. Moreover, for very large values of $%
\gamma$ (or when $q=0$), the temperature approaches zero as $%
r_{+}\rightarrow 0$ (i.e., $\lim_{r_{+}\longrightarrow 0}T\longrightarrow 0$%
).

3) The right panel in Fig. \ref{Fig6} shows that for $k=-1$, there is one
root for the temperature, expressed in the following form 
\begin{equation}
r_{+_{T=0}}=\sqrt{\frac{2+2\sqrt{1+4\Lambda q^{2}e^{-\gamma }}}{-\Lambda }},
\end{equation}%
provided $q^{2}e^{-\gamma }>\frac{-1}{4\Lambda }$. This root increases as $%
\gamma$ decreases (or $q$ increases). The temperature is negative for $r_{+}
< r_{+_{T=0}}$ and positive for $r_{+} > r_{+_{T=0}}$. Consequently,
the temperature of the small ModAMax AdS black hole can attain both negative
and positive values, a behavior dependent on $\gamma$.

Using Gauss's law, we can express the electric charge of a black hole per
unit volume ($\mathcal{V}$) in the following way 
\begin{equation}
Q=\frac{\widetilde{Q}}{\mathcal{V}}=\frac{F_{tr}}{4\pi }\int_{0}^{2\pi
}\int_{0}^{\pi }\sqrt{g_{k}}d\theta d\varphi =\frac{q}{4\pi }.  \label{Q}
\end{equation}%
where $F_{tr}=\frac{q}{r^{2}}$, and for case $t=constant$ and $r=constant$,
the determinant of metric tensor $g_{k}$ is $r^{4}\det \left( d\Omega
_{k}^{2}\right) $ (i.e., $g_{k}=\det \left( g_{k}\right) =r^{4}\det \left(
d\Omega _{k}^{2}\right) $). Furthermore, $\mathcal{V}=\int_{0}^{2\pi
}\int_{0}^{\pi }\sqrt{\det \left( d\Omega _{k}^{2}\right) }d\theta d\varphi $%
, where is the area of a unit volume of constant ($t$, $r$) space. For
example, $\mathcal{V}$ is $4\pi $ for $k=1$.

The electric potential at the event horizon ($U$), relative to the reference
point at infinity ($r \rightarrow \infty$), is given by 
\begin{equation}
U=A_{\mu }\chi ^{\mu }\left\vert _{r\rightarrow \infty }\right. -A_{\mu
}\chi ^{\mu }\left\vert _{r\rightarrow r_{+}}\right. =\frac{qe^{-\gamma }}{%
r_{+}},  \label{elcpo}
\end{equation}%
where the gauge potential is zero when $r\rightarrow \infty $.

Applying the area law, we can get the entropy of the topological Mod(A)Max
black holes per unit volume ($\mathcal{V}$), which leads to 
\begin{equation}
S=\frac{\widetilde{S}}{\mathcal{V}}=\frac{\mathcal{A}}{4\mathcal{V}}=\frac{%
\left. \int_{0}^{2\pi }\int_{0}^{\pi }\sqrt{g_{\theta \theta }g_{\varphi
\varphi }}\right\vert _{r=r_{+}}}{4}=\frac{r_{+}^{2}}{4},  \label{S}
\end{equation}%
where $\mathcal{A}$ is the horizon area.

The Ashtekar-Magnon-Das (AMD) approach \cite{AMD1,AMD2} for
four-dimensional spacetime is 
\begin{equation}
\widetilde{M}=\frac{-\ell }{8\pi }\oint_{S}^{{}}\hat{E}_{t}^{t}dS,
\label{AMD1}
\end{equation}%
where $\ell ^{2}=-3/\Lambda $ is AdS radius. Generally, the quantity $\hat{E}%
_{~\nu }^{\mu }=\Omega ^{-1}C_{~\rho \nu \sigma }^{\mu }n^{\rho }n^{\sigma }$
is defined as the electric part of the Weyl tensor with respect to the unit
normal $n^{\rho }$ at conformal infinity. Here, $\Omega $ is the conformal
factor bringing the AdS boundary to a finite coordinate location ($\Omega =%
\frac{1}{r}$ for AdS). Also, $C_{~~\rho \nu \sigma }^{\mu }$ denotes the
Weyl tensor. Then, $\hat{E}_{~t}^{t}=\Omega ^{-1}C_{~\rho t\sigma
}^{t}n^{\rho }n^{\sigma }$, which, for the given spacetime (\ref{Metric}),
reduces to $\hat{E}_{~t}^{t}=\Omega ^{-1}C_{~rtr}^{t}n^{r}n^{r}\propto
\Omega ^{-1}C_{~rtr}^{t}$. Now, the Weyl tensor component $C_{~rtr}^{t}$ for
these black holes is found to be 
\begin{equation}
C_{~rtr}^{t}\sim \frac{-m}{r^{3}}+\frac{\eta q^{2}e^{-\gamma }}{r^{4}}%
+O\left( \frac{1}{r^{5}}\right) ,  \label{Weyl}
\end{equation}%
so that $\hat{E}_{~~t}^{t}$ takes the following asymptotic form 
\begin{equation}
\hat{E}_{~~t}^{t}=\frac{-m}{r^{2}}+\frac{\eta q^{2}e^{-\gamma }}{r^{3}}.
\label{partADM}
\end{equation}

Substituting Eqs. (\ref{Weyl}), (\ref{partADM}), and $dS=r^{2}d\Omega
_{k}^{2}$ into the AMD intergral (\ref{AMD1}), we obtain 
\begin{equation}
\widetilde{M}=\frac{-\ell }{8\pi }\underset{r\rightarrow \infty }{\lim }%
\oint_{S}^{{}}\left( \frac{-m}{r^{2}}+O(\frac{1}{r^{3}})\right) dS=\frac{%
-\ell }{8\pi }\left( r^{2}\mathcal{V}\right) \left( \frac{-m}{r^{2}}\right) =%
\frac{m\mathcal{V}}{8\pi }.  \label{AAMD}
\end{equation}

Hence, the total mass of the topological Mod(A)Max black holes per unit
volume $\mathcal{V}$, is given by 
\begin{equation}
M=\frac{\widetilde{M}}{\mathcal{V}}=\frac{m}{8\pi }=\frac{1}{8\pi }\left(
kr_{+}-\frac{\Lambda r_{+}^{3}}{3}+\frac{\eta q^{2}e^{-\gamma }}{r_{+}}%
\right) =\left\{ 
\begin{array}{ccc}
\frac{1}{8\pi }\left( kr_{+}-\frac{\Lambda r_{+}^{3}}{3}+\frac{%
q^{2}e^{-\gamma }}{r_{+}}\right) , &  & \text{ModMax} \\ 
&  &  \\ 
\frac{1}{8\pi }\left( kr_{+}-\frac{\Lambda r_{+}^{3}}{3}-\frac{%
q^{2}e^{-\gamma }}{r_{+}}\right) , &  & \text{ModAMax}%
\end{array}%
\right. ,  \label{MM}
\end{equation}%
in the above equation, we use from the geometrical mass (\ref{mm}).

\begin{figure}[tbph]
\centering
\includegraphics[width=0.325\linewidth]{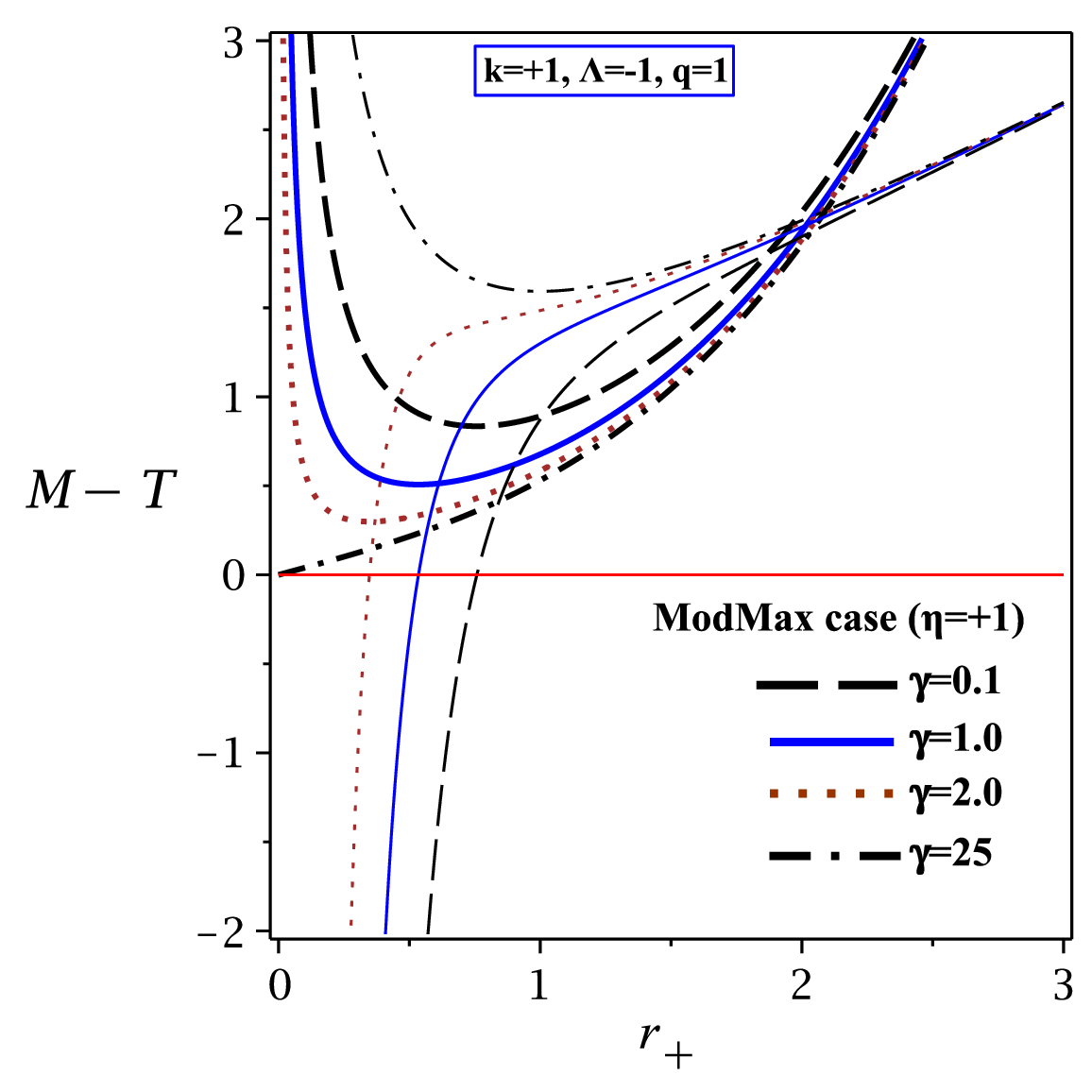} %
\includegraphics[width=0.325\linewidth]{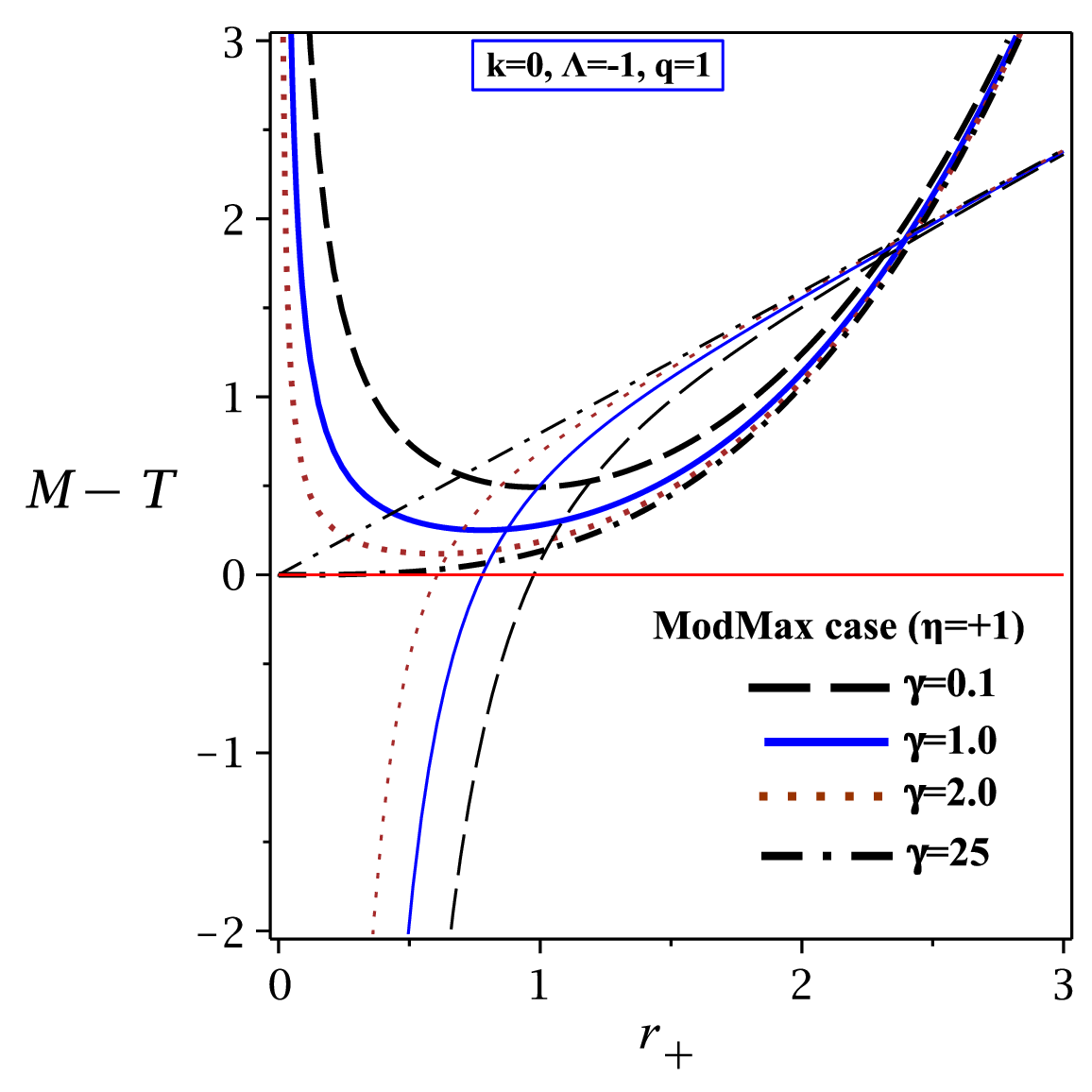} %
\includegraphics[width=0.325\linewidth]{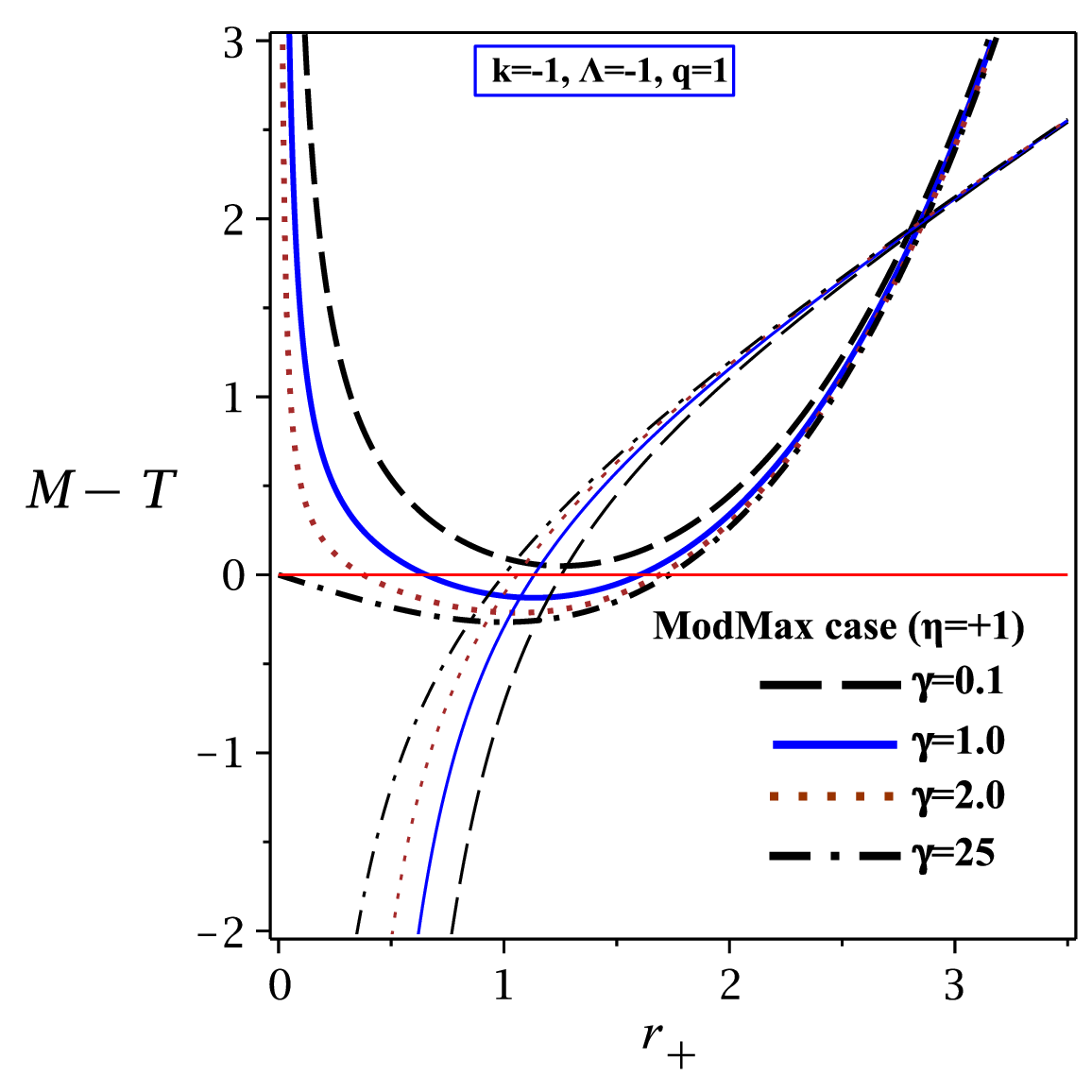} \newline
\caption{The total mass $M$ (thick lines) and the Hawking
temperature $T$ (thin lines) versus $r_{+}$ for $k=+1$ (left panel), $k=0$
(middle panel), and $k=-1$ (right panel) by considering the ModMax field ($%
\protect\eta =+1$).}
\label{Fig7}
\end{figure}

To ensure that the total mass remains positive, the following
constraint must be satisfied 
\begin{equation}
M>0~\Longrightarrow ~3kr_{+}^{2}-\Lambda r_{+}^{4}+3\eta q^{2}e^{-\gamma }>0,
\end{equation}%
where, for Mod(A)Max AdS black holes, this leads to the following two
conditions 
\begin{equation}
\left\{ 
\begin{array}{ccc}
3kr_{+}^{2}-\Lambda r_{+}^{4}+3q^{2}e^{-\gamma }>0, &  & \text{ModMax} \\ 
&  &  \\ 
3kr_{+}^{2}-\Lambda r_{+}^{4}-3q^{2}e^{-\gamma }>0, &  & \text{ModAMax}%
\end{array}%
\right. .
\end{equation}%
Moreover, for different values of the topological constant, the above
conditions in the ModMax case reduce to 
\begin{equation}
M_{\text{ModMax}}>0~\Longrightarrow \left\{ 
\begin{array}{ccc}
3r_{+}^{2}-\Lambda r_{+}^{4}+3q^{2}e^{-\gamma }>0 &  & k=+1 \\ 
&  &  \\ 
-\Lambda r_{+}^{4}+3q^{2}e^{-\gamma }>0 &  & k=0 \\ 
&  &  \\ 
-3r_{+}^{2}-\Lambda r_{+}^{4}+3q^{2}e^{-\gamma }>0 &  & k=-1%
\end{array}%
\right. ,  \label{MModMax}
\end{equation}%
where $M_{\text{ModMax}}$ denotes the total mass of ModMax black holes. On
the other hand, since the cosmological constant is negative ($\Lambda <0$)
for AdS black holes, the total mass of ModMax AdS black holes depends on the
topological constant ($k$), the electric charge ($q$), and the ModMax
parameter ($\gamma $), leading to three distinct behaviors: 

i) For the case $k=+1$ (where $3r_{+}^{2}-\Lambda r_{+}^{4}+3q^{2}e^{-\gamma
}>0$), $M_{\text{ModMax}}$ is always positive across all radii (as shown by
the thick lines in the left panel of Fig. \ref{Fig7}). Notably, $\underset{%
r_{+}\rightarrow 0}{\lim }M_{\text{ModMax}}\rightarrow 0$ when $q=0$ (or
equivalently, when $\gamma \rightarrow $ very large value), which is
illustrated by the thick, dotted-dashed line in the left panel of Fig. \ref%
{Fig7}.

ii) For the case $k=0$ (under the condition $-\Lambda
r_{+}^{4}+3q^{2}e^{-\gamma }>0$), similar to the preceding case, $M_{\text{%
ModMax}}$ is consistently positive across all radii (as shown by the thick
lines in the middle panel of Fig. \ref{Fig7}). Notably, the total mass of
ModMax AdS black holes is zero at $r_{+}=0$, meaning $\underset{%
r_{+}\rightarrow 0}{\lim }M_{\text{ModMax}}\rightarrow 0$ when $q=0$ (or
when $\gamma \rightarrow $ very large value), which is illustrated by the
thick, dotted-dashed line in the middle panel of Fig. \ref{Fig7}.

iii) For the case $k=-1$, ensuring a positive value for $M_{\text{ModMax}}$
requires the satisfaction of the condition $\frac{-\Lambda r_{+}^{2}}{3}+%
\frac{q^{2}e^{-\gamma }}{r_{+}^{2}}>1$. This condition implies that the
total mass of ModMax AdS black holes cannot be positive at any radius (as
demonstrated by the thick lines in the right panel of Fig. \ref{Fig7}).

\begin{figure}[tbph]
\centering
\includegraphics[width=0.325\linewidth]{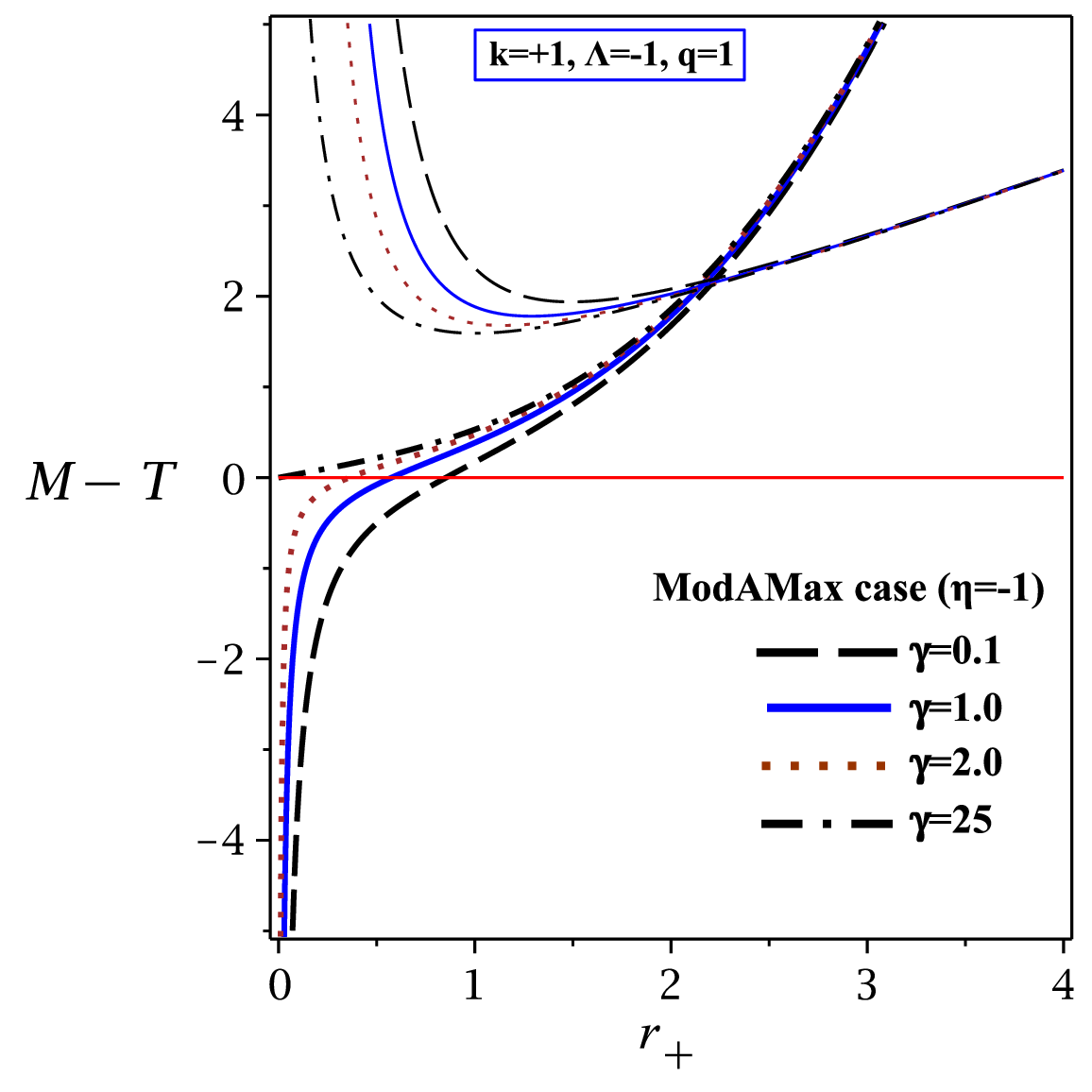} %
\includegraphics[width=0.325\linewidth]{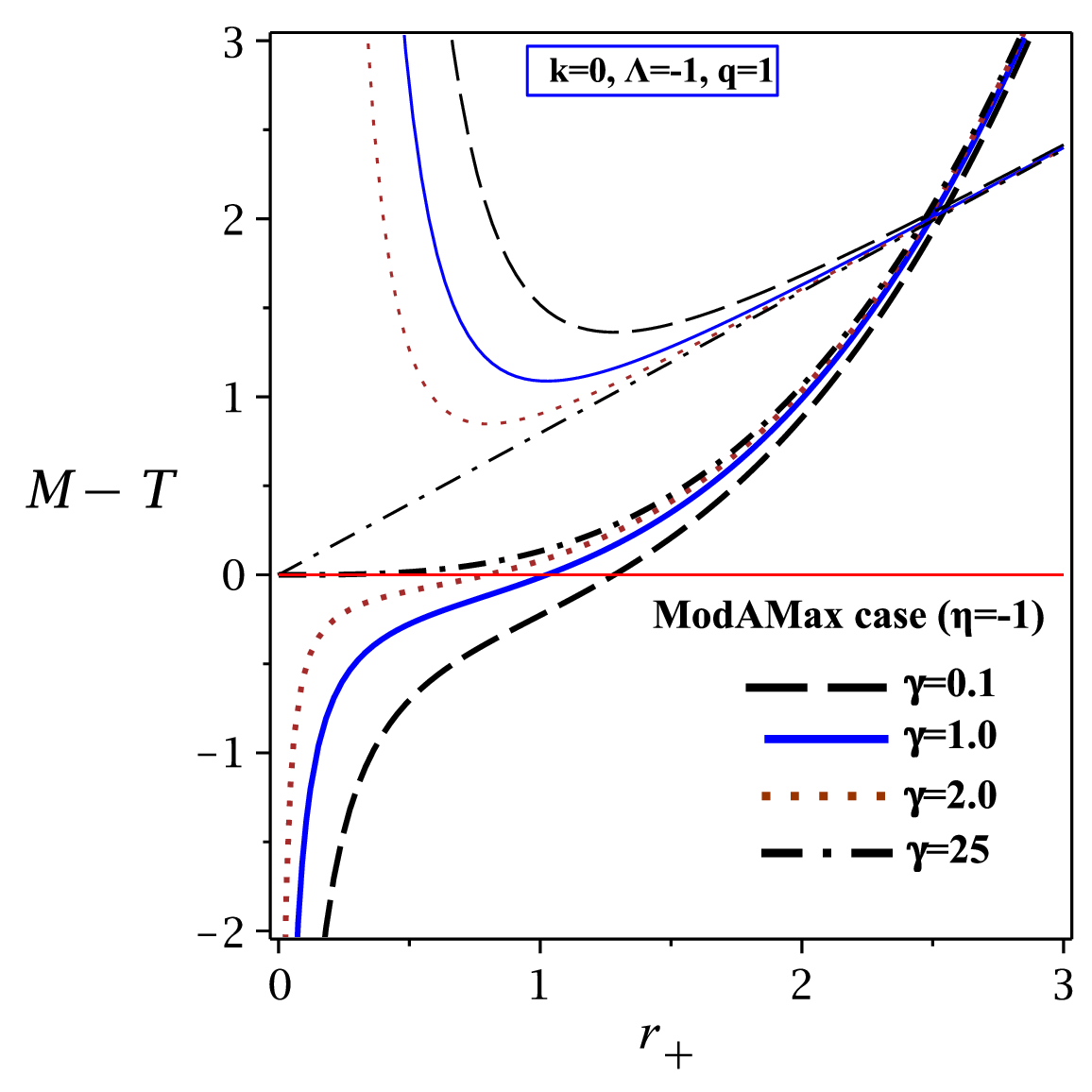} %
\includegraphics[width=0.325\linewidth]{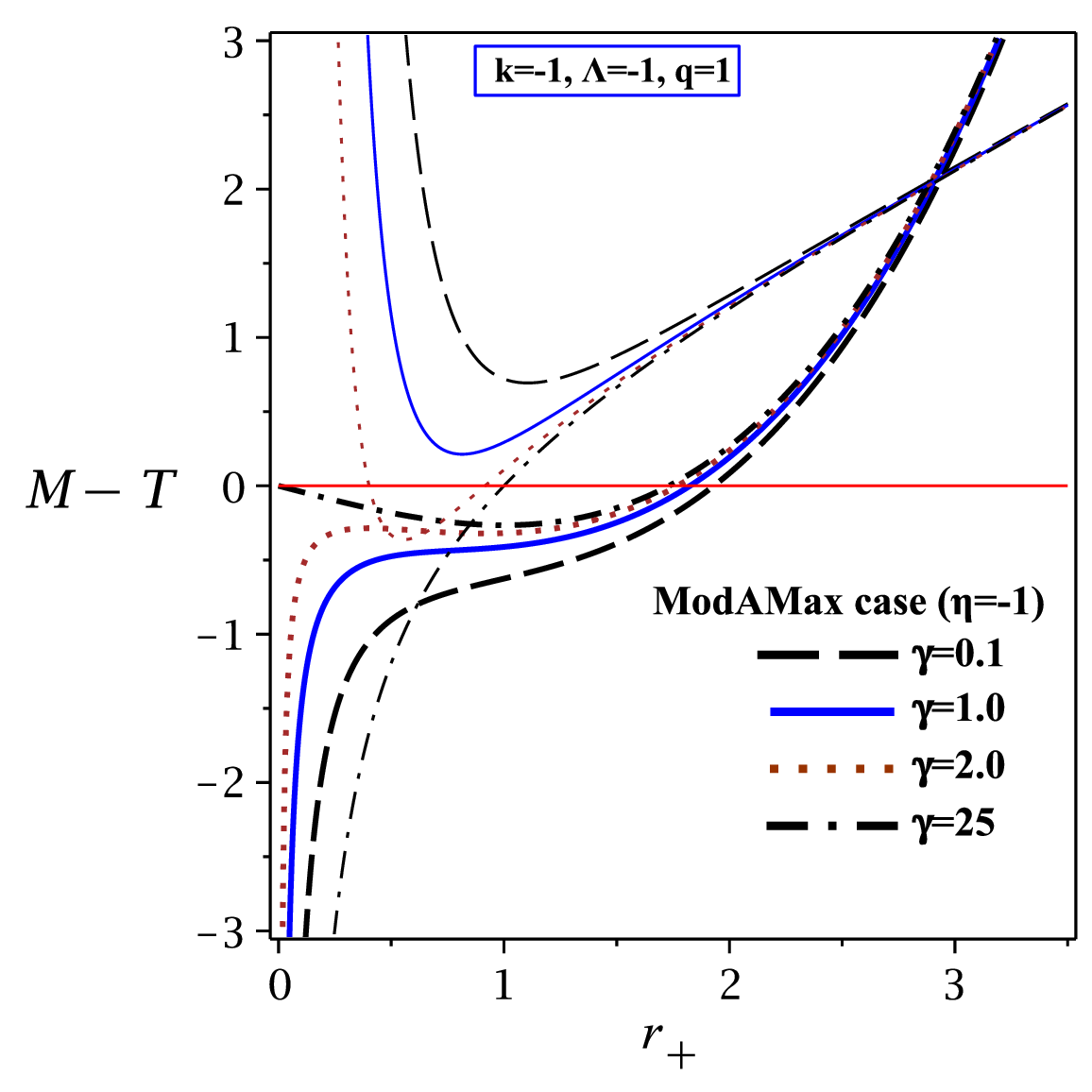} \newline
\caption{The total mass $M$ (thick lines) and the Hawking
temperature $T$ (thin lines) versus $r_{+}$ for $k=+1$ (left panel), $k=0$
(middle panel), and $k=-1$ (right panel) by considering the \textbf{ModAMax
field ($\protect\eta =-1$)}.}
\label{Fig8}
\end{figure}

To ensure that the total mass of ModAMax AdS black holes is positive, we
find that 
\begin{equation}
M_{\text{ModAMax}}>0~\Longrightarrow \left\{ 
\begin{array}{ccc}
3r_{+}^{2}-\Lambda r_{+}^{4}-3q^{2}e^{-\gamma }>0 &  & k=+1 \\ 
&  &  \\ 
-\Lambda r_{+}^{4}-3q^{2}e^{-\gamma }>0 &  & k=0 \\ 
&  &  \\ 
-3r_{+}^{2}-\Lambda r_{+}^{4}-3q^{2}e^{-\gamma }>0 &  & k=-1%
\end{array}%
\right. ,  \label{MModAMax}
\end{equation}%
where $M_{\text{ModAMax}}$ represents the total mass of ModAMax black holes.
The total mass of ModAMax AdS black holes is determined by the topological
constant $k$, the electric charge $q$, and the ModAMax parameter $\gamma $;
these dependencies give rise to three distinct physical behaviors:

i) For $k=+1$, $M_{\text{ModAMax}}$ remains positive when $r_{+}^{2}\left(
3-\Lambda r_{+}^{2}\right) >3q^{2}e^{-\gamma }$. This condition indicates
that the total mass of ModAMax AdS black holes cannot be positive at any
radius (see the thick lines in the left panel of Fig. \ref{Fig8}).
Furthermore, in the limit $r_{+}\rightarrow 0$, $M_{\text{ModAMax}%
}\rightarrow 0$ when $q=0$ (or when $\gamma $ approaches a very large
value), see the thick dotted-dashed lines in the left panel of Fig. \ref%
{Fig8}.

ii) For $k=0$, $M_{\text{ModAMax}}$ becomes positive under the condition $%
-\Lambda r_{+}^{4}>3q^{2}e^{-\gamma }$. Moreover, as $r_{+}\rightarrow 0$,
the total mass tends to zero for $q=0$ or when $\gamma$ approaches very
large values, consistent with the thick dotted-dashed lines in the middle
panel of Fig. \ref{Fig8}.

iii) To ensure a positive value of $M_{\text{ModAMax}}$ for $k=-1$, the
condition $\frac{-\Lambda r_{+}^{2}}{3}>1+\frac{q^{2}e^{-\gamma }}{r_{+}^{2}}
$ must be satisfied. Moreover, in the absence of electric charge, the total
mass of ModAMax AdS black holes becomes positive when $-\Lambda r_{+}^{2}>3$
(see the thick lines in the right panel of Fig. \ref{Fig8}).

We study the high energy limit and the asymptotic limit of the total mass to
evaluate the effects of various parameters. The high energy limit of the
total mass is given by 
\begin{equation}
\underset{r_{+}\rightarrow 0}{\lim }M\propto \frac{\eta q^{2}e^{-\gamma }}{%
r_{+}}=\left\{ 
\begin{array}{ccc}
\frac{q^{2}e^{-\gamma }}{8\pi r_{+}}, &  & \text{ModMax} \\ 
&  &  \\ 
-\frac{q^{2}e^{-\gamma }}{8\pi r_{+}}, &  & \text{ModAMax}%
\end{array}%
\right. ,  \label{HighM}
\end{equation}%
where for small black holes, the total mass is determined solely by the
electrodynamic field. It is evident that the mass of small ModMax AdS black
holes can be positive. In contrast, the total mass of small ModAMax black
holes cannot be positive. This is the main difference between ModMax and
ModAMax fields in the high energy limit of the total mass of black holes.

The asymptotic limit of the total mass depends solely on the cosmological
constant, expressed as 
\begin{equation}
\underset{r_{+}\rightarrow \infty }{\lim }M\propto -\frac{\Lambda r_{+}^{3}}{%
24\pi },  \label{AsyM}
\end{equation}%
where indicates that the total mass of the large Mod(A)Max AdS black hole is
always positive.

The term $\frac{kr_{+}}{8\pi}$ in Eq. (\ref{MM}) is significant for medium
black holes. Specifically, the total mass of medium ModMax AdS black holes
can be negative when $k=-1$. Additionally, in the high energy limit (or for
small black holes), the total mass, as shown in Eq. (\ref{HighM}), is
influenced by the electrodynamic field and is expressed as $\frac{\eta q^{2}
e^{-\gamma}}{8\pi r_{+}}$. This mass increases as $r_{+}$ approaches $0$,
implying that small black holes may possess an implausibly large mass. To
identify an appropriate range for the total mass, we present the total mass
and temperature together in Figs. \ref{Fig7} and \ref{Fig8}.

In Fig. \ref{Fig7}, we examine the simultaneous effects of the ModMax
parameter and the topological constant on total mass (thick lines)
and temperature (thin lines). Our analysis indicates that:

1) In the left panel of Fig. \ref{Fig7}, for $k=+1$, increasing $r_{+}$
initially causes the total mass, $M$, to decrease until it reaches a
critical point where the slope of $M$ is zero (i.e., $\frac{dM}{dr_{+}}=0$).
Beyond this point, $M$ begins to increase as $r_{+}$ continues to rise. This
critical point corresponds to the zero point of the temperature, as
indicated by the relation $T=\frac{dM}{dr_{+}}/\frac{dS}{dr_{+}}$.
Specifically, there exists a critical radius for the total mass, denoted as $%
r_{+_{T=0}}$. For values of $r_{+} < r_{+_{T=0}}$, black holes cannot be
defined because the temperature in this region is negative, rendering the
thermodynamic system non-physical. Conversely, for $r_{+} > r_{+_{T=0}}$,
the temperature becomes positive, and the total mass increases with $r_{+}$.
This indicates that the mini ModMax AdS black holes (i.e., those with $%
r_{+}<r_{+_{T=0}}$) cannot be considered physical objects (refer to the
dashed, continuous, and dotted thin lines in the left panel of Fig. %
\ref{Fig7}). Additionally, as $\gamma$ increases, the critical point for
total mass can be eliminated, allowing small ModMax AdS black holes to be
considered physical objects. In fact, for sufficiently large values of $%
\gamma$, the critical point disappears entirely, permitting black holes of
any radius to exist (see the dotted-dashed line in the left panel of Fig. %
\ref{Fig7}). Thus, the ModMax parameter significantly influences the
physical area of black holes, highlighting that the physical area of ModMax
AdS black holes expands with increasing $\gamma$.

2) In the middle panel of Fig. \ref{Fig7}, for $k=0$, the total mass behaves
similarly to the previous case with $k=+1$. However, when comparing the left
and middle panels, there are two notable differences between $k=0$ and $k=+1$%
: i) For large values of $\gamma$, very small ModMax AdS black holes cannot
be considered physical objects for $k=0$, as their temperature becomes
negative while the total mass approaches zero (see the dashed-dotted line in
the middle panel of Fig. \ref{Fig7}). In contrast, ModMax AdS black holes
are physical objects at any radius for $k=+1$ (see the dashed-dotted line in
the left panel of Fig. \ref{Fig7}). ii) The physical area for $k=+1$ is
larger than that for $k=0$.

3) In the right panel of Fig. \ref{Fig7}, for $k = -1$, there exists a
critical value for the ModMax parameter ($\gamma_{\text{critical}}$). When $%
\gamma > \gamma_{\text{critical}}$, the total mass of ModMax black holes
with $r_{+} > r_{+_{T=0}}$ becomes negative, even though the temperature
remains positive. Conversely, when $\gamma < \gamma_{\text{critical}}$, the
total mass of these black holes is positive in the same region ($r_{+} >
r_{+_{T=0}}$), as indicated by the dashed line in the right panel of Fig. %
\ref{Fig7}. Additionally, the physical area decreases as $\gamma$ increases.

By comparing the behavior of the total mass and the Hawking temperature, we
observe two notable results:

i) For ModMax AdS black holes with $k=+1$, the physical area (defined as the
area where both $M>0$ and $T>0$) is superior to that of other topological
cases.

ii) As $\gamma$ increases, the physical area increases for both $k=+1$ and $%
k=0$; however, for $k=-1$, the physical area decreases with an increase in $%
\gamma$.

In Fig. \ref{Fig8}, we examine the effects of the ModMax parameter and
topological constant by analyzing the impact of a phantom field on both
total mass (thick lines) and temperature (thin lines) simultaneously. Our findings reveal that:

1) In the left panel of Fig. \ref{Fig8}, for $k = +1$, we observe that the
total mass of the ModAMax AdS black hole is negative before a certain root
and becomes positive thereafter, consistent with our expectations from Eq. (%
\ref{HighM}). Furthermore, as $\gamma$ increases, the region of negative
mass decreases, and for sufficiently large values of $\gamma$, the negative
mass region is eliminated (as indicated by the dashed-dotted line in the
left panel of Fig. \ref{Fig8}). Additionally, the temperature of ModAMax
black holes remains positive at all radii, which means that increasing $%
\gamma$ leads to an increase in the physical area.

2) In the middle panel of Fig. \ref{Fig8}, for $k=0$, the behavior of the
mass of small black holes resembles that of the previous case ($k=+1$).
Specifically, there is a root, and the small black holes cannot meet the
positive mass condition. Additionally, as the value of $\gamma$ increases,
the negative area of the mass decreases. However, for very large values of
the ModMax parameter, the negative area of the total mass can be eliminated
(as shown by the dashed-dotted line in the left panel of Fig. \ref{Fig8}).
Similar to the previous case, the temperature of these black holes remains
positive throughout, resulting in an increase in the physical area as $%
\gamma $ increases.

3) In the right panel of Fig. \ref{Fig8}, for $k=-1$, the large ModAMax
black holes satisfy the physical conditions, as both mass and temperature
are positive. Additionally, the physical area of these large black holes
increases with increasing $\gamma$. In contrast, for small ModAMax black
holes, both temperature and mass are negative for every value of $\gamma$.
Therefore, small ModAMax AdS black holes cannot be considered physical
objects.

We can now state that the conserved and thermodynamic quantities obtained in
Eqs. (\ref{TemII}), (\ref{Q}), (\ref{elcpo}), (\ref{S}), and (\ref{MM})
satisfy the first law of thermodynamics in the following form 
\begin{equation}
dM=TdS+\eta UdQ=\left\{ 
\begin{array}{ccc}
TdS+UdQ &  & \text{ModMax} \\ 
&  &  \\ 
TdS-UdQ &  & \text{ModAMax}%
\end{array}%
\right. ,
\end{equation}%
where $T=\left( \frac{\partial M}{\partial S}\right) _{Q}$, and $U=\left( 
\frac{\partial M}{\partial Q}\right) _{S}$ are, respectively, in agreement
with those of calculated in Eqs. (\ref{TemII}) and (\ref{elcpo}).

\section{Local and Global Stabilities}

This section is dedicated to the investigation of the local and global
stabilities of the topological Mod(A)Max AdS black holes, utilizing the
thermodynamic variables of heat capacity, as well as the Helmholtz and Gibbs
free energies. Specifically, we analyze the influence of both the
topological constant ($k$) and the ModMax parameter ($\gamma $) on these
stability criteria.

\subsection{Local Stability}

In the framework of the canonical ensemble, the local stability of a
thermodynamical system can be analyzed through its heat capacity. The heat
capacity encodes essential information about the thermal structure of black
holes and provides three particularly significant insights, which are; i)
Discontinuities in the heat capacity indicate possible thermal phase
transitions that the system may experience. ii) The sign of the heat
capacity determines the thermal stability of the system; a positive value
corresponds to stability, whereas a negative one signifies instability. iii)
The zeros (roots) of the heat capacity are also of interest, as they may
signal transitions between stable and unstable phases or represent boundary
points. Motivated by these considerations, we calculate the heat capacity of
the obtained solutions and analyze the local thermodynamic stability of the
corresponding black holes based on this quantity.

Prior to calculating the heat capacity, we first rewrite the expression for
the total mass of the black hole (\ref{MM}) in terms of the electric charge
(Eq. (\ref{Q})), and the entropy (Eq. (\ref{S})) in the following form 
\begin{equation}
M\left( S,Q\right) =\frac{3kS-4\Lambda S^{2}+12\eta \pi ^{2}Q^{2}e^{-\gamma }%
}{12\pi \sqrt{S}},  \label{MSQ}
\end{equation}%
using Equation (\ref{MSQ}), the temperature is rewritten in the following
form 
\begin{equation}
T=\left( \frac{\partial M\left( S,Q\right) }{\partial S}\right) _{Q}=\frac{%
kS-4\Lambda S^{2}-4\eta \pi ^{2}Q^{2}e^{-\gamma }}{8\pi S^{3/2}},
\label{TSQ}
\end{equation}

The heat capacity is defind by the relation 
\begin{equation}
C_{Q}=\frac{T}{\left( \frac{\partial T}{\partial S}\right) _{Q}}=\frac{%
\left( \frac{\partial M\left( S,Q\right) }{\partial S}\right) _{Q}}{\left( 
\frac{\partial ^{2}M\left( S,Q\right) }{\partial S^{2}}\right) _{Q}},
\label{C}
\end{equation}

by considering Eqs. (\ref{MSQ}) and (\ref{TSQ}) within Eq. (\ref{C}), the
resulting expression for the heat capacity is 
\begin{equation}
C_{Q}=\frac{2\left( 4\Lambda S^{2}-kS+4\eta \pi ^{2}Q^{2}e^{-\gamma }\right)
S}{kS+4\Lambda S^{2}-12\eta \pi ^{2}Q^{2}e^{-\gamma }}.
\end{equation}

Within the context of black hole thermodynamics, the roots of the heat
capacity, where $C_{Q}=0$ (or $T=0$), are generally considered to represent
the boundary between physically admissible black holes (where temperature $%
T>0$) and non-physical solutions (where $T<0$). We refer to these points as
physical limitation points. It is noteworthy that at these specific points,
the sign of the heat capacity changes. Furthermore, the divergences
(singularities) in the heat capacity are understood to correspond to the
critical points of thermal phase transitions for these black holes.
Consequently, the critical points for the phase transitions and the
aforementioned limitation points, as determined by the heat capacity
analysis, are calculated using the following relations: 
\begin{equation}
\left\{ 
\begin{array}{ccc}
T=\left( \frac{\partial M\left( S,Q\right) }{\partial S}\right) _{Q}=0, &  & 
\text{physical limitation points} \\ 
&  &  \\ 
\left( \frac{\partial ^{2}M\left( S,Q\right) }{\partial S^{2}}\right) _{Q}=0,
&  & \text{phase transition critical points}%
\end{array}%
\right. .  \label{PhysBound}
\end{equation}

Now, we obtain physical limitation points by solving (\ref{TSQ}) in terms of
the entropy as 
\begin{equation}
\left\{ 
\begin{array}{c}
S_{root_{1}}=\frac{k+\sqrt{k^{2}-64\pi ^{2}\Lambda \eta Q^{2}e^{-\gamma }}}{%
8\Lambda }, \\ 
\\ 
S_{root_{2}}=\frac{k-\sqrt{k^{2}-64\pi ^{2}\Lambda \eta Q^{2}e^{-\gamma }}}{%
8\Lambda },%
\end{array}%
\right. .
\end{equation}

To have the real root(s), we have to respect $k^{2}\geq 64\pi ^{2}\Lambda
\eta Q^{2}e^{-\gamma }$. By considering this constraint and this fact that
we consider AdS case ($\Lambda <0$), we find that there is one real positive
root for the heat capacity and it is related to $S_{root_{2}}$. Indeed $%
S_{root_{1}}$ leads to negative value and $S_{root_{2}}$ includes a real
positive value and determine the physical limitation point of the
thermodynamical system. In addition, the physical limitation point ($%
S_{root_{2}}$) depends on various parameters such as $k$, $Q$, $\eta $, $%
\Lambda $, and $\gamma $. We study the effects of these parameters on the
physical limitation point of Mod(A)Max AdS black holes ($S_{root_{2}}$), and
we find that

\begin{description}
\item[\textbf{ModMax field ($\protect\eta =+1$):}] Considering $\eta =+1$,
and for different topological constant we find that 
\begin{equation}
\text{physical limitation point for ModMax case}\Longrightarrow \left\{ 
\begin{array}{ccc}
S_{root_{2}}=\frac{-1+\sqrt{1-64\pi ^{2}\Lambda Q^{2}e^{-\gamma }}}{%
-8\Lambda }, &  & k=+1 \\ 
&  &  \\ 
S_{root_{2}}=\frac{\sqrt{-64\pi ^{2}\Lambda Q^{2}e^{-\gamma }}}{-8\Lambda },
&  & k=0 \\ 
&  &  \\ 
S_{root_{2}}=\frac{1+\sqrt{1-64\pi ^{2}\Lambda Q^{2}e^{-\gamma }}}{-8\Lambda 
}, &  & k=-1%
\end{array}%
\right. ,
\end{equation}

\item which indicates that the largest root belongs to $k=-1$, and the
smallest root is related to $k=+1$. In other words, by fixing other values
of parameter (such as $\Lambda $, $Q$, and $\gamma $), we find that a order
for the physical limitation point which depends on the topological constant
as $S_{root_{2_{k=+1}}}<S_{root_{2_{k=0}}}<S_{root_{2_{k=-1}}}$. On the
other hand, by increasing $\gamma $ the physical limitation point decreases
(see thick dotted-dashed and dashed lines in Fig. \ref{Fig9}). Also, for
very large values of $\gamma $, the physical limitation point disappear for $%
k=+1$ and $k=0$, because $\underset{\gamma \rightarrow \infty }{\lim }%
S_{root_{2}}\rightarrow 0$. However, for $k=-1$, the physical limitation
point goes to $\frac{-1}{4\Lambda }$ ($\underset{\gamma \rightarrow \infty }{%
\lim }S_{root_{2}}\rightarrow \frac{-1}{4\Lambda }$ when $k=-1$).

\item[\textbf{ModAMax field ($\protect\eta =-1$):}] By replacing $\eta =-1$
in the root $S_{root_{2}}$, we find that the following coditions for
different values of the topological constants 
\begin{equation}
\text{physical limitation point for ModAMax case}\Longrightarrow \left\{ 
\begin{array}{ccc}
S_{root_{2}}=\frac{1-\sqrt{1+64\pi ^{2}\Lambda Q^{2}e^{-\gamma }}}{8\Lambda }%
, &  & k=+1 \\ 
&  &  \\ 
S_{root_{2}}=\frac{\sqrt{64\pi ^{2}\Lambda Q^{2}e^{-\gamma }}}{-8\Lambda },
&  & k=0 \\ 
&  &  \\ 
S_{root_{2}}=\frac{1+\sqrt{1+64\pi ^{2}\Lambda Q^{2}e^{-\gamma }}}{-8\Lambda 
}, &  & k=-1%
\end{array}%
\right. ,
\end{equation}%
there is no the real positive physical limitation point for ModAMax AdS
black holes for $k=+1$ and $k=0$ (see thick lines in Fig. \ref{Fig10}, for
more details). However, for $k=-1$, one real positive root for the heat
capacity exists when $\gamma \rightarrow $ very large values (see the thick
constinuous line in the right panel of Fig. \ref{Fig10}). Also, similar to
the previous case, for $k=-1$ and in the limit $\gamma \rightarrow \infty $, 
$S_{root_{2}}\rightarrow \frac{-1}{4\Lambda }$.
\end{description}

In order to study the phase transition critical points (or divergence points
of the heat capacity), we have to solve the relation $\left( \frac{\partial
^{2}M\left( S,Q\right) }{\partial S^{2}}\right) _{Q}=0$. So, we have 
\begin{equation}
\left\{ 
\begin{array}{c}
S_{div_{1}}=\frac{-k+\sqrt{k^{2}+192\pi ^{2}\Lambda \eta Q^{2}e^{-\gamma }}}{%
8\Lambda }, \\ 
\\ 
S_{div_{2}}=\frac{-k-\sqrt{k^{2}+192\pi ^{2}\Lambda \eta Q^{2}e^{-\gamma }}}{%
8\Lambda },%
\end{array}%
\right. ,  \label{rdivHeat}
\end{equation}%
where indicate that we have to respect $k^{2}\geq -192\pi ^{2}\Lambda \eta
Q^{2}e^{-\gamma }$, for having the real divergent point(s). Our analysis
shows that the heat capacity of these black holes has one divergent point
and it related to $S_{div_{2}}$. This divergent point depends on various
parameters such as $k$, $Q$, $\eta $, $\Lambda $, and $\gamma $. Now, we
evaluate the effects of these parameters on the phase transition critical
point of Mod(A)Max AdS black holes ($S_{div_{2}}$), and we find that

\begin{description}
\item[\textbf{ModMax field ($\protect\eta =+1$):}] Considering $\eta =+1$, $%
\Lambda <0$, and different values of topological constant within Eq. (\ref%
{TSQ}), we find that there is no real positive divergence point for heat
capacity for $k=0$, and $k=-1$ (see the thick lines in the middel and right
panels in Fig. \ref{Fig9}). However, for $k=+1$, it is possible to find two
the phase transition critical points which are 
\begin{equation}
\text{phase transition critical points for ModMax }\Longrightarrow \left\{ 
\begin{array}{ccc}
S_{div_{1}}=\frac{1-\sqrt{1+192\pi ^{2}\Lambda Q^{2}e^{-\gamma }}}{-8\Lambda 
}, &  & k=+1 \\ 
&  &  \\ 
S_{div_{2}}=\frac{1+\sqrt{1+192\pi ^{2}\Lambda Q^{2}e^{-\gamma }}}{-8\Lambda 
}, &  & k=+1%
\end{array}%
\right. ,
\end{equation}%
and $S_{div_{2}}$ is located in larger entropy than $S_{div_{1}}$ (i.e., $%
S_{div_{2}}>S_{div_{1}}$). It is notable that in the limit $\gamma
\rightarrow \infty $, the phase transition critical points reduce from two
to one real root, because $\underset{\gamma \rightarrow \infty }{\lim }%
S_{div_{1}}\rightarrow 0$, and $\underset{\gamma \rightarrow \infty }{\lim }%
S_{div_{2}}\rightarrow \frac{-1}{4\Lambda }$ (see the thick continuouse line
in the left panel of Fig. \ref{Fig9}).

\item[\textbf{ModAMax field ($\protect\eta =-1$):}] By replacing $\eta =-1$
in the roots of $S_{div_{1}}$, and $S_{div_{2}}$, we find that there is only
one the physical limitation point for ModAMax AdS black holes and it is
related to $S_{div_{2}}$. In other words, there is one divergence point for
the heat capacity when $\eta =-1$. In addition, for different values of the
topological constants $S_{div_{2}}$ turns to the following three conditions 
\begin{equation}
\text{phase transition critical points for ModAMax}\Longrightarrow \left\{ 
\begin{array}{ccc}
S_{div_{2}}=\frac{-1-\sqrt{1-192\pi ^{2}\Lambda Q^{2}e^{-\gamma }}}{8\Lambda 
}, &  & k=+1 \\ 
&  &  \\ 
S_{div_{2}}=\frac{-\sqrt{-192\pi ^{2}\Lambda Q^{2}e^{-\gamma }}}{8\Lambda },
&  & k=0 \\ 
&  &  \\ 
S_{div_{2}}=\frac{1-\sqrt{1-192\pi ^{2}\Lambda Q^{2}e^{-\gamma }}}{8\Lambda }%
, &  & k=-1%
\end{array}%
\right. ,
\end{equation}%
which indicates that the largest root belongs to $k=+1$, and the smallest
root is related to $k=-1$. In other words, by fixing other values of
parameter (such as $\Lambda $, $Q$, and $\gamma $), we find that a order for
the phase transition critical points which depends on the topological
constant as $S_{div_{2_{k=-1}}}<S_{div_{2_{k=0}}}<S_{div_{2_{k=+1}}}$.
Furthermore, in the limit $\gamma \rightarrow \infty $, the phase transition
critical point only exist for $k=+1$, because 
\begin{equation}
\left\{ 
\begin{array}{ccc}
\underset{\gamma \rightarrow \infty }{\lim }S_{div_{2}}\rightarrow \frac{-1}{%
4\Lambda } &  & k=+1 \\ 
&  &  \\ 
\underset{\gamma \rightarrow \infty }{\lim }S_{div_{2}}\rightarrow 0 &  & k=0
\\ 
&  &  \\ 
\underset{\gamma \rightarrow \infty }{\lim }S_{div_{2}}\rightarrow 0 &  & 
k=-1%
\end{array}%
\right. .
\end{equation}
\end{description}

Now we can evaluate the effects of various parameters on the local stability
by using the behavior of temperature and heat capacity, simoultenatously.
For this purpose, we plot Figs. \ref{Fig9} and \ref{Fig10}. Our findings
reveal some interesting behaviors which are:

\begin{description}
\item[\textbf{ModMax field ($\protect\eta =+1$):}] Based on different values
of the topological constants and for $\eta =+1$, we find three categories;

i) For $k=+1$ (left panel in Fig. \ref{Fig9}), and by considering small
values of $\gamma $, there are two different areas for the heat capacity.
The large ModMax AdS black holes satisfy the local stability condition
because in the range $S>S_{root_{2}}$, the heat capacity and the temperature
are positive. However, the small ModMax AdS black holes are instable because
the heat capacity is negative in the range $S<S_{root_{2}}$ (see dotted and
dotted-dashed lines in the left panel of Fig. \ref{Fig9}). It is notable
that, the local stability area increases when the ModMax parameter
increases. There is an interesting behavior for the heat capacity when $%
\gamma $ is big. Indeed, there is a critical value for $\gamma $ ($\gamma
_{critical}$), which leads to two divergence points for the heat capacity
when $k=+1$, and $\gamma $ is big but follow the condition $\gamma <$ $%
\gamma _{critical}$. In this case, there are three areas for the heat
capacity of ModMax AdS black holes (see dashed line in the left panel of
Fig. \ref{Fig9}); the first area satisfies the local stability condition and
is located in the range $S>S_{div_{2}}$. The second area is located in the
range $S_{div_{1}}<S<S_{div_{2}}$, and the thermodynamical system is
unstable due to the negative value of the heat capacity. The third area
belongs to stable area and is located in the range $S<S_{div_{1}}$. On the
other hand, for $\gamma >\gamma _{critical}$, there is one divergence point
(which is related to $S_{div_{2}}$, and in the limit $\gamma \rightarrow
\infty $, $S_{div_{2}}\rightarrow \frac{-1}{4\Lambda }$) for the heat
capacity which large ModMax AdS black holes can satisfy the local stability
condition. In other words, the heat capacity is posive (negatvie) in the
range $S>S_{div_{2}}$ ($S<S_{div_{2}}$), see continuouos line in the left
panel of Fig. \ref{Fig9}.

ii) For $k=0$ (middle panel in Fig. \ref{Fig9}), there is one real positive
root of the heat capacity ($S_{root_{2}}$) which determines stable and
unstable areas. Indeed, for $k=0$, the ModMax AdS black holes satisfy the
local stability condition when $S>S_{root_{2}}$, whereas the small black
holes ($S<S_{root_{2}}$) are unstable thermal objects because the heat
capacity is negative when $S<S_{root_{2}}$. Also, there is a critical value
for $\gamma $ ($\gamma _{critical}$), which by considering $\gamma <\gamma
_{critical}$, the root of the heat capacity decreases by increasing $\gamma $
which leads to increase the stable area. But for $\gamma >\gamma _{critical}$
this root disapears and ModMax AdS black holes are stable thermodynanics
objects with any radius.

iii) For $k=-1 $, (right panel in Fig. \ref{Fig9}), there is always one real
positive root of the heat capacity ($S_{root_{2}}$) even in the limit $%
\gamma \rightarrow \infty $ (i.e., $\underset{\gamma \rightarrow \infty }{%
\lim }S_{root_{2}}\rightarrow \frac{-1}{4\Lambda }$). By considering $k=-1$,
the ModMax AdS AdS black holes are stable and unstabel thermal objects when $%
S>S_{root_{2}}$ and $S<S_{root_{2}}$, respectively. It is notable that, the
local stability area increases by increasing $\gamma $.

\begin{figure}[tbph]
\centering
\includegraphics[width=0.32\linewidth]{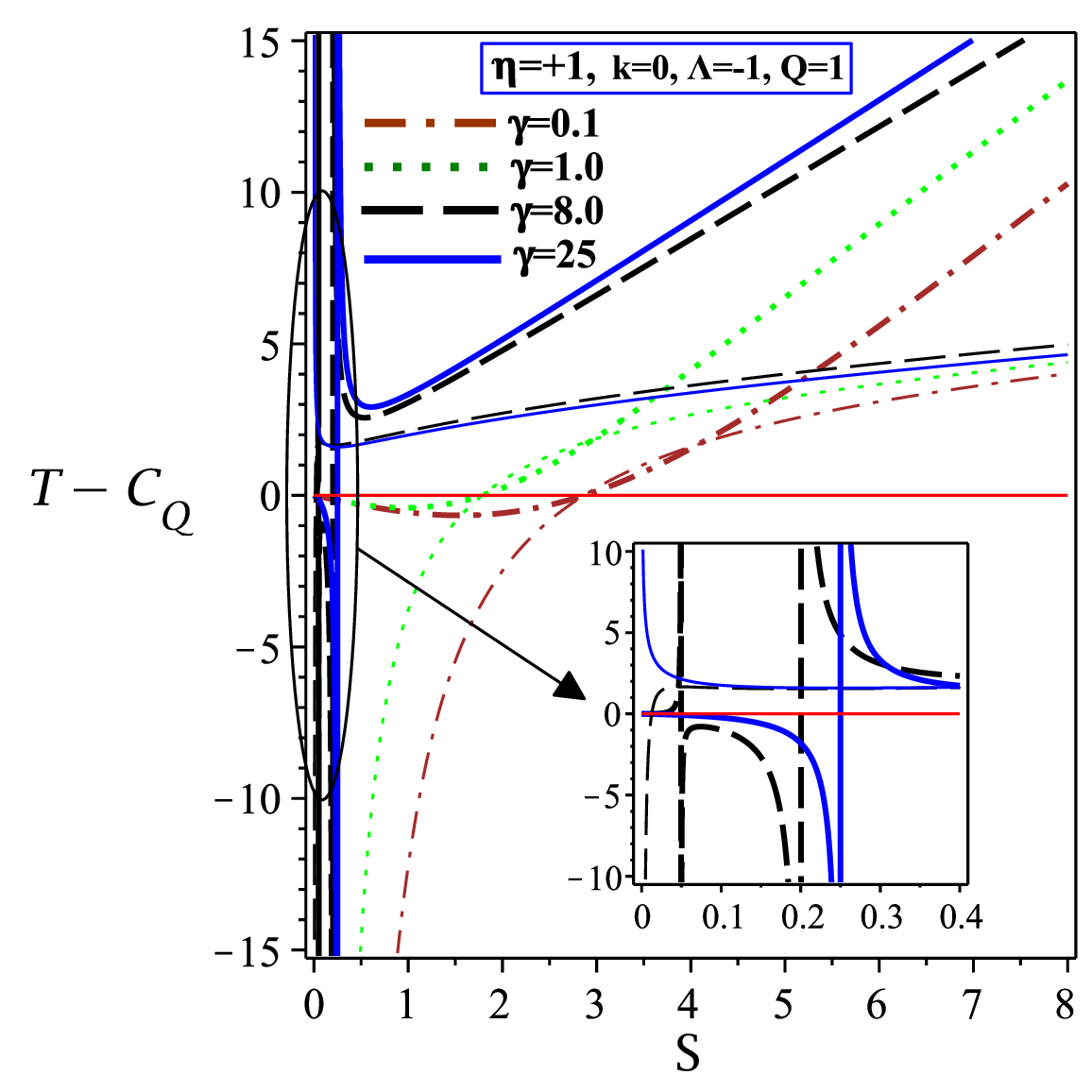} %
\includegraphics[width=0.32	\linewidth]{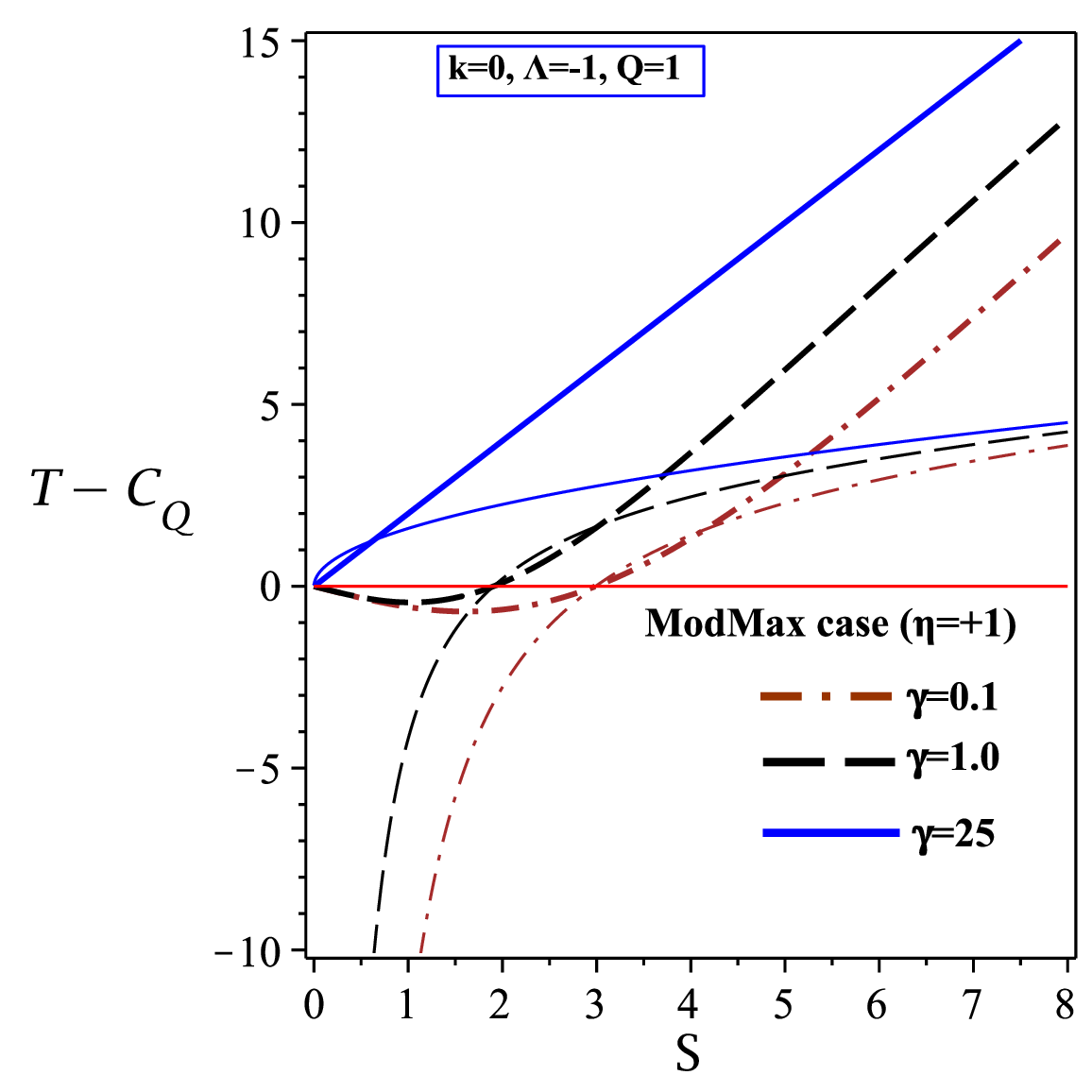} %
\includegraphics[width=0.32\linewidth]{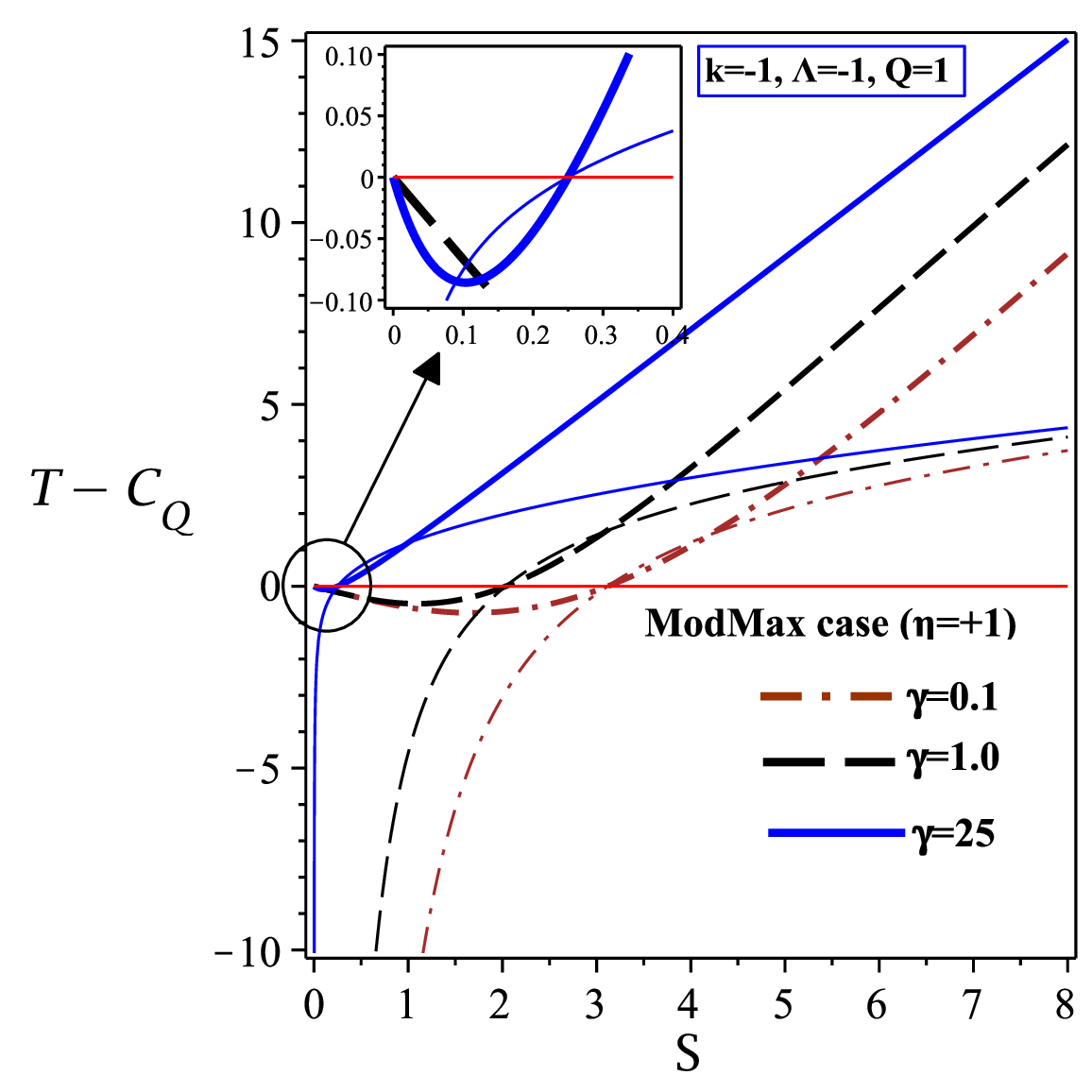} \newline
\caption{The Hawking temperature $T$ (thin lines) and the heat capacity $%
C_{Q}$ (thick lines) versus entropy $S$ for $k=+1$ (left panel), $k=0$
(middle panel), and $k=-1$ (right panel) by considering the ModMax field ($%
\protect\eta =+1$).}
\label{Fig9}
\end{figure}

\item[\textbf{ModAMax field ($\protect\eta =-1$):}] There exists a critical
value for the ModMax parameter, denoted as $\gamma _{critical}$, which
dictates a qualitative change in the behavior of the heat capacity ($C_{Q}$).

i) For $\gamma <\gamma _{critical}$: A single divergence point for the heat
capacity, $S_{div_{2}}$, exists, contingent on the topological constant ($k$%
). In the regime $S<S_{div_{2}}$, the heat capacity $C_{Q}$ is negative,
indicating thermodynamic instability. These configurations correspond to
small ModAMax AdS black holes. In the regime $S>S_{div_{2}}$, the heat
capacity is positive, implying local thermodynamic stability (as illustrated
by the dashed and dash-dotted lines in Fig. \ref{Fig10}). As $\gamma$
increases within this range, the divergence point $S_{div_{2}}$ shifts
toward lower entropy values. This effectively means that increasing $\gamma$
enhances the local stability of the system.

ii) For $\gamma >\gamma _{critical}$: The thermal behavior is strongly
dependent on the topological constant $k$. For $k=+1$, the divergence point $%
S_{div_{2}}$ persists, with $\underset{\gamma \rightarrow \infty }{\lim }%
S_{div_{2}}\rightarrow \frac{-1}{4\Lambda }$. For $k=0$, the heat capacity $%
C_{Q}$ remains positive across all entropy values (as shown by the
continuous line in the middle panel of Fig. \ref{Fig10}). Consequently,
ModAMax AdS black holes in this configuration always satisfy the local
stability criterion, irrespective of their entropy. For $k=-1$, a root for
the heat capacity, $S_{_{root_{2}}}$, appears where $C_{Q}$ changes sign.
Specifically, $C_{Q}<0$ for $S<S_{_{root_{2}}}$ (unstable small black
holes), and $C_{Q}>0$ for $S>S_{_{root_{2}}}$ (locally stable large black
holes), as depicted by the solid line in the right panel of Fig. \ref{Fig10}%
. Thus, for $k=-1$, local stability is achieved once the black hole radius
(or entropy) is sufficiently large.
\end{description}

\begin{figure}[tbph]
\centering
\includegraphics[width=0.32\linewidth]{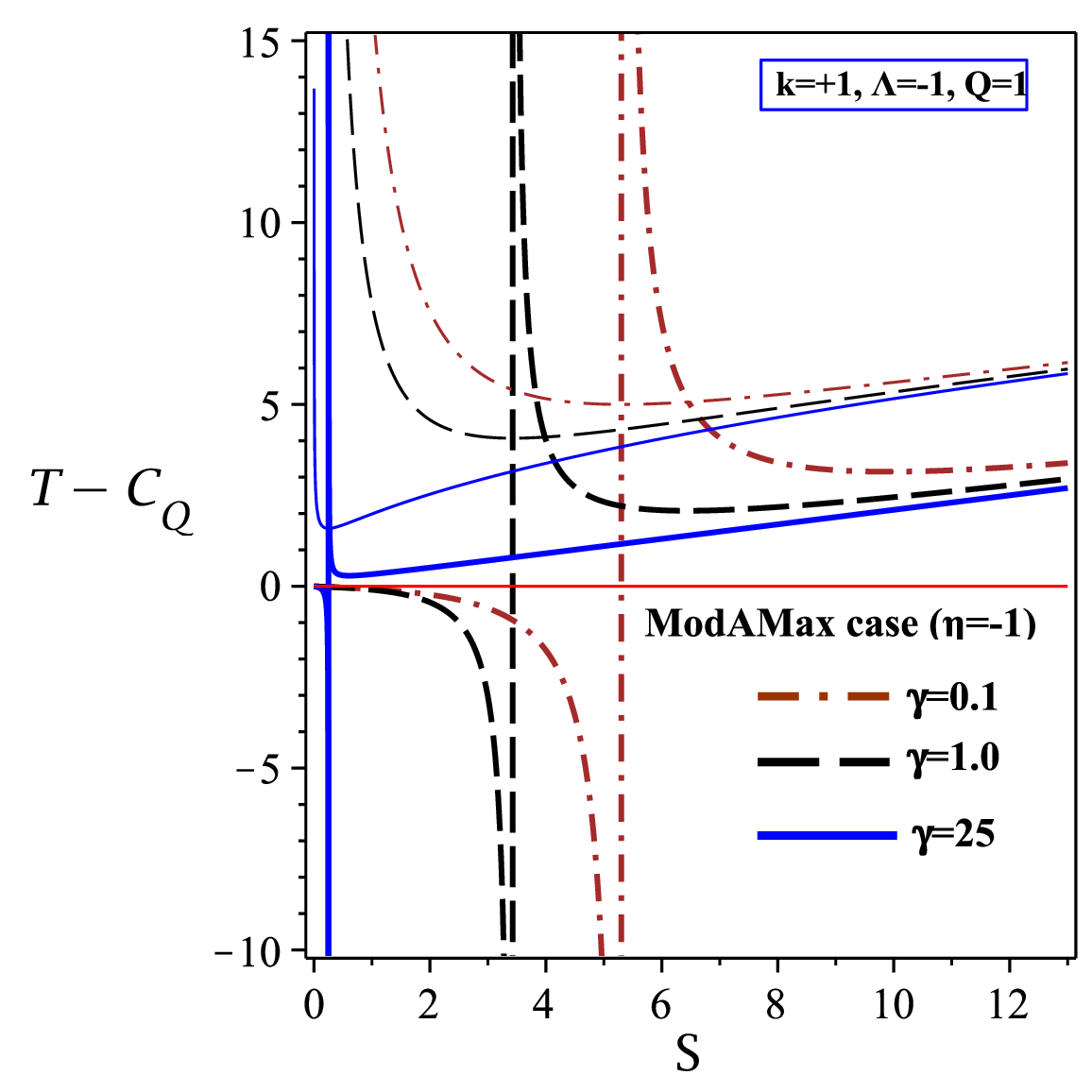} %
\includegraphics[width=0.32\linewidth]{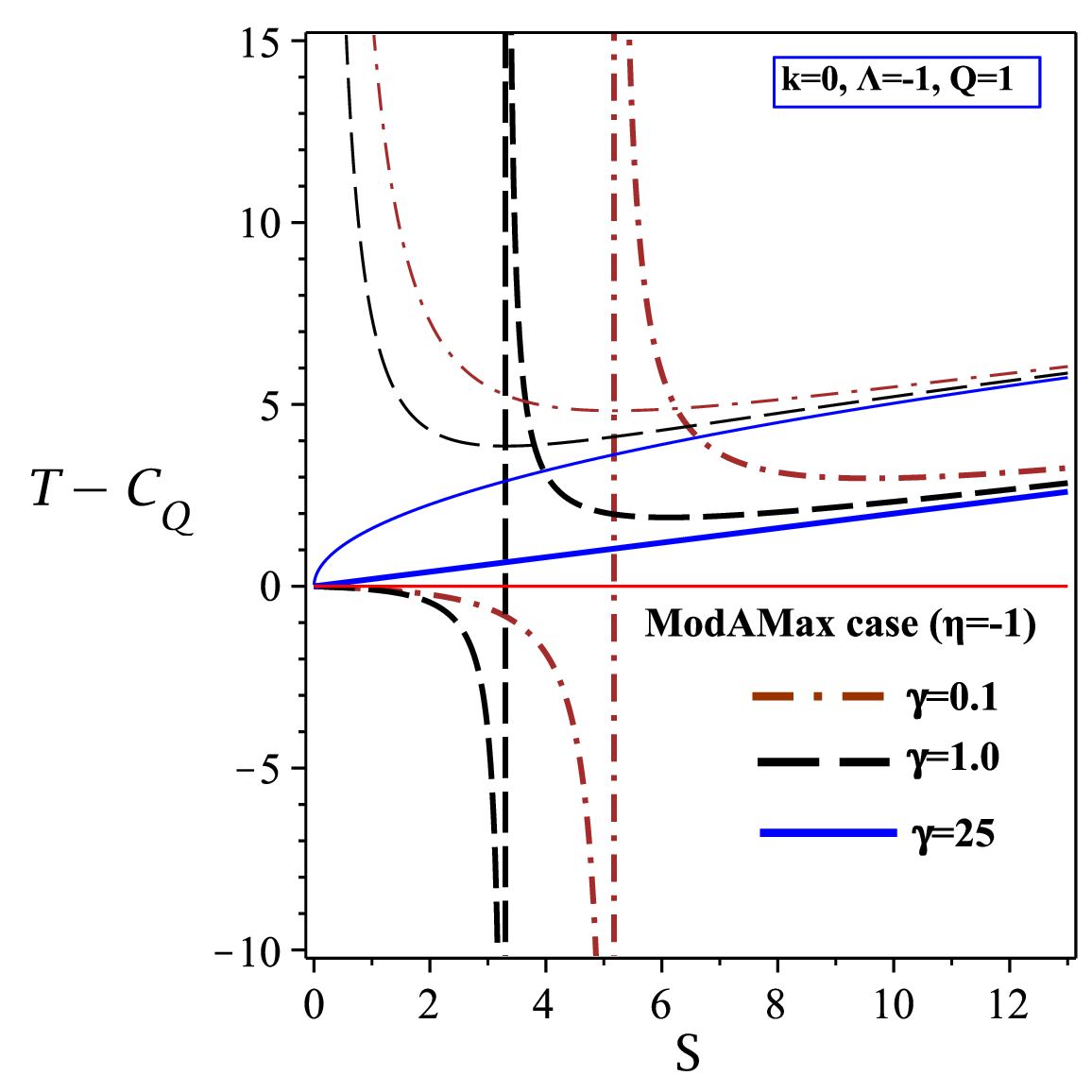} %
\includegraphics[width=0.32\linewidth]{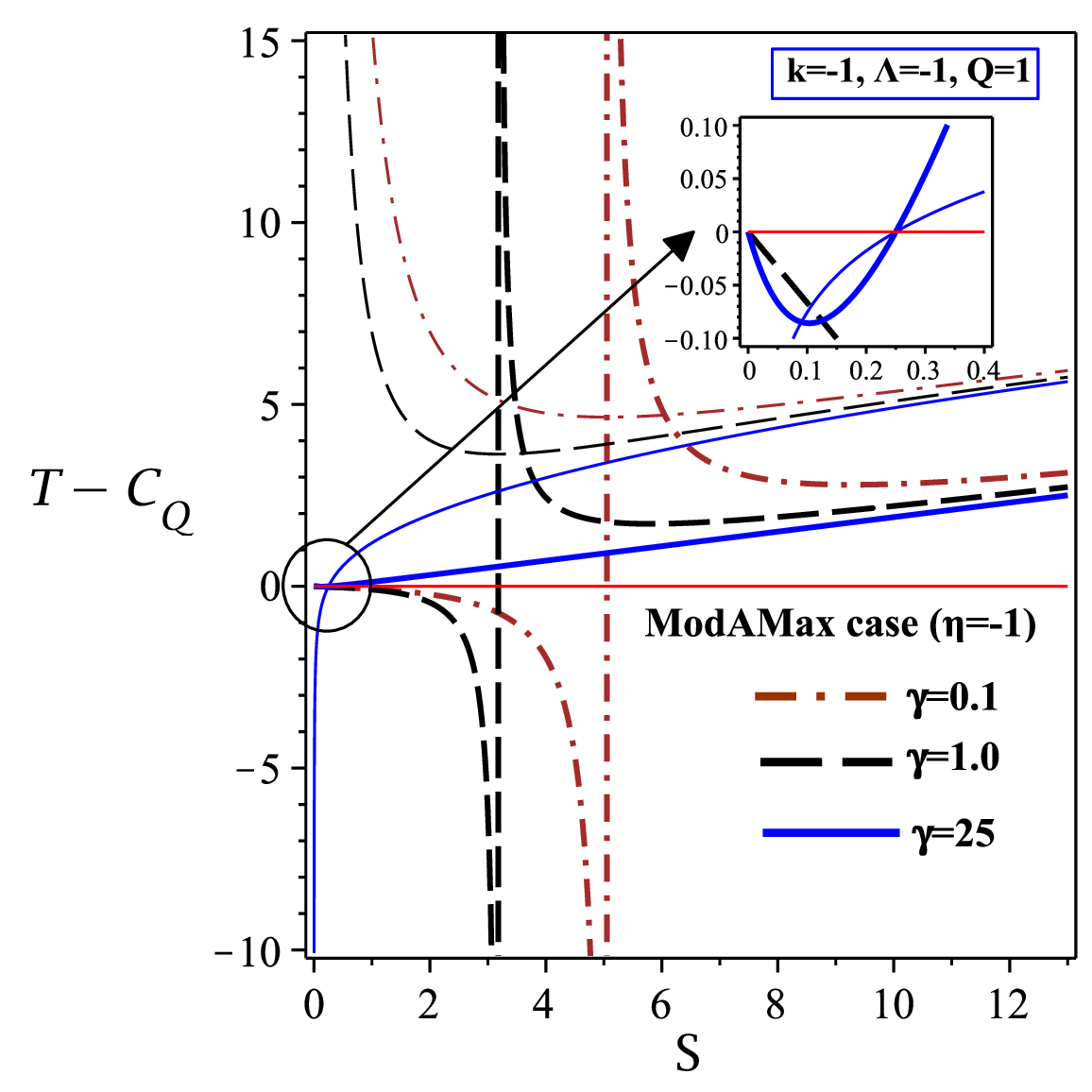} \newline
\caption{The Hawking temperature $T$ (thin lines) and the heat capacity $%
C_{Q}$ (thick lines) versus entropy $S$ for $k=+1$ (left panel), $k=0$
(middle panel), and $k=-1$ (right panel) by considering the ModAMax field ($%
\protect\eta =-1$).}
\label{Fig10}
\end{figure}

\subsection{Global Stability}

In the grand-canonical ensemble, the global thermodynamic stability is
ascertained by examining the Gibbs free energy potential ($G$);
specifically, a negative sign for this potential ensures global stability.
Concurrently, within the canonical ensemble, the global stability criterion
is satisfied by the negative of the Helmholtz free energy ($F$).
Consequently, we employ the Gibbs and Helmholtz free energies to
systematically evaluate the global stability properties of the topological
Mod(A)Max AdS black holes.

It is notable that in the usual case of thermodynamics the Helmholtz free
energy is given by $F=U-TS$. However, in the context of the black holes,
Helmholtz free energy is defined in the following form

While the conventional thermodynamic definition for the Helmholtz free
energy is given by $F=U-TS$, it is crucial to note that within the context
of black hole thermodynamics, this potential is redefined as follows 
\begin{equation}
F=M\left( S,Q\right) -TS,  \label{F}
\end{equation}%
by using Eqs. (\ref{MSQ}) and (\ref{TSQ}) within Eq. (\ref{F}), we get the
Helmholtz free energy in the following form 
\begin{equation}
F=\frac{4\Lambda S^{2}+3kS+36\pi ^{2}\eta Q^{2}e^{-\gamma }}{24\pi \sqrt{S}},
\end{equation}%
and by solving $F=0$, we can obtain the real positive root of the Helmholtz
free energy which is as 
\begin{equation}
S_{F=0}=\frac{3k+3\sqrt{k^{2}-64\pi ^{2}\Lambda \eta Q^{2}e^{-\gamma }}}{%
-8\Lambda },
\end{equation}%
which for different topological constant we have 
\begin{equation}
S_{F=0}=\left\{ 
\begin{array}{ccc}
\frac{3+3\sqrt{1-64\pi ^{2}\Lambda \eta Q^{2}e^{-\gamma }}}{-8\Lambda }, & 
& k=+1 \\ 
&  &  \\ 
\frac{3\sqrt{-64\pi ^{2}\Lambda \eta Q^{2}e^{-\gamma }}}{-8\Lambda }, &  & 
k=0 \\ 
&  &  \\ 
\frac{-3+3\sqrt{1-64\pi ^{2}\Lambda \eta Q^{2}e^{-\gamma }}}{-8\Lambda }, & 
& k=-1%
\end{array}%
\right. .
\end{equation}

\begin{figure}[tbph]
\centering
\includegraphics[width=0.32\linewidth]{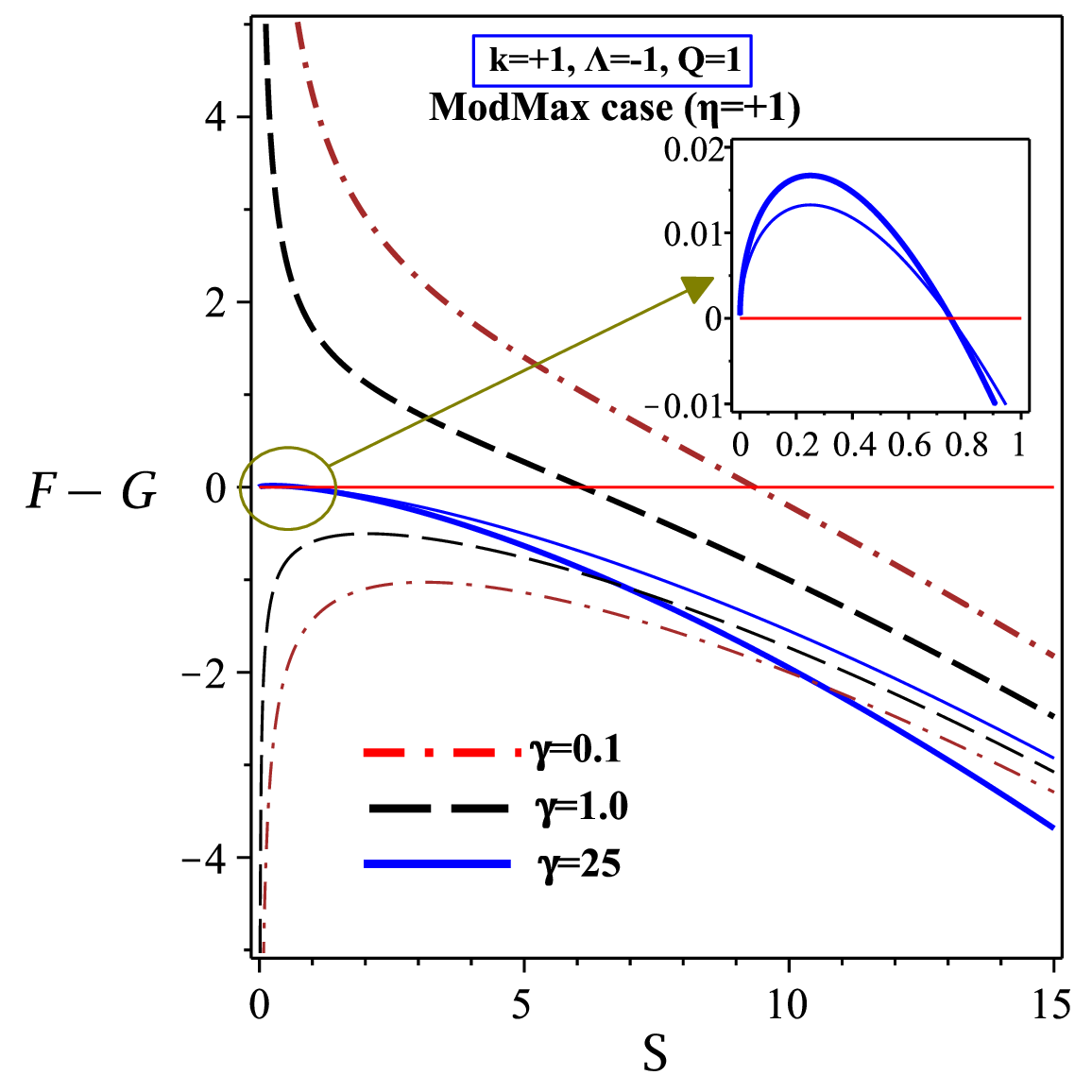} %
\includegraphics[width=0.32\linewidth]{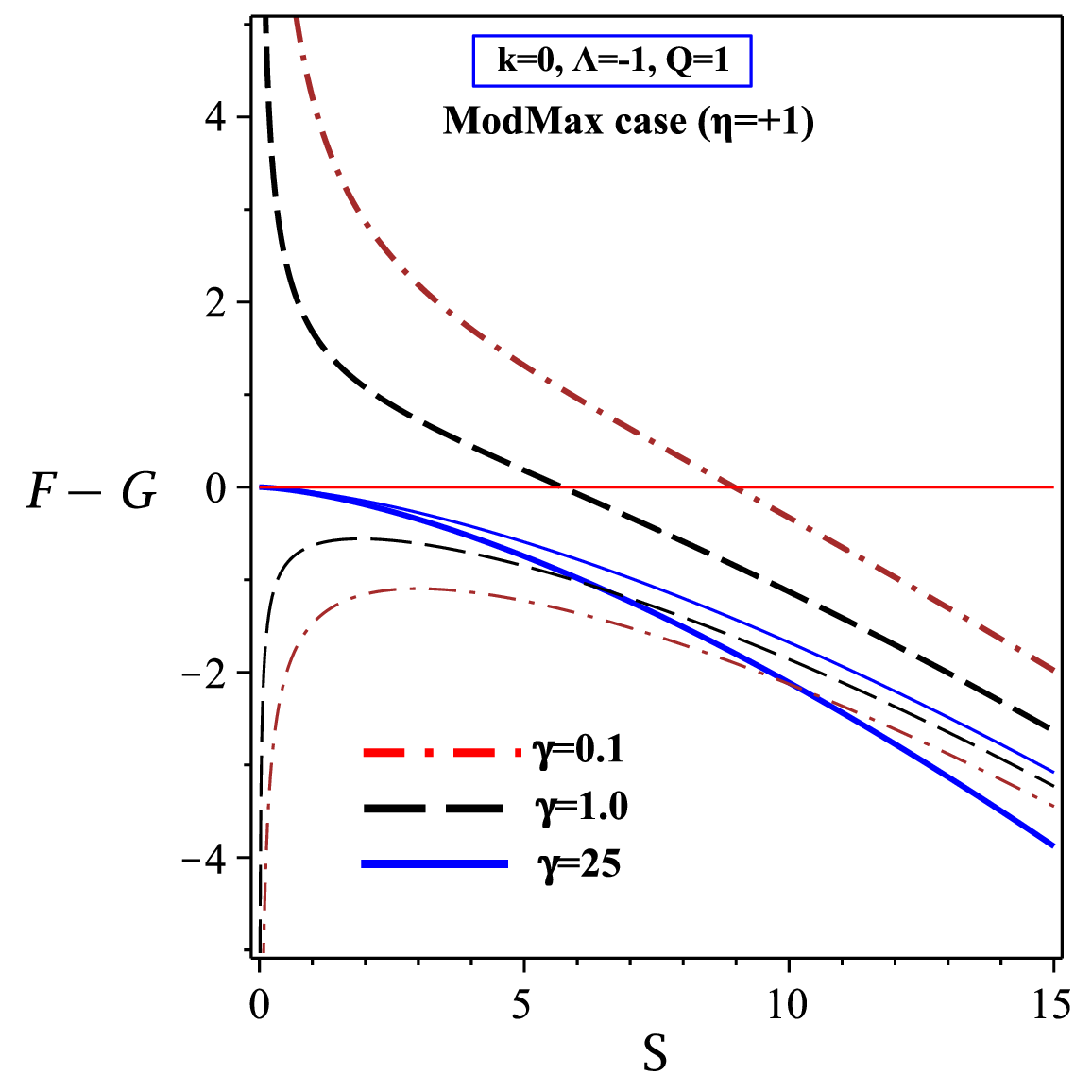} %
\includegraphics[width=0.32\linewidth]{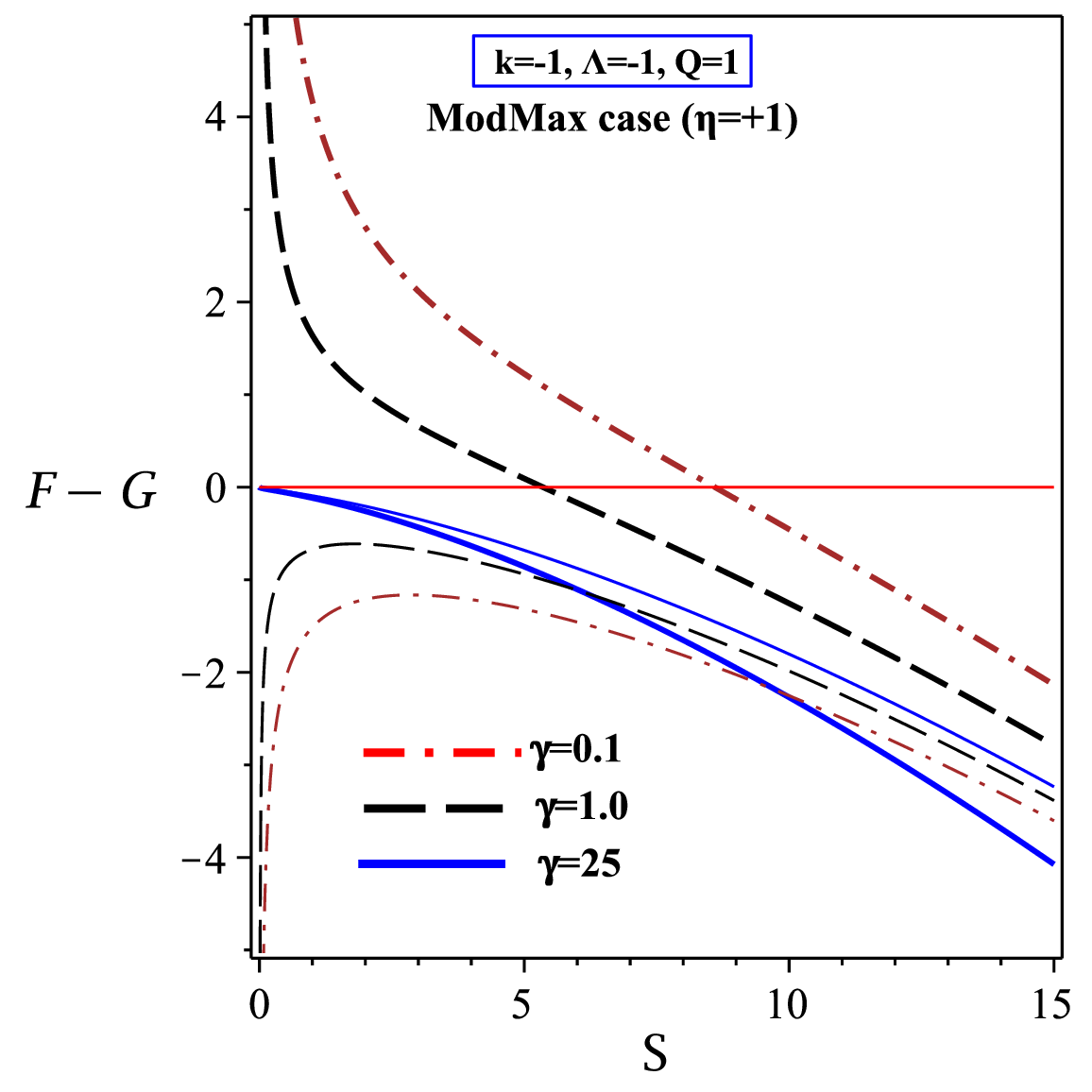} \newline
\caption{The Helmholtz free energy $F$ (thick lines) and the Gibbs free
energy $G$ (thin lines) versus entropy $S$ for $k=+1$ (left panel), $k=0$
(middle panel), and $k=-1$ (right panel) by considering the ModMax field ($%
\protect\eta =+1$).}
\label{Fig11}
\end{figure}

The largest and smallest real positive roots associated with the Helmholtz
free energy correspond to the topological indices $k=+1$, and $k=-1$,
respectively. Under the condition of a negative cosmological constant ($%
\Lambda <0$), the existence of roots for the ModAMax AdS black holes is
determined by the specific thermodynamic potential being analyzed. More
specifically, for the case $\eta =+1$, exactly one real, positive root is
identified for the Helmholtz free energy ($F=0$). To characterize the area
of global thermodynamic stability for the Mod(A)Max AdS black holes, we
present a plot of $F$ versus the entropy ($S$) in Fig. \ref{Fig11}. Our
findings reveal two critical points related to the effect of $\gamma $,
which are

\begin{figure}[tbph]
\centering
\includegraphics[width=0.32\linewidth]{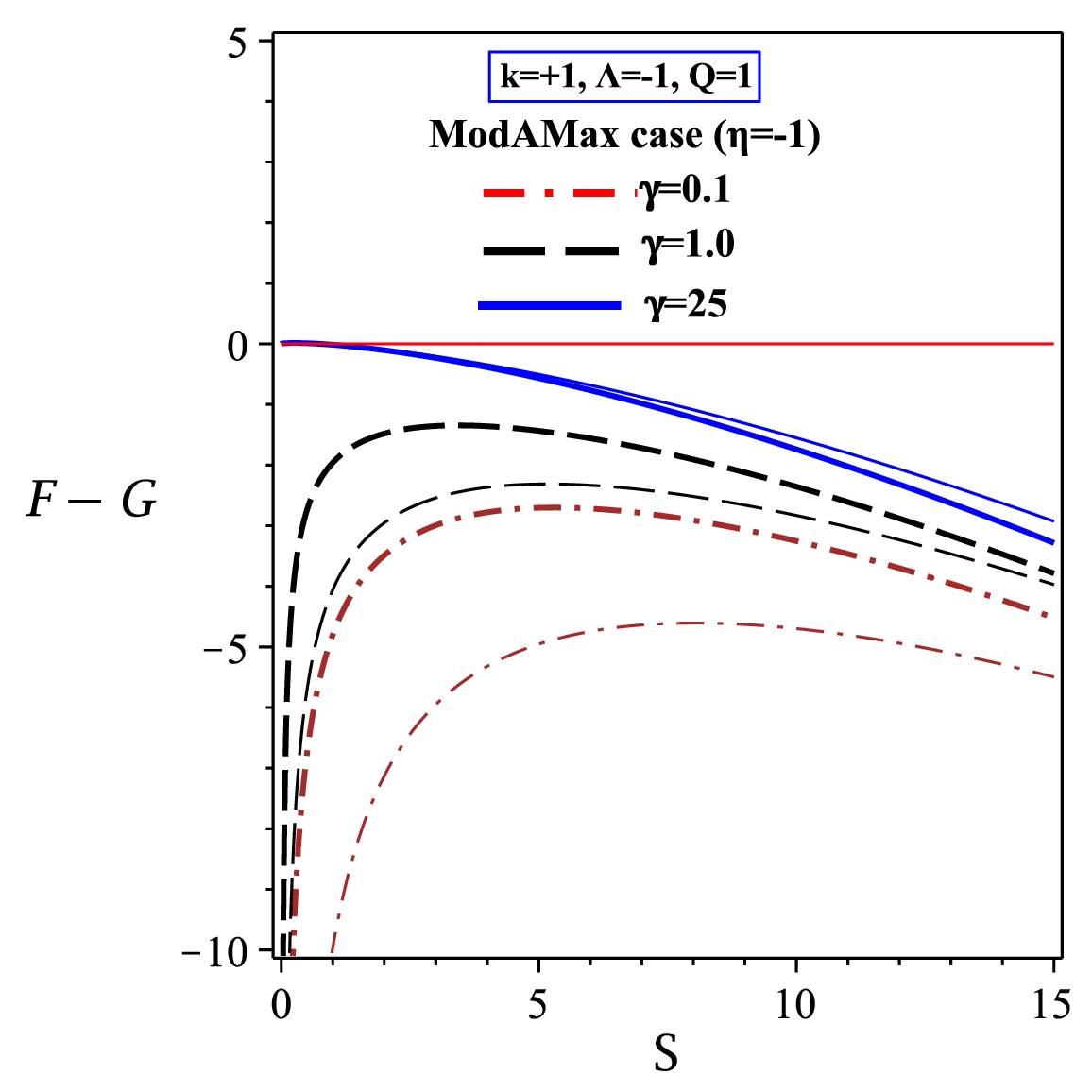} %
\includegraphics[width=0.32\linewidth]{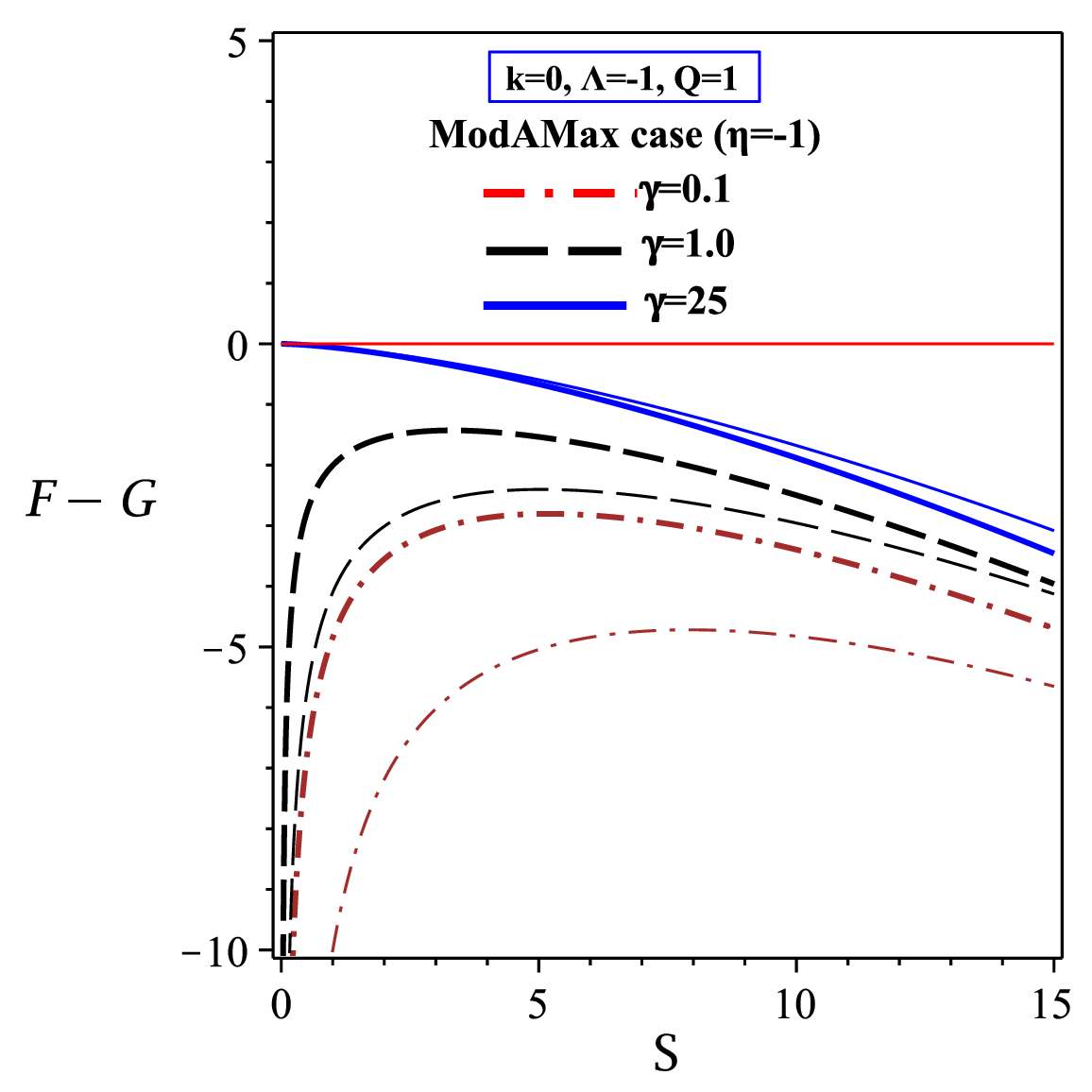} %
\includegraphics[width=0.32\linewidth]{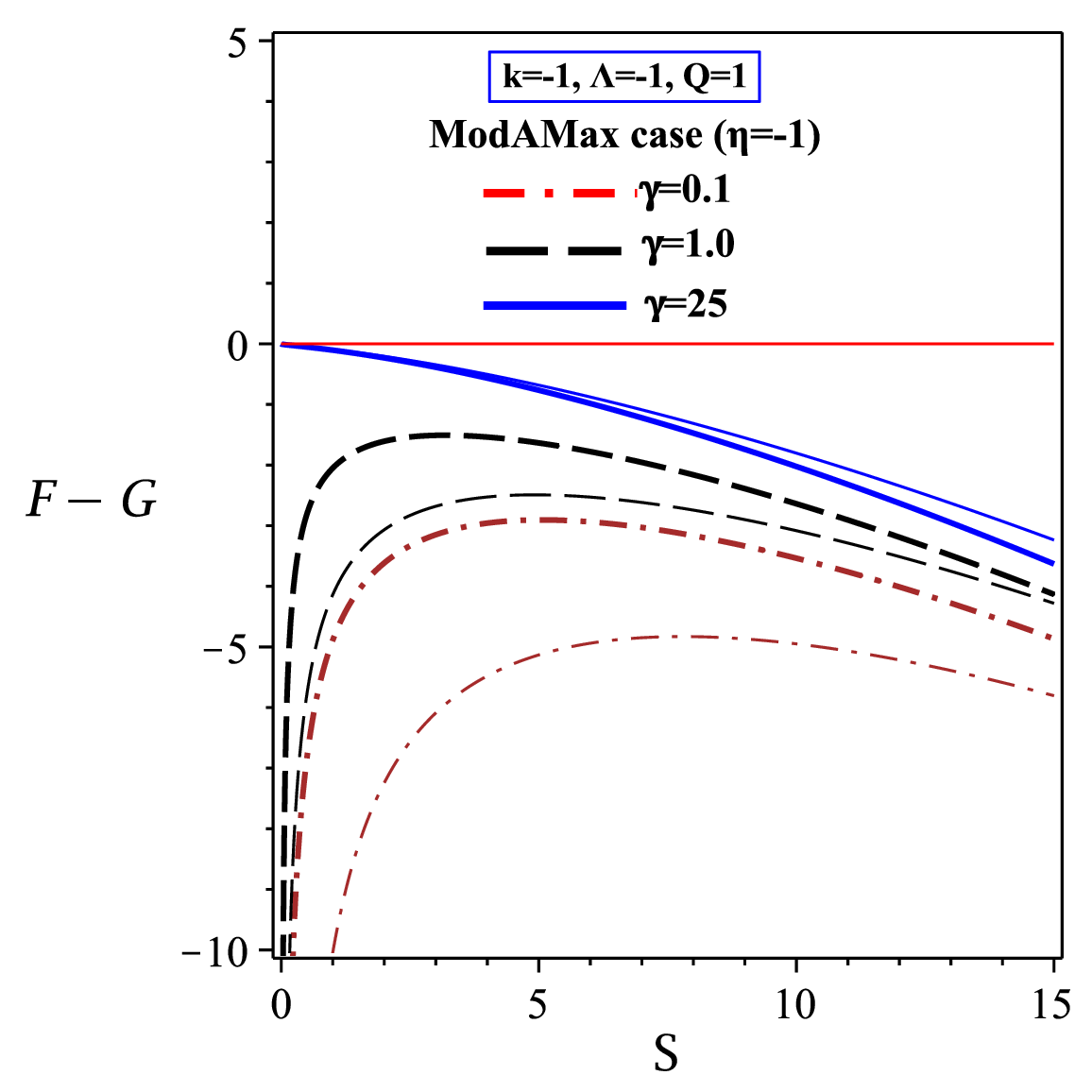} \newline
\caption{The Helmholtz free energy $F$ (thick lines) and the Gibbs free
energy $G$ (thin lines) versus entropy $S$ for $k=+1$ (left panel), $k=0$
(middle panel), and $k=-1$ (right panel) by considering the ModAMax field ($%
\protect\eta =-1$).}
\label{Fig12}
\end{figure}

\begin{description}
\item[\textbf{ModMax field ($\protect\eta =+1$):}] Similar to preceding
analyses, a critical value for the ModMax parameter ($\gamma _{critical}$)
demarcates two distinct stability behaviors: i) For $\gamma <\gamma
_{critical}$: The Helmholtz free energy exhibits a sign change at $S_{F=0}$
, being negative for $S>S_{F=0}$ and positive for $S<S_{F=0}$. Consequently,
large ModMax AdS black holes satisfy the global stability condition ($F<0$)
when $S>S_{F=0}$ (as indicated by the dotted-dashed and dashed bold lines in
Fig. \ref{Fig11}). Furthermore, increasing the ModMax parameter expands this
region of global stability. ii) For $\gamma >\gamma _{critical}$ and $k=+1$:
Only one real root exists for the Helmholtz free energy. Global stability is
satisfied for ModMax AdS black holes when $S>S_{F=0}$ (shown by the
continuous thick line in the left panel of Fig. \ref{Fig11}). Cases $k=0$,
and $k=-1$: No real root exists for $F$. In these scenarios, $F$ remains
strictly negative across all relevant physical horizons. Therefore, ModMax
AdS black holes with zero ($k=0$) and negative ($k=-1$) topological indices
inherently respect the global stability criterion ($F<0$) (illustrated in
the thick lines in the middle and right panels of Fig. \ref{Fig11}).

\item[\textbf{ModAMax field ($\protect\eta =-1$):}] In this regime, the
Helmholtz free energy for the ModAMax AdS black holes exhibits no real root (%
$F\neq 0$); consequently, this quantity is strictly negative across all
relevant horizons. Thus, the ModAMax AdS black holes intrinsically satisfy
the global stability condition. In other words, $F$ remains negative ($F<0$)
for ModAMax AdS black holes characterized by various topological indices (as
shown by the thick lines in Fig. \ref{Fig12}).
\end{description}

The Gibbs free energy (or Gibbs potential) is defined in the following form 
\begin{equation}
G=M\left( S,Q\right) -TS-\eta UQ,  \label{G}
\end{equation}%
by applying $U=\left( \frac{\partial M\left( S,Q\right) }{\partial Q}\right)
_{S}$ , Eqs. (\ref{MSQ}) and (\ref{TSQ}) within Eq. (\ref{G}), we can obtain
the Gibbs free energy as 
\begin{equation}
G=\frac{4\Lambda S^{2}+3kS+48\pi ^{2}\left( \frac{3}{4}-\eta \right)
Q^{2}e^{-\gamma }}{24\pi \sqrt{S}}.
\end{equation}

To determine the point of thermodynamic equilibrium, the equation $G=0$ must
be solved. Analysis of the resulting solutions indicated that there is only
one real positive root for $\eta =+1$ (ModMax field), and $k=+1$, which is
in the following form 
\begin{equation}
S_{G=0}=\frac{3+\sqrt{9+192\pi ^{2}\Lambda Q^{2}e^{-\gamma }}}{-8\Lambda },
\end{equation}%
this root is positive when $\gamma $ includes very large values (see the
continuous thin line in the left panel of Fig. \ref{Fig11}). It is notable
that, the global thermodynamic stability is therefore established by the
condition $G<0$. To rigorously evaluate the global stability criterion for
these black holes, we present the results in Figs. \ref{Fig11} and \ref%
{Fig12}, which are organized into six distinct panels. Our findings reveals
that:

\begin{description}
\item[\textbf{ModMax field ($\protect\eta =+1$):}] For the case $k=+1$, a
critical value for the ModMax parameter, $\gamma _{critical}$, dictates the
existence of a real, positive root for the Gibbs free energy ($G=0$). When $%
\gamma <\gamma _{critical}$, the Gibbs free energy remains negative ($G<0$);
consequently, the ModMax AdS black holes always satisfy the global stability
condition. Conversely, when $\gamma >\gamma _{critical}$, a single real,
positive root for $G=0$ emerges, which defines the transition between
negative and positive regions of $G$. Specifically, $G$ is negative
(positive) when $S>S_{G=0}$ ($S<S_{G=0}$), as illustrated by the continuous
thin line in the left panel of Fig. \ref{Fig11}. For the topological
constants $k=0$, and $k=-1$, the Gibbs free energy is invariably negative ($%
G<0$), ensuring that the ModMax AdS black holes satisfy the global stability
criterion across all relevant parameter values (see the thin lines in the
middle and right panels of Fig. \ref{Fig11}).

\item[\textbf{ModAMax field ($\protect\eta =-1$):}] In this scenario, no
real root for the Gibbs free energy ($G=0$) exists, and the potential
remains negative ($G<0$) across all physically relevant radii. Consequently,
the ModAMax AdS black holes inherently satisfy the global stability
condition (as depicted by the thin lines in Fig. \ref{Fig12}).
\end{description}

\section{Joule--Thomson expansion}

During a Joule--Thomson expansion, a gas flows from a high-pressure area
through a porous plug or small valve into a lower-pressure section of a
thermally insulated tube, maintaining a constant enthalpy throughout the
process. The resulting temperature change as a function of pressure is
described by 
\begin{equation}
\mu _{J}=\left( \frac{\partial T}{\partial P}\right) _{H},  \label{4}
\end{equation}%
where $\mu _{J}$ is a Joule--Thomson coefficient. One can predict whether
the gas will cool or heat by examining the sign of $\mu _{J}$. In a
Joule--Thomson expansion, the pressure always decreases, meaning the change
in pressure is negative; however, the temperature change can be either
positive or negative. If the temperature decreases, the Joule--Thomson
coefficient is positive, indicating that the gas cools.

The Joule--Thomson coefficient can be derived by expressing the fundamental
thermodynamic relations in terms of volume and heat capacity at constant
pressure. For a system with a fixed number of particles $N$, the first law
of thermodynamics gives 
\begin{equation}
dM=dH=TdS+VdP+\eta UdQ.  \label{dh}
\end{equation}

In a Joule-Thomson expansion, enthalpy is constant ($dH=0$) and any
non-mechanical work is assumed to be zero, which leads to 
\begin{equation}
T\left( \frac{\partial S}{\partial P}\right) _{H}+V=0.  \label{tv}
\end{equation}

Because the entropy $S$ is a state function, its differential can be written
as 
\begin{equation}
dS=\left( \frac{\partial S}{\partial P}\right) _{T}dT+\left( \frac{\partial S%
}{\partial T}\right) _{P}dP.
\end{equation}%
Thus, the partial derivative of entropy with respect to pressure at constant
enthalpy can be written: 
\begin{equation}
\left( \frac{\partial S}{\partial P}\right) _{H}=\left( \frac{\partial S}{%
\partial P}\right) _{T}\left( \frac{\partial T}{\partial P}%
\right)_{H}+\left( \frac{\partial S}{\partial T}\right) _{P}.  \label{a8}
\end{equation}

Substituting Eq. (\ref{a8}) into Eq. (\ref{tv}), gives 
\begin{equation}
T\left[ \left( \frac{\partial S}{\partial P}\right) _{T}+\left( \frac{%
\partial S}{\partial T}\right) _{P}\left( \frac{\partial T}{\partial P}
\right) _{H}\right] +V=0.
\end{equation}

Applying the Maxwell relation $\frac{C_{P}}{T}=\left( \frac{\partial S}{%
\partial T}\right) _{P}$ and the identity $\left( \frac{\partial S}{\partial
T}\right) _{T}=-\left( \frac{\partial V}{\partial T}\right) _{P}$, we obtain

\begin{equation}
-T\left( \frac{\partial V}{\partial T}\right) _{P}+C_{P}\left( \frac{%
\partial T}{\partial P}\right) _{H}+V=0,
\end{equation}%
which can be rearranged to give the Joule-Thomson coefficient 
\begin{equation}
\mu _{J}=\left( \frac{\partial T}{\partial P}\right) _{H}=\frac{1}{C_{p}}%
\left[ T\left( \frac{\partial V}{\partial T}\right) _{P}-V\right] .
\label{muJTexpl}
\end{equation}
Eq. (\ref{muJTexpl}) shows that the sign of $\mu _{J}$ is determined by the
bracketed term, which yields the classical inversion curve where $\mu _{J}=0$%
, while the magnitude and possible divergence of $\mu _{J}$ are controlled
by the factor $1/C_{P}$. The notation $C_{P}$ may also be written as $%
C_{P,Q} $ to emphasize that, during a Joule--Thomson expansion, the heat
capacity is evaluated at constant pressure and fixed electric charge. The
appearance of $C_{P}$ follows directly from the isoenthalpic condition of
the process and from the first law of thermodynamics; other heat capacities,
such as $C_{V}$ do not enter the analysis of the Joule-Thomson effect.

The expression (\ref{muJTexpl}) makes two features explicit:

\begin{itemize}
\item $C_{P}$\textbf{$\rightarrow $}$0$ : this occurs when $T\rightarrow 0$.
In this regime $|\mu _{J}|\rightarrow \infty $, as expected near extremality.

\item $C_{P}$\textbf{$\rightarrow $}$\infty $ : this occurs at second-order
phase transitions. In this limit, $\mu _{J}\rightarrow 0$ \textit{regardless}
of the value of the bracketed term in Eq.~\ref{muJTexpl}. Hence, the
inversion condition $\mu _{J}=0$ can be satisfied trivially due to the
divergence of $C_{P}$, a mechanism distinct from the classical inversion
curve where $T(\partial V/\partial T)_{P}=V$. This behavior underscores that
near critical points, the dominant thermodynamic anomaly is caloric rather
than volumetric.
\end{itemize}

Setting $T\left( \frac{\partial V}{\partial T}\right) _{P}-V=0$, yields the
condition for the inversion temperature, beyond which the sign of $\mu _{J}$
changes and the transition between heating and cooling occurs. The inversion
temperature is therefore expressed as 
\begin{equation}
T_{i}=V\left( \frac{\partial T}{\partial V}\right) _{P}.
\end{equation}

As is well known, the thermodynamic pressure is related to the cosmological
constant through%
\begin{equation}
P=-\frac{\Lambda }{8\pi }.  \label{12}
\end{equation}

By substituting this relation into the expression for the black hole mass,
one can rewrite the mass in the form%
\begin{equation}
M=\frac{kr_{+}}{8\pi }+\frac{Pr_{+}^{3}}{3}+\frac{\eta q^{2}e^{-\gamma }}{
8\pi r_{+}}=\left\{ 
\begin{array}{ccc}
\frac{kr_{+}}{8\pi }+\frac{Pr_{+}^{3}}{3}+\frac{q^{2}e^{-\gamma }}{8\pi r_{+}%
}, &  & \text{ModMax} \\ 
&  &  \\ 
\frac{kr_{+}}{8\pi }+\frac{Pr_{+}^{3}}{3}-\frac{q^{2}e^{-\gamma }}{8\pi r_{+}%
}, &  & \text{ModAMax}%
\end{array}%
\right. .
\end{equation}%
which is taken as the enthalpy $H$ of the system. Next, by substituting the
pressure relation (\ref{12}) into the expression for the temperature, we
obtain

\begin{equation*}
P=\frac{T}{2r_{+}}-\frac{k}{8\pi r_{+}^{2}}+\frac{\eta q^{2}e^{-\gamma }}{
8\pi r_{+}^{4}}=\left\{ 
\begin{array}{ccc}
\frac{T}{2r_{+}}-\frac{k}{8\pi r_{+}^{2}}+\frac{q^{2}e^{-\gamma }}{8\pi
r_{+}^{4}}, &  & \text{ModMax} \\ 
&  &  \\ 
\frac{T}{2r_{+}}-\frac{k}{8\pi r_{+}^{2}}-\frac{\eta q^{2}e^{-\gamma }}{8\pi
r_{+}^{4}}, &  & \text{ModAMax}%
\end{array}%
\right. .
\end{equation*}

With these expressions at hand, we now employ the definition of the
Joule--Thomson coefficient (\ref{4}) to determine its explicit form.%
\begin{equation}
\mu _{J}=\frac{\frac{4k}{r_{+}^{2}}+\frac{16\pi P}{3}-\frac{6\eta
q^{2}e^{-\gamma }}{r_{+}^{4}}}{4\pi \left( \frac{3k}{8\pi r_{+}^{3}}+\frac{3P%
}{r_{+}}-\frac{3\eta q^{2}e^{-\gamma }}{8\pi r_{+}^{5}}\right) }=\left\{ 
\begin{array}{ccc}
\frac{\frac{4k}{r_{+}^{2}}+\frac{16\pi P}{3}-\frac{6q^{2}e^{-\gamma }}{%
r_{+}^{4}}}{4\pi \left( \frac{3k}{8\pi r_{+}^{3}}+\frac{3P}{r_{+}}-\frac{%
3q^{2}e^{-\gamma }}{8\pi r_{+}^{5}}\right) }, &  & \text{ModMax} \\ 
&  &  \\ 
\frac{\frac{4k}{r_{+}^{2}}+\frac{16\pi P}{3}+\frac{6q^{2}e^{-\gamma }}{%
r_{+}^{4}}}{4\pi \left( \frac{3k}{8\pi r_{+}^{3}}+\frac{3P}{r_{+}}+\frac{%
3q^{2}e^{-\gamma }}{8\pi r_{+}^{5}}\right) }, &  & \text{ModAMax}%
\end{array}%
\right. .  \label{mup}
\end{equation}

To investigate the divergence of the Joule--Thomson coefficient in terms of
the black hole mass, we eliminate the thermodynamic pressure using the
enthalpy relation. Substituting $P\left( P\rightarrow M,r_{+}\right) $ into
the above expression yields

\begin{equation}
\mu _{J}=\frac{\frac{2k}{r_{+}^{2}}+\frac{16\pi M}{r_{+}^{3}}-\frac{8\eta
q^{2}e^{-\gamma }}{r_{+}^{4}}}{4\pi \left( \frac{9M}{r_{+}^{4}}-\frac{6k}{%
8\pi r_{+}^{3}}-\frac{3\eta q^{2}e^{-\gamma }}{2\pi r_{+}^{5}}\right) }%
=\left\{ 
\begin{array}{ccc}
\frac{\frac{2k}{r_{+}^{2}}+\frac{16\pi M}{r_{+}^{3}}-\frac{8q^{2}e^{-\gamma }%
}{r_{+}^{4}}}{4\pi \left( \frac{9M}{r_{+}^{4}}-\frac{6k}{8\pi r_{+}^{3}}-%
\frac{3q^{2}e^{-\gamma }}{2\pi r_{+}^{5}}\right) }, &  & \text{ModMax} \\ 
&  &  \\ 
\frac{\frac{2k}{r_{+}^{2}}+\frac{16\pi M}{r_{+}^{3}}+\frac{8q^{2}e^{-\gamma }%
}{r_{+}^{4}}}{4\pi \left( \frac{9M}{r_{+}^{4}}-\frac{6k}{8\pi r_{+}^{3}}+%
\frac{3q^{2}e^{-\gamma }}{2\pi r_{+}^{5}}\right) }, &  & \text{ModAMax}%
\end{array}%
\right. .
\end{equation}%
The divergence of $\mu _{J}$ occurs when the denominator vanishes. Solving
this condition for $r_{+}$ at fixed mass leads to the quadratic equation: 
\begin{equation}
kr_{+}^{2}-12\pi Mr_{+}+2\eta q^{2}e^{-\gamma }=0.
\end{equation}

The real and positive roots of this equation define the physical horizon
radii at which $\mu_{J}$ becomes singular. This indicates that the presence
of a divergence is highly sensitive to both the topology $k$ and the NED
parameter $\eta$. Since the equation is quadratic in $r_{+}$, it can yield
up to two mathematical roots; however, only real and positive solutions
correspond to physical black holes. The specific roots depend explicitly on $%
k$ and $\eta$. For spherical topology ($k=+1$), the discriminant indicates
whether a divergence exists. In flat topology ($k=0$), a single positive
root emerges for ModMax black holes. In the hyperbolic case ($k=-1$), ModMax
($\eta =+1$) consistently yields a real root, while ModAMax ($\eta =-1$)
results in negative or complex solutions, thus preventing a physical
divergence. These findings are summarized in the table below.

\begin{table}[]
\caption{Solutions of the divergence condition.}
\label{tab:Jmp}%
\begin{ruledtabular}
\begin{tabular}{|c|c|c|}
Model & Topology ($k$) & $r_{+{div}_{\mu _{J}}}$ \\
\hline
ModMax & $+1$ & $6\pi M+\sqrt{36\pi ^{2}M^{2}-2q^{2}e^{-\gamma }}$ \\
($\eta=+1$) & $0$ & $\frac{q^{2}e^{-\gamma }}{6\pi M}$ \\
& $-1$ & $-6\pi M+\sqrt{36\pi ^{2}M^{2}+2q^{2}e^{-\gamma }}$\\
\hline
ModAMax & $+1$ & $6\pi M+\sqrt{36\pi ^{2}M^{2}+2q^{2}e^{-\gamma }}$ \\
($\eta=-1$) & $0$ & Does Not Exist \\
& \(-1\) & Does Not Exist \\
\end{tabular}
\end{ruledtabular}
\end{table}

\begin{figure}[]
\centering
\includegraphics[width=0.325\linewidth]{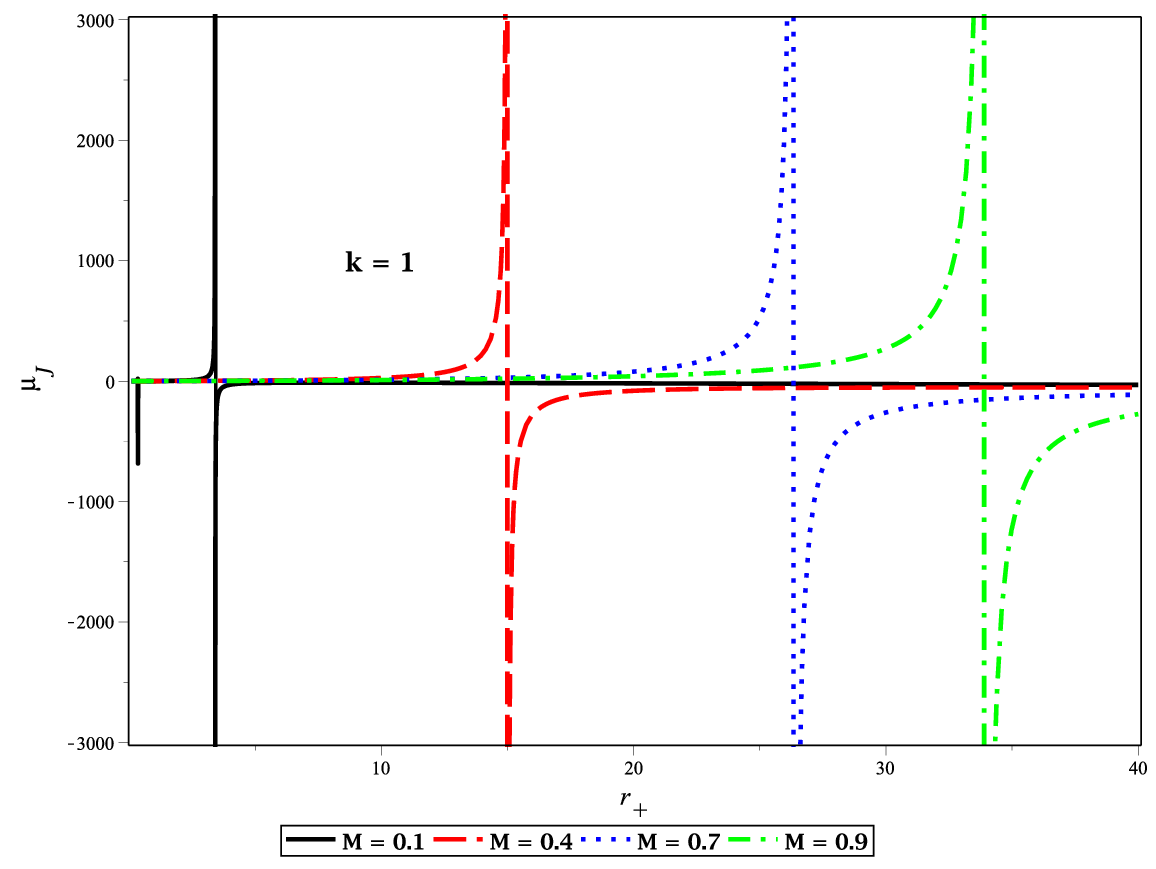} %
\includegraphics[width=0.325\linewidth]{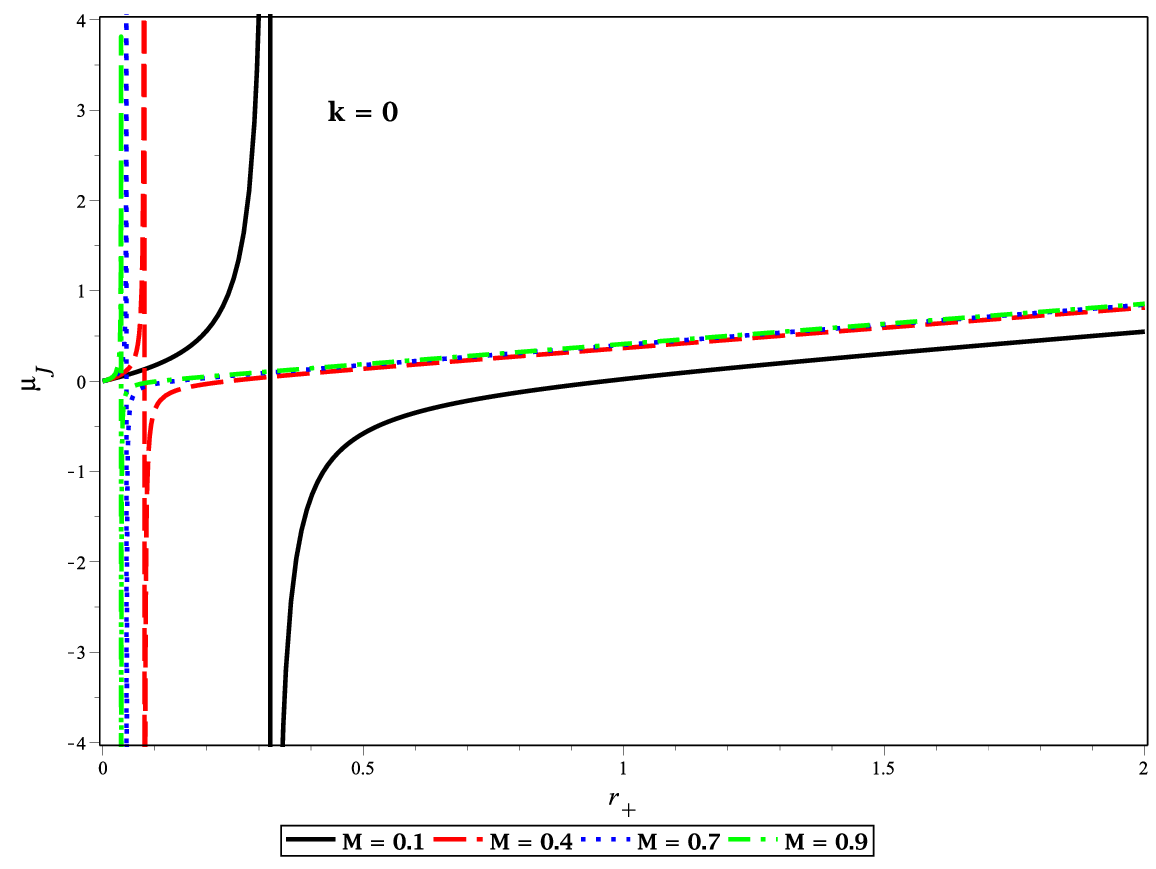} %
\includegraphics[width=0.325\linewidth]{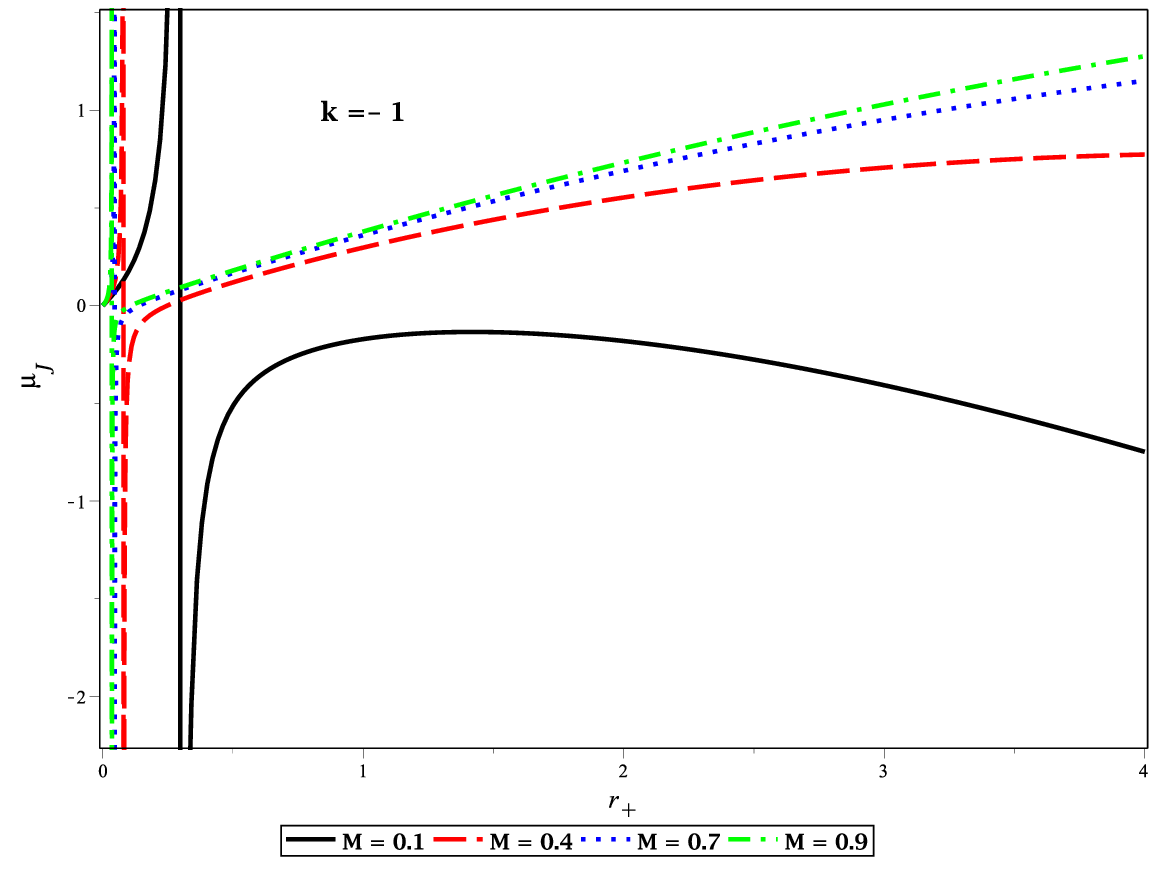} \newline
\caption{Joule-Thomson coefficient $\protect\mu _{J}$ with respect $r_{+}$,
and different values of the ModMax's parameters. Here we take: $q=1$ and $%
\protect\gamma=0.5$}
\label{FigmuMOD1}
\end{figure}

\begin{figure}[]
\centering
\includegraphics[width=0.325\linewidth]{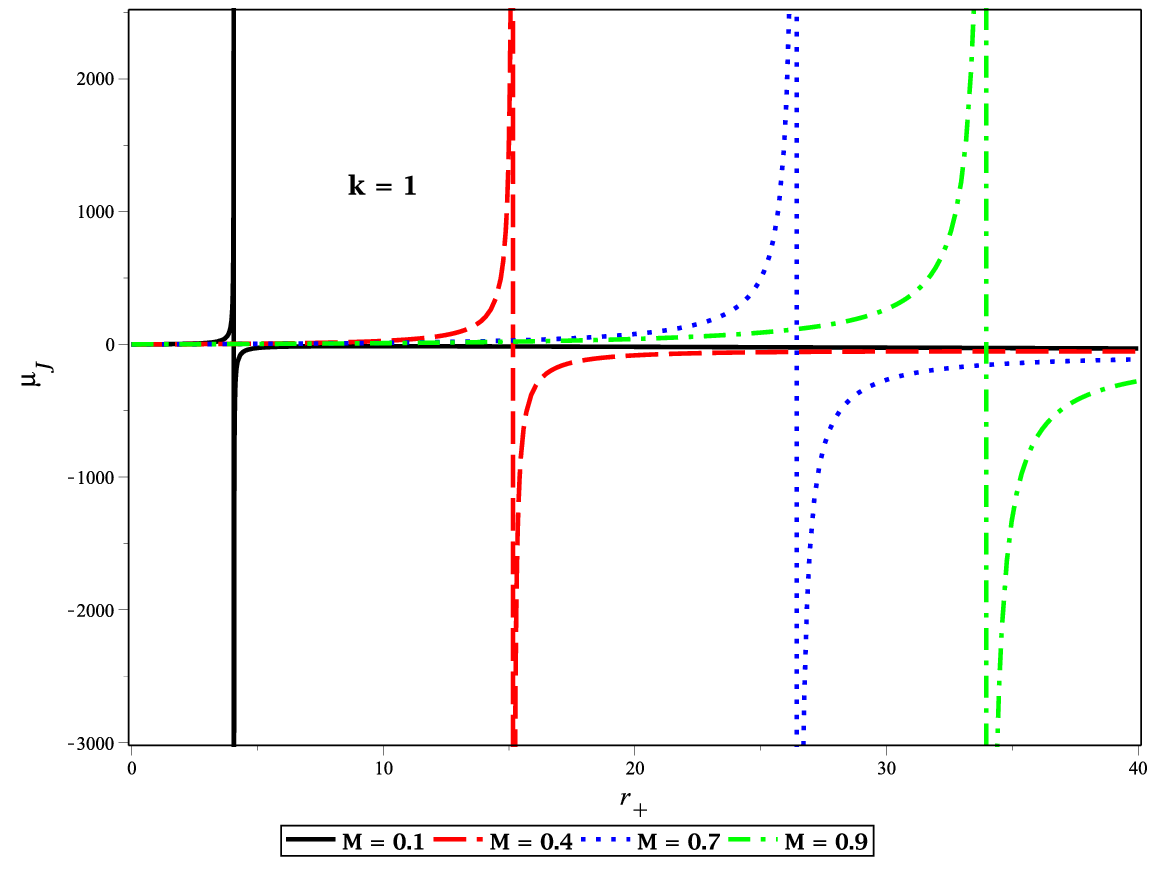} %
\includegraphics[width=0.325\linewidth]{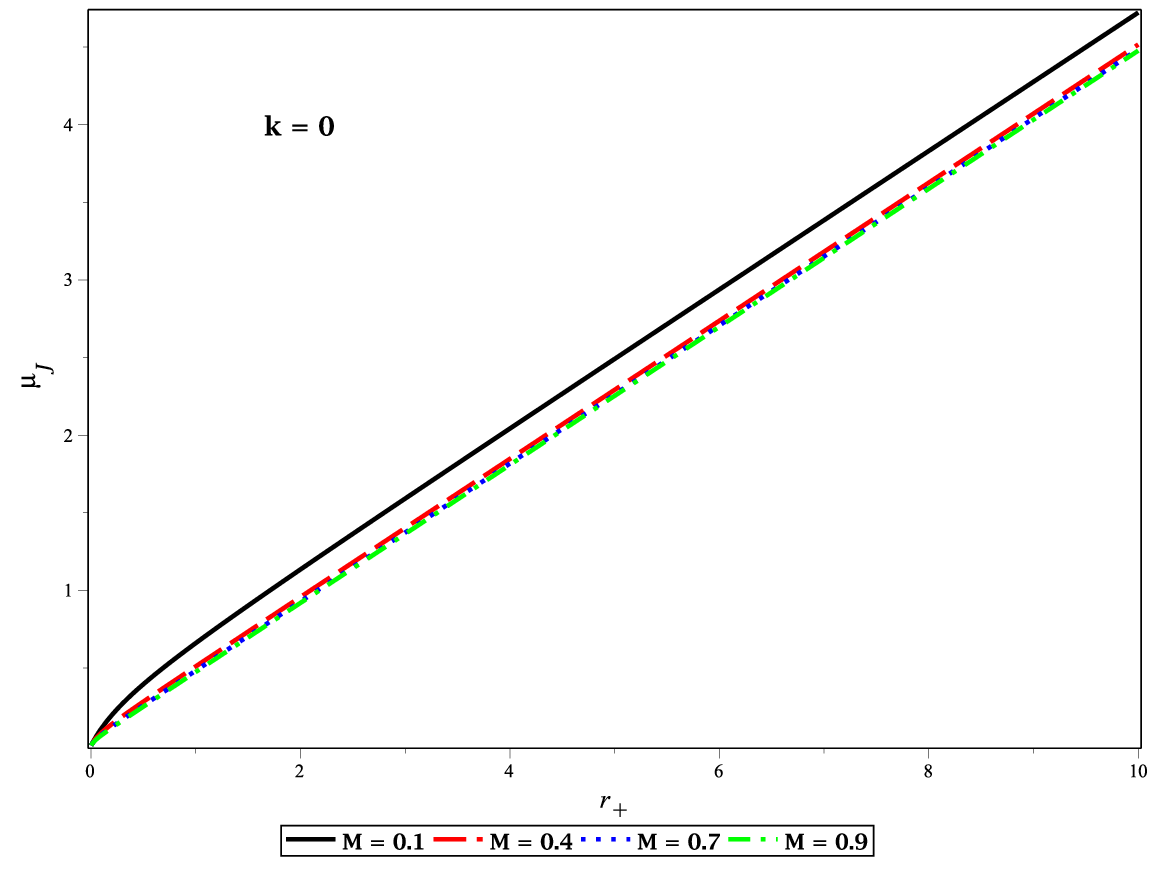} %
\includegraphics[width=0.325\linewidth]{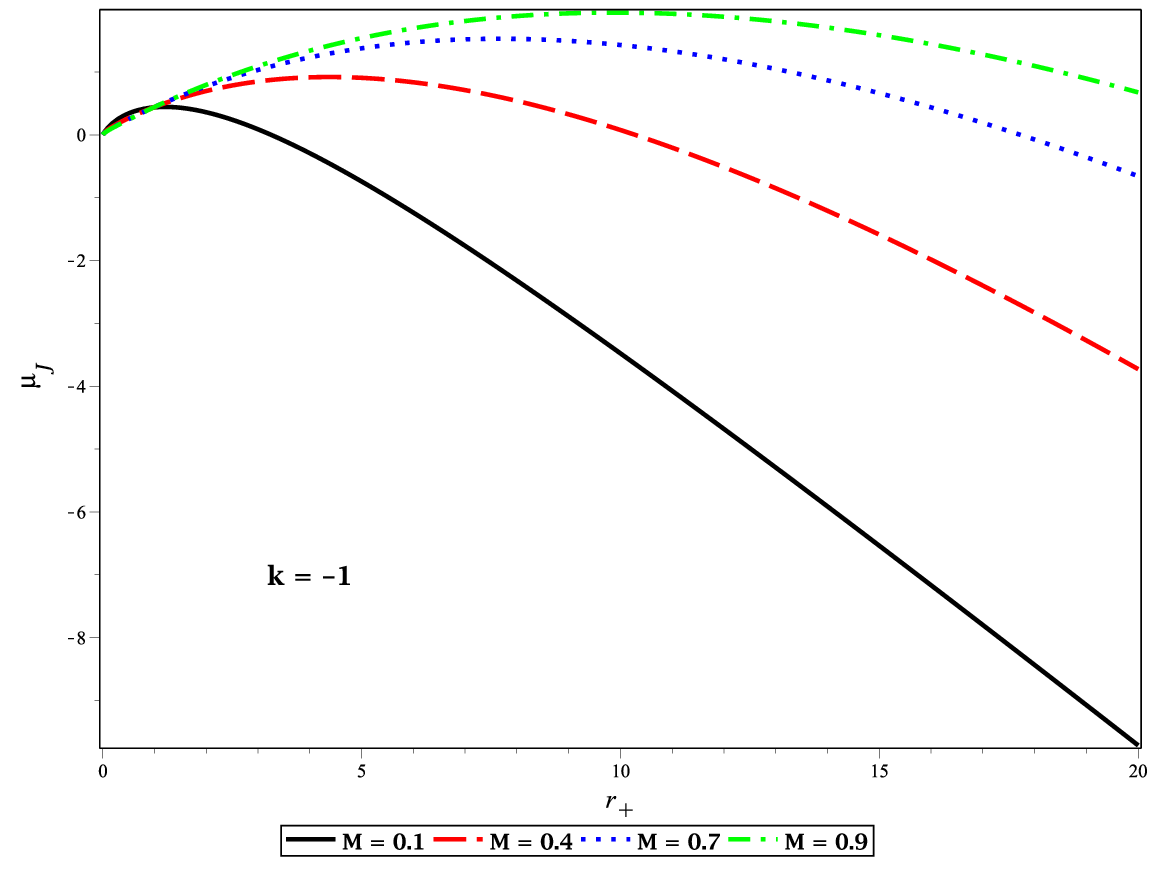} \newline
\caption{Joule-Thomson coefficient $\protect\mu _{J}$ with respect $r_{+}$,
and different values of the ModAMax's parameters. Here we take: $q=1$ and $%
\protect\gamma=0.5$}
\label{FigmuMODA1}
\end{figure}

We aim to investigate how the Joule--Thomson coefficient $\mu_{J}$ varies
with the horizon radius $r_{+}$ for different black hole masses $M$ within
the Mod(A)Max framework. To demonstrate this relationship, $\mu_{J}$ is
plotted against $r_{+}$ in Figs. \ref{FigmuMOD1} and \ref{FigmuMODA1}.

\begin{description}
\item[\textbf{ModMax ($\protect\eta =+1$):}] In this case, the
Joule--Thomson coefficient $\mu _{J}$ also exhibits a well-defined inversion
point separating the cooling and heating regimes. However, the \textit{\
critical distinction among the topological classes} is governed by the
influence of the black hole mass $M$. For $k=+1$, increasing the mass
systematically shifts the inversion point toward larger critical horizon
radii, reflecting the enhancement of the gravitational field strength.
Conversely, for $k=0$ and $k=-1$, the trend is inverted: increasing the mass
shifts the inversion point toward smaller critical horizon radii.

\item[\textbf{ModAMax ($\protect\eta =-1$):}] For the spherical topology ($%
k=+1$), $\mu _{J}$ exhibits a well-defined inversion point, separating the
cooling and heating regions in a manner similar to the ModMax case. In the
flat topology ($k=0$), $\mu _{J}$ remains positive for all values of $r_{+}$
, indicating that the system stays in a continuous cooling phase with no
inversion point. However, for the hyperbolic topology ($k=-1$), $\mu _{J}$
is initially positive for small $r_{+}$, corresponding to a cooling regime,
and then becomes negative as $r_{+}$ increases, signaling a transition from
cooling to heating. This reveals that hyperbolic ModAMax black holes exhibit
a mixed thermodynamic behavior, where both cooling and heating phases
coexist depending on the horizon radius. Overall, the results demonstrate
that the topology parameter $k$ strongly influences the thermal
characteristics of ModAMax AdS black holes and that the transition behavior
for $k=-1$ is qualitatively distinct from the other topological cases.
\end{description}

The condition $\mu _{J}=0$ applied to Eq.(\ref{mup}) , gives 
\begin{equation}
8\pi P_{i}r_{+}^{4}+6kr_{+}^{2}-9\eta q^{2}e^{-\gamma }=0.  \label{29}
\end{equation}

Solving this quartic equation for $r_{+}$ yields four roots; however, only
one of them is physically acceptable, as the others are either complex or
negative. The positive real root is given by 
\begin{equation}
r_{+}=r_{i}=\sqrt{-\frac{3k}{8\pi P_{i}}+\frac{3\sqrt{k^{2}+8\pi
P_{i}q^{2}\eta e^{-\gamma }}}{8\pi P_{i}}}=\left\{ 
\begin{array}{ccc}
\sqrt{-\frac{3k}{8\pi P_{i}}+\frac{3\sqrt{k^{2}+8\pi P_{i}q^{2}e^{-\gamma }} 
}{8\pi P_{i}}}, &  & \text{ModMax} \\ 
&  &  \\ 
\sqrt{-\frac{3k}{8\pi P_{i}}+\frac{3\sqrt{k^{2}-8\pi P_{i}q^{2}e^{-\gamma }} 
}{8\pi P_{i}}}, &  & \text{ModAMax}%
\end{array}%
\right. ,
\end{equation}%
where $P_{i}$ manifests the inversion pressure. $\allowbreak \allowbreak $%
Furthermore, the corresponding expression for the black hole inversion
temperature is :

\begin{equation}
T_{i}=\frac{2Pr_{i}}{3}-\frac{k}{12\pi r_{i}}+\frac{\eta e^{-\gamma }q^{2}}{
4\pi r_{i}^{3}}=\left\{ 
\begin{array}{ccc}
\frac{2Pr_{i}}{3}-\frac{k}{12\pi r_{i}}+\frac{e^{-\gamma }q^{2}}{4\pi
r_{i}^{3}}, &  & \text{ModMax} \\ 
&  &  \\ 
\frac{2Pr_{i}}{3}-\frac{k}{12\pi r_{i}}-\frac{e^{-\gamma }q^{2}}{4\pi
r_{i}^{3}}, &  & \text{ModAMax}%
\end{array}%
\right. .
\end{equation}

The minimum inversion temperature is obtained by substituting $r_{+}=r_{i}$
from Eq. (\ref{29}) and taking $P_{i}=0$ in the above relation, which gives 
\begin{equation*}
T_{i}^{\min }=\frac{k}{12\pi }\sqrt{\frac{2k}{3\eta q^{2}e^{-\gamma }}}.
\end{equation*}

Correspondingly, the minimum inversion mass is given by 
\begin{equation}
M_{i}^{\min }=\frac{5k}{24\pi }\sqrt{\frac{3\eta q^{2}e^{-\gamma }}{2k}}.
\end{equation}

Table. \ref{tab:cmp} presents a comparative summary of the Joule--Thomson
inversion behavior and the corresponding minimum quantities for the
topological Mod(A)Max AdS black holes. The results show that the existence
of a minimum inversion temperature $T_{i}^{min}$ and mass $M_{i}^{min}$
strongly depends on both the electrodynamic model and the horizon topology.
In the ModMax case ($\eta =+1$), well-defined and positive values of $%
T_{i}^{min}$ and $M_{i}^{min}$ appear for the spherical topology ($k=+1$),
indicating that these black holes can undergo a cooling phase during the
Joule--Thomson expansion. For the flat topology ($k=0$), the inversion
occurs only at zero temperature and mass, while for the hyperbolic case ($%
k=-1$), the resulting values are unphysical. In contrast, for the ModAMax
model ($\eta =-1$), the inversion characteristics are notably different.
Although spherical and hyperbolic topologies still exhibit sign changes in $%
\mu _{J}$, the corresponding minimum inversion quantities are unphysical.
For $k=0$, there is no inversion point, and the system remains in a
continuous cooling phase.

\begin{table}[]
\caption{Summary of Joule--Thomson inversion behavior and minimum quantities
for topological Mod(A)Max AdS black holes.}
\label{tab:cmp}%
\begin{ruledtabular}
\begin{tabular}{|c|c|c|c|c|c|}
Model & Topology ($k$) & $\mu_J$ Sign Change & Inversion Point? & $T_i^{\text{min}}$ & $M_i^{\text{min}}$ \\ \hline
ModMax & $+1$ & Yes & Yes & \(\dfrac{1}{12\pi}\sqrt{\dfrac{2}{3q^{2}e^{-\gamma}}}\) & \(\dfrac{5}{24\pi}\sqrt{\dfrac{3q^{2}e^{-\gamma}}{2}}\) \\ ($\eta=+1$) & $0$ & Yes & Yes & 0 & 0 \\ & $-1$ & Yes & Yes & unphysical & unphysical \\ \hline
ModAMax & $+1$ & Yes & Yes & unphysical & unphysical \\
($\eta=-1$) & $0$ & No & No & Does Not Exist & Does Not Exist \\
& \(-1\) & Yes & Yes & unphysical & unphysical \\
\end{tabular}
\end{ruledtabular}
\end{table}

\begin{figure}[]
\centering
\includegraphics[width=0.4\linewidth]{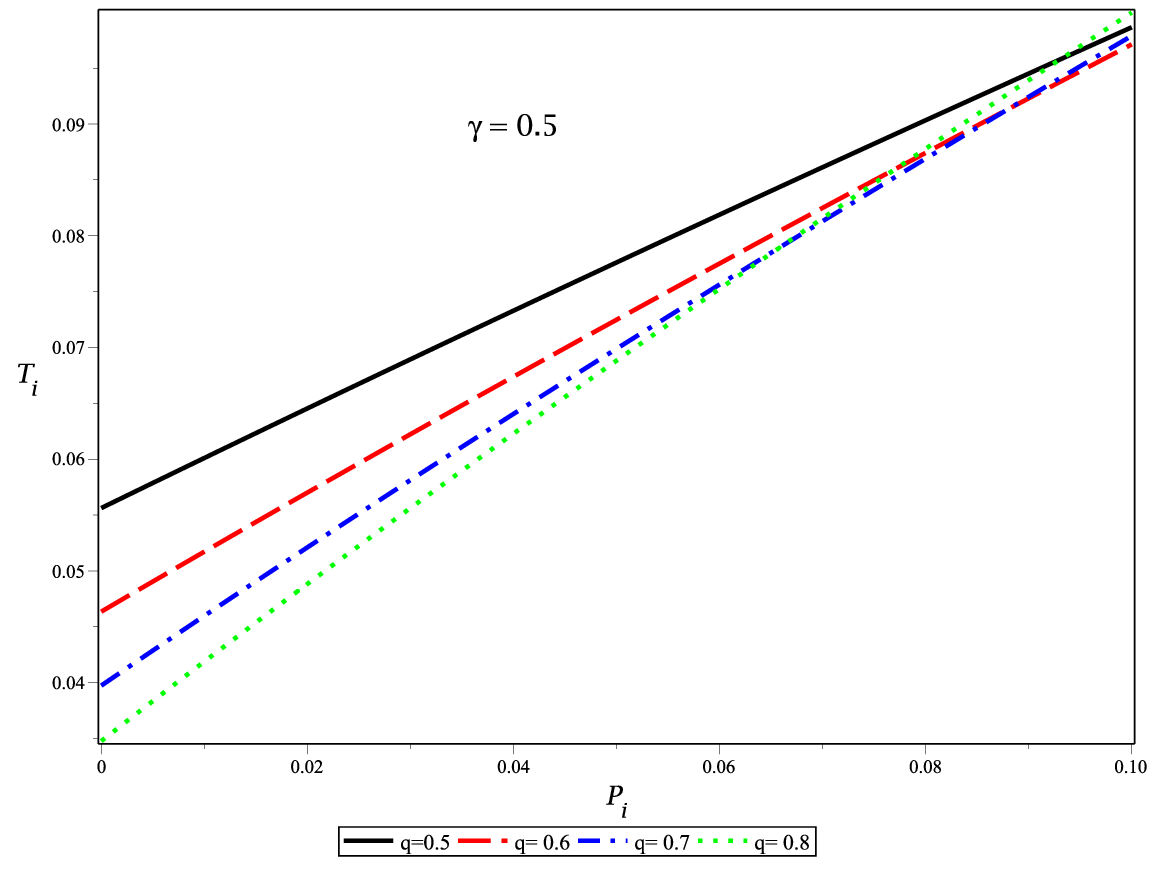} %
\includegraphics[width=0.4\linewidth]{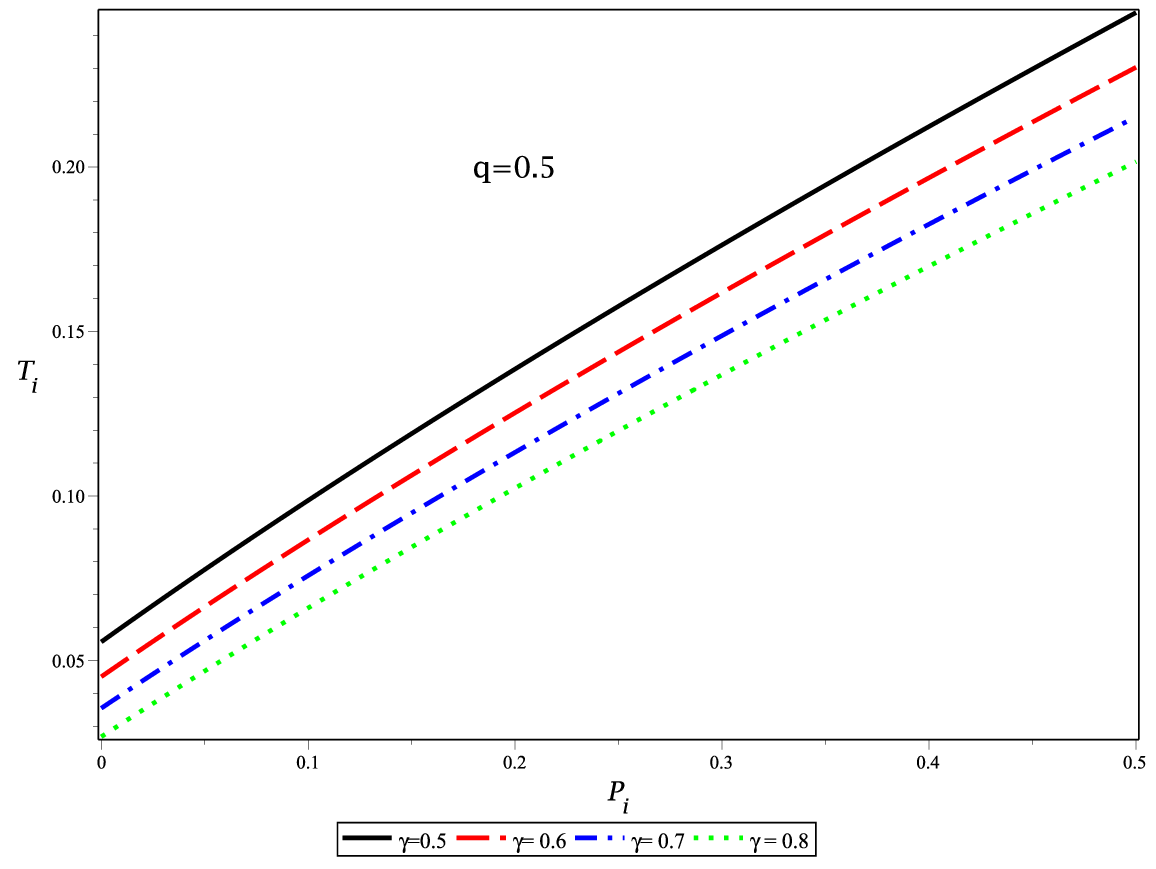} \newline
\caption{Inversion temperature $T_{i}$ of the ModMax black hole as a
function of inversion pressure $P_{i}$ for $k=+1$. }
\label{Figt11}
\end{figure}

In Fig. \ref{Figt11} we depicted the inversion temperature $T_{i}$ of the
ModMax black hole as a function of the inversion pressure $P_{i}$ for $k=1$.
From this plot, we observe that the inversion temperature increases
monotonically with pressure. Notably, unlike conventional thermodynamic
systems such as the van der Waals gas, these black holes do not exhibit a
maximum inversion temperature. Each curve possesses a minimum inversion
temperature $T_{i}^{\min }$, below which inversion does not occur. The
region above each curve represents the cooling regime ($\mu _{J}>0$), while
the area below corresponds to the heating regime ($\mu _{J}<0$) 

\section{Heat Engines}

\begin{figure}[]
\centering
\includegraphics[width=0.45\linewidth]{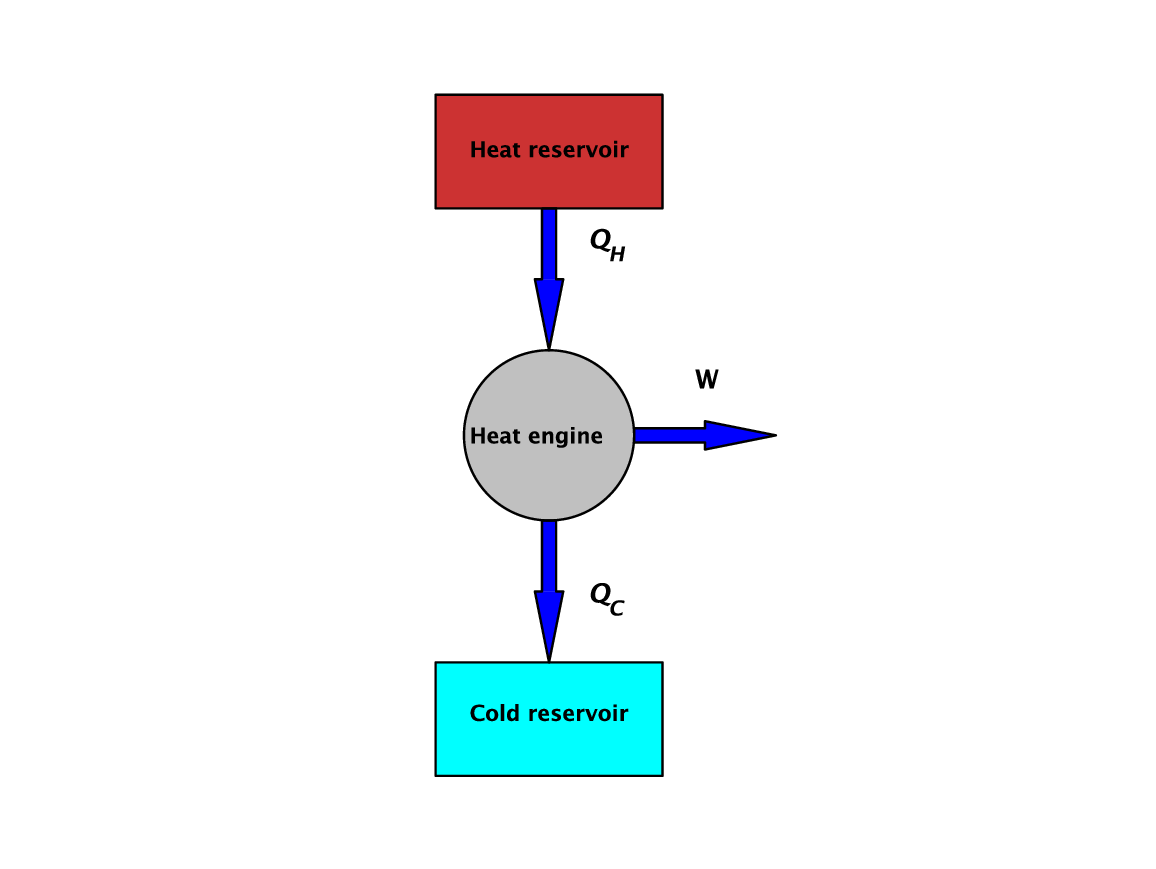} %
\includegraphics[width=0.35	\linewidth]{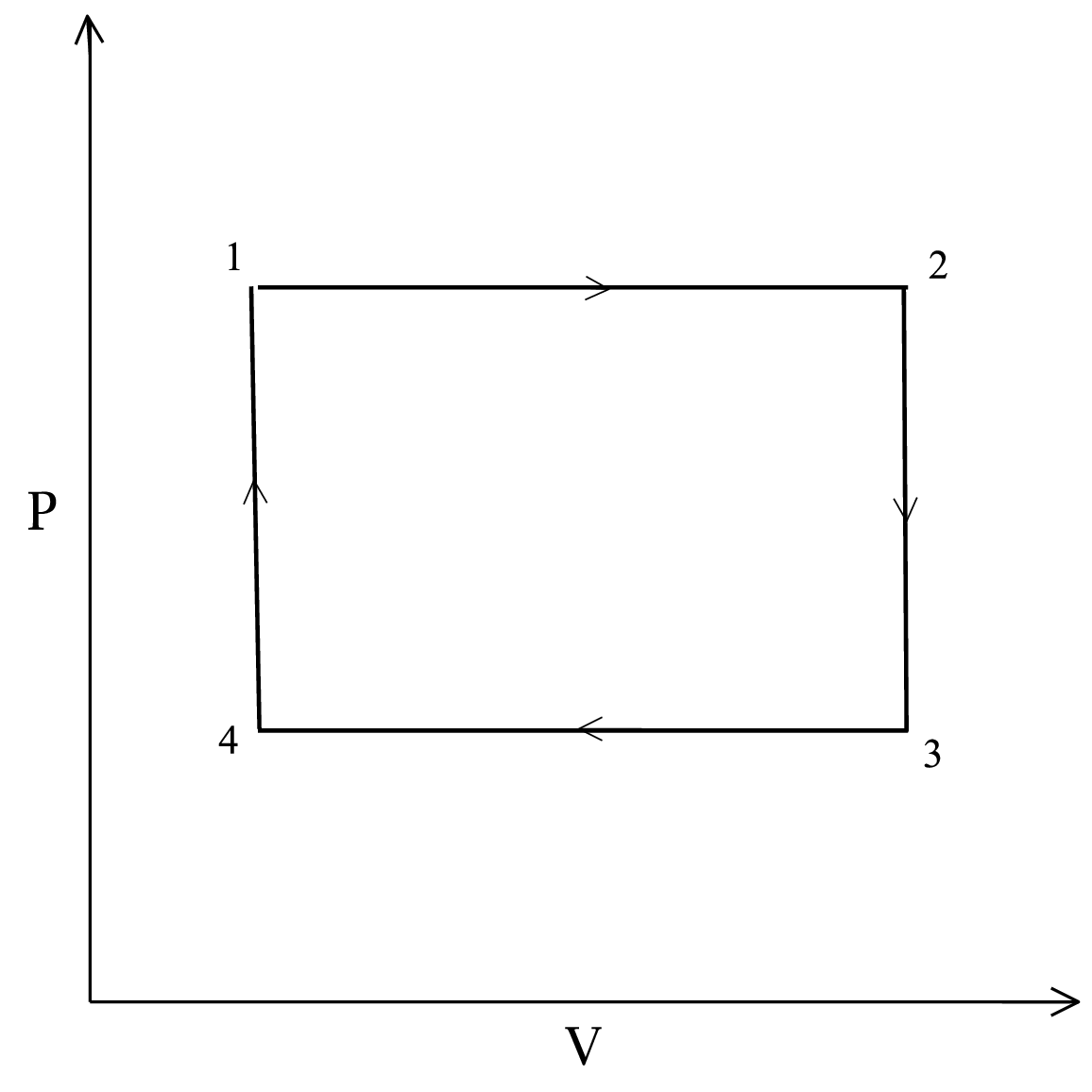} \newline
\caption{The figure on the left shows a schematic representation of the heat
engine, while the figure on the right illustrates its thermodynamic cycle.}
\label{FigH1}
\end{figure}

A black hole heat engine can be mathematically described by a closed $P-V$
cycle. In this cycle, the system absorbs heat $Q_{H}$ from a
high-temperature reservoir. As the process unfolds, part of this absorbed
heat is converted into mechanical work $W$, while the remaining heat $Q_{C}$
is released to a lower-temperature reservoir (Fig. \ref{FigH1}). Thus, the
efficiency $\Gamma $ of the heat engine is defined as 
\begin{equation}
\Gamma =\frac{W}{Q_{H}}.
\end{equation}%

The efficiency of a Carnot engine represents the highest theoretical
efficiency that any heat engine can achieve. It is expressed as 
\begin{equation}
\Gamma _{C}=1-\frac{Q_{C}}{Q_{H}}=1-\frac{T_{C}}{T_{H}},
\end{equation}%
where $T_{C}$ and $T_{H}$ represent the temperatures of the cold and hot
reservoirs, respectively. As shown in Fig. \ref{FigH1}, the black hole
absorbs a quantity of heat $Q_{H}$ during isothermal expansion and releases $%
Q_{C}$ during isothermal compression.

To model the Topological Mod(A)Max AdS black hole as a heat engine, we
analyze a rectangular thermodynamic cycle in the $P-V$ plane. This cycle
consists of two isobaric and two isochoric processes, illustrated in Fig. %
\ref{FigH1}. The rectangular cycle is favored for its analytical simplicity,
as the work done during the isochoric segments is zero. Consequently, the
total work output is directly proportional to the area enclosed by the
isobaric paths in the $P-V$ diagram.

Since the paths $1 \rightarrow 2$ and $3 \rightarrow 4$ represent isobaric
processes, we have $P_{1} = P_{2}$ and $P_{3} = P_{4}$. The total work done
during one complete cycle is given by 
\begin{equation}
W_{tot}=\oint PdV.
\end{equation}

Since the segments $2\rightarrow 3$ and $4\rightarrow 1$ are isochores,
their corresponding work contributions are zero. As a result, the net output
work, which geometrically represents the area enclosed by the rectangular
cycle, is 
\begin{eqnarray}
W_{tot} &=&W_{1\rightarrow 2}+W_{3\rightarrow 4}=P_{1}\left(
V_{1}-V_{2}\right) +P_{4}\left( V_{3}-V_{4}\right) ,  \notag \\
&&  \notag \\
&=&\frac{8}{3}\left( P_{1}-P_{4}\right) \left(
S_{2}^{3/2}-S_{1}^{3/2}\right) .  \label{w}
\end{eqnarray}

Conversely, the net inflow of heat during the upper isobaric process can be
determined from 
\begin{equation}
Q_{H}=\int\limits_{T}^{T_{1}}C_{P}dT=M_{2}-M_{1},
\end{equation}%
which yields 
\begin{equation}
Q_{H}=\frac{k\left( \sqrt{S_{2}}-\sqrt{S_{1}}\right) }{4\pi }+\frac{
8P_{1}\left( S_{2}^{3/2}-S_{1}^{3/2}\right) }{3}+\frac{\eta q^{2}e^{-\gamma
}\left( \frac{1}{\sqrt{S_{2}}}-\frac{1}{\sqrt{S_{1}}}\right) }{16\pi }.
\label{qh}
\end{equation}

Using Eqs. (\ref{w}) and (\ref{qh}), we can express the efficiency of this
rectangular cycle as follows 
\begin{equation}
\Gamma =\frac{\left( 1-\frac{P_{4}}{P_{1}}\right) }{1+\frac{3k\left( \sqrt{
S_{2}}-\sqrt{S_{1}}\right) }{32\pi P_{1}\left(
S_{2}^{3/2}-S_{1}^{3/2}\right) }+\frac{3\eta q^{2}e^{-\gamma }\left( \frac{1 
}{\sqrt{S_{2}}}-\frac{1}{\sqrt{S_{1}}}\right) }{128\pi P_{1}\left(
S_{2}^{3/2}-S_{1}^{3/2}\right) }}.
\end{equation}

The efficiency of the black hole heat engine, denoted as $\Gamma$, can be
compared to that of the Carnot engine, $\Gamma_{C}$. By associating the
higher temperature $T_{H}$ with $T_{2}$ and the lower temperature $T_{C}$
with $T_{4}$, the Carnot efficiency is expressed as follows 
\begin{equation}
\Gamma _{C}=1-\frac{\left( 1+\frac{k}{32\pi P_{1}S_{2}}-\frac{\eta
q^{2}e^{-\gamma }}{128\pi P_{1}S_{2}^{2}}\right) \frac{P_{1}}{P_{4}}\left( 
\frac{S_{2}}{S_{1}}\right) ^{\frac{1}{2}}}{\left( 1+\frac{k}{32\pi
P_{4}S_{1} }-\frac{\eta q^{2}e^{-\gamma }}{128\pi P_{4}S_{1}^{2}}\right) }.
\end{equation}

We have derived the expressions for the efficiency and Carnot efficiency of
the black hole heat engine. Now, we are ready to explore how the engine's
efficiency and effective efficiency are affected by the various parameters
that define the underlying theory.

\begin{figure}[]
\centering
\includegraphics[width=0.325\linewidth]{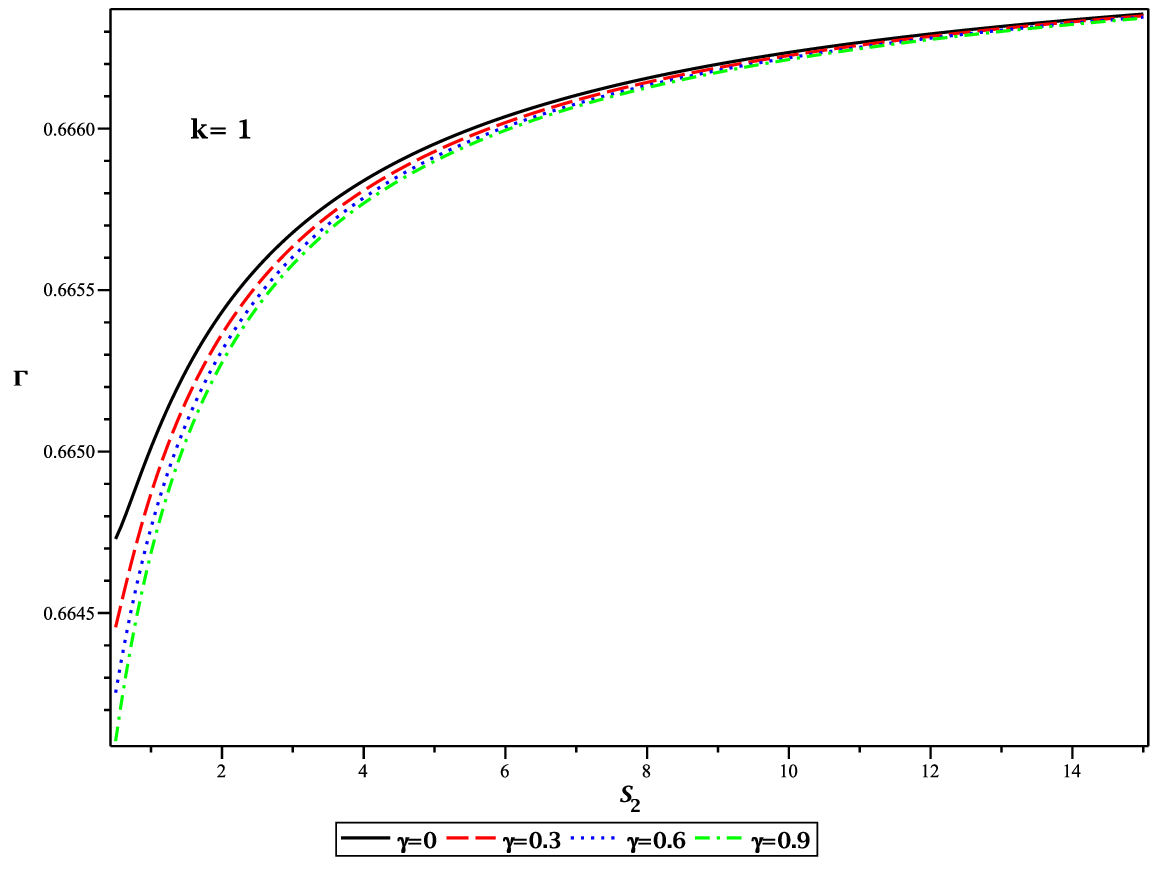} %
\includegraphics[width=0.325\linewidth]{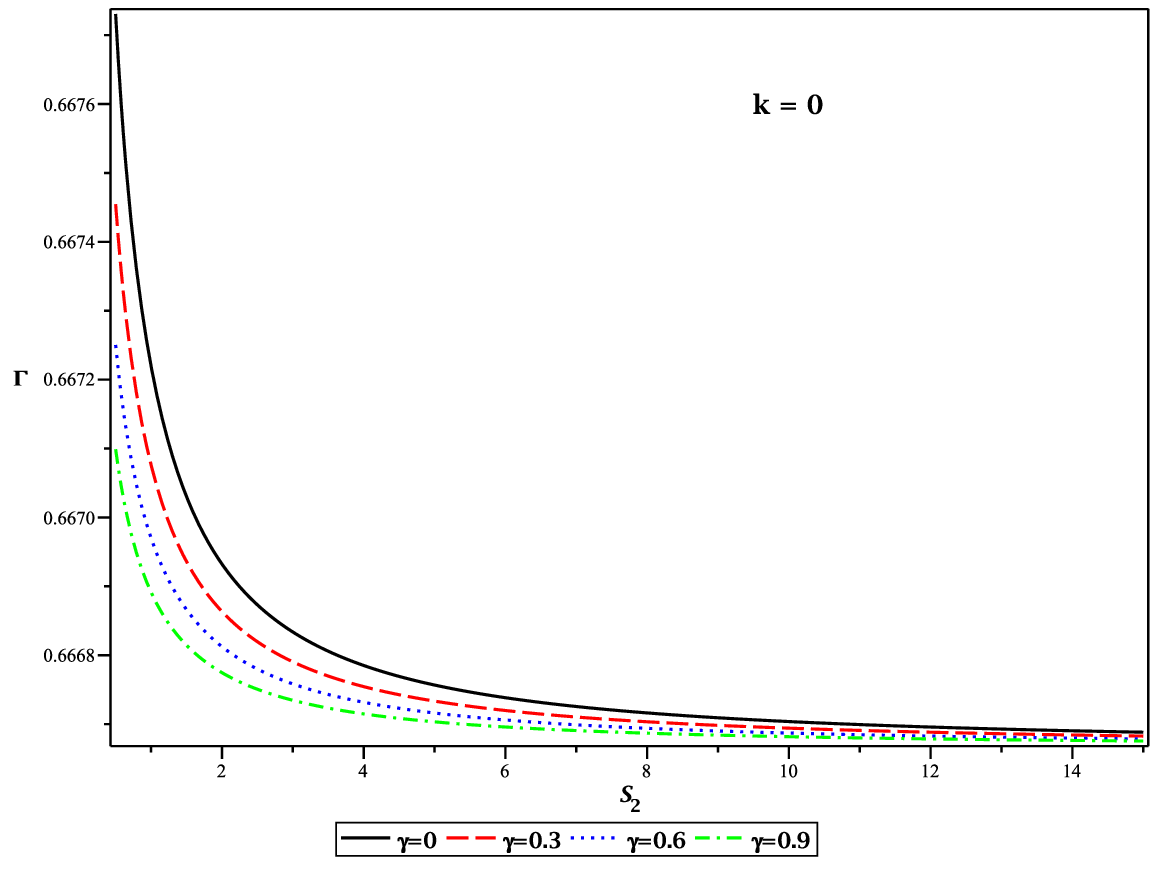} %
\includegraphics[width=0.325\linewidth]{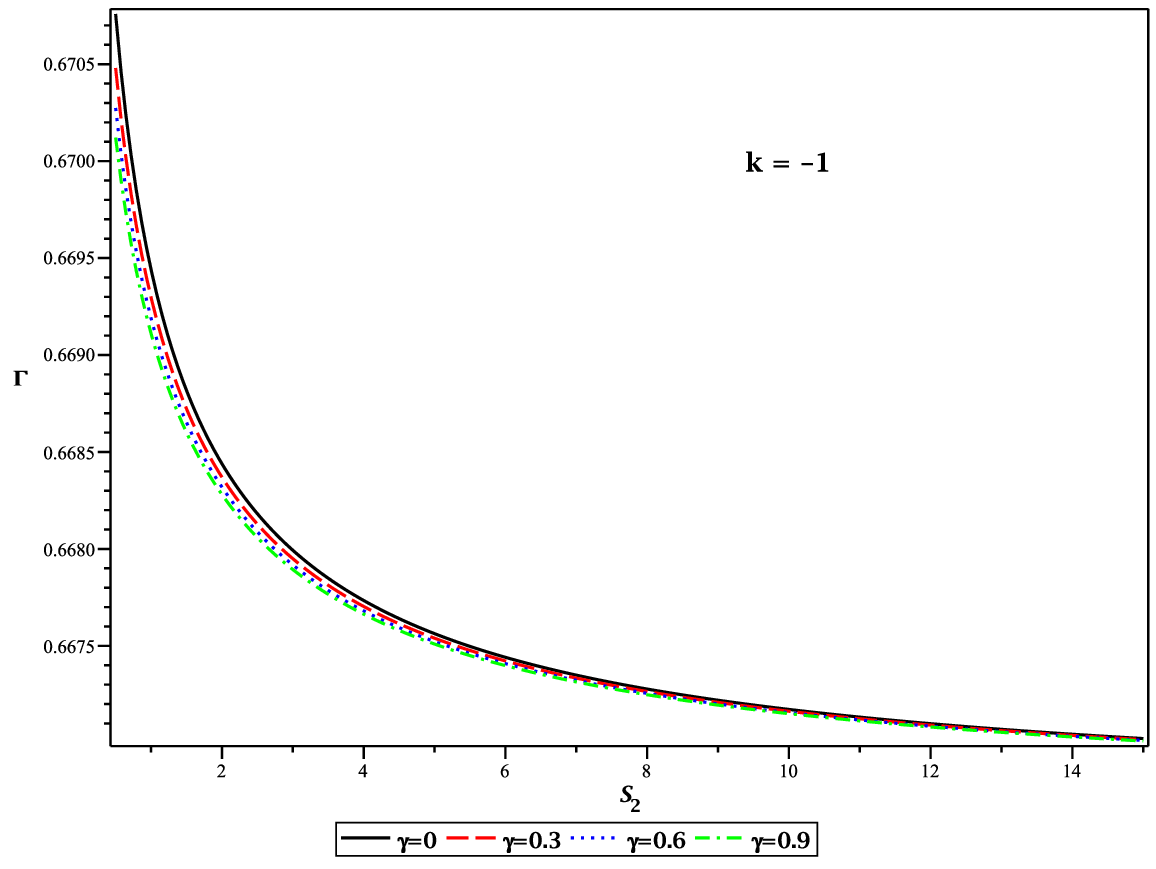} \newline
\caption{Plots of the heat engine efficiency $\Gamma$ as a function of
entropy $S_{2}$ for the ModMax AdS black hole are presented for various
values of $\protect\gamma$. In this analysis, we have set $P_{1} = 3$, $%
P_{4} = 1$, $S_{1} = 1$, and $q = 1$.}
\label{FigE1}
\end{figure}

\begin{figure}[]
\centering
\includegraphics[width=0.325\linewidth]{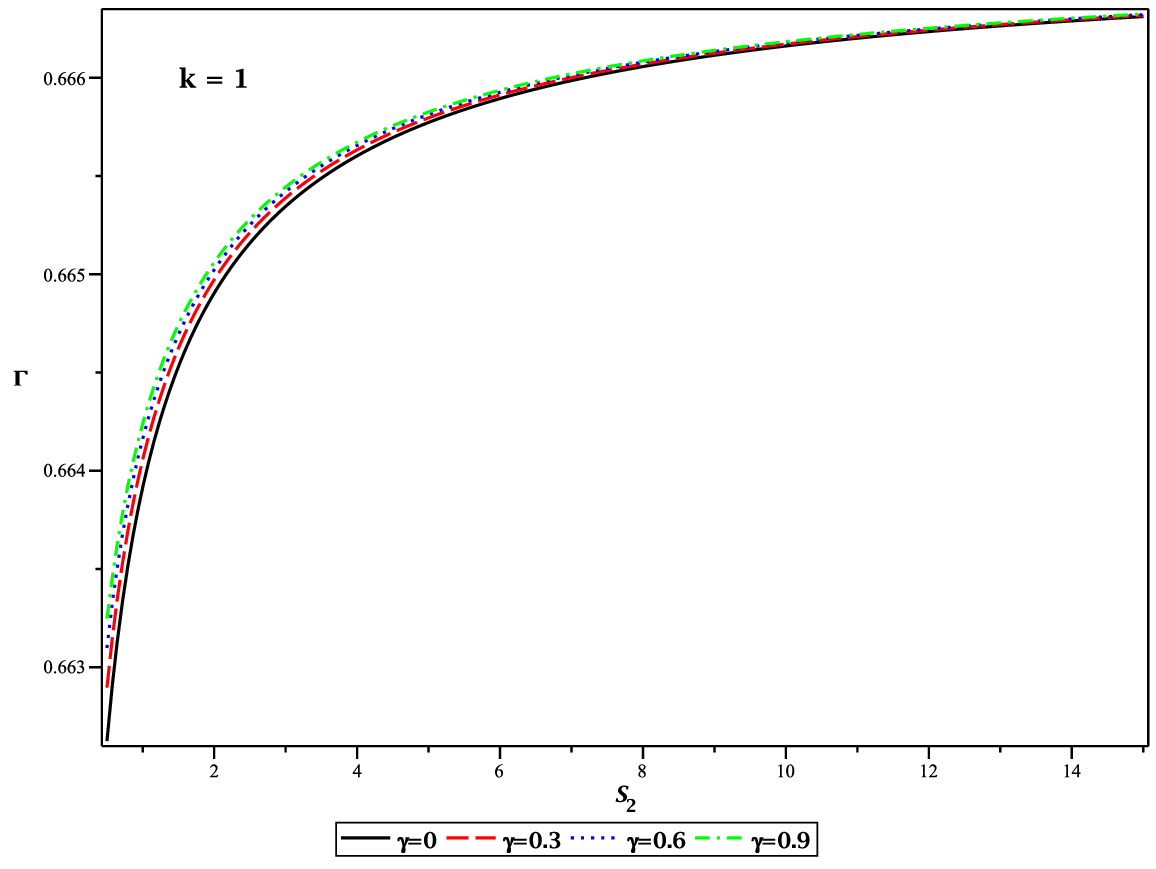} %
\includegraphics[width=0.325\linewidth]{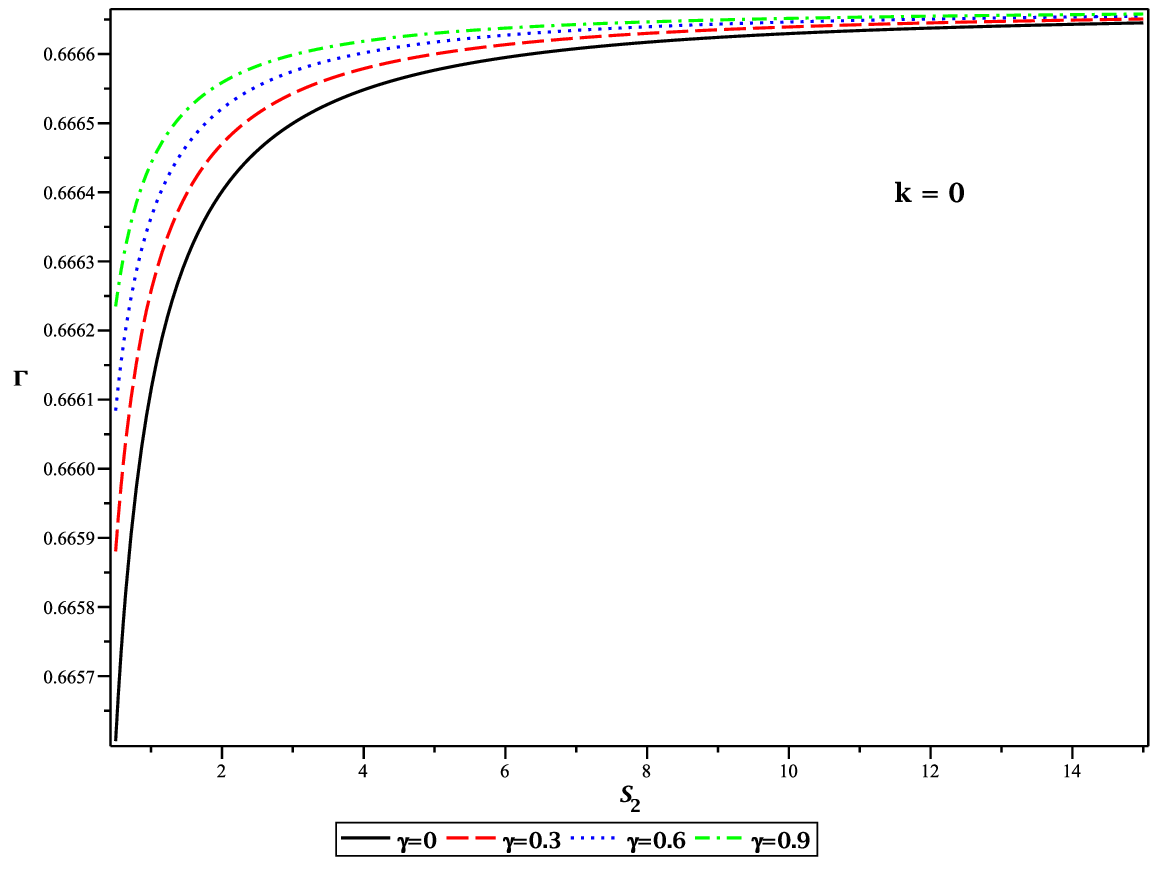} %
\includegraphics[width=0.325\linewidth]{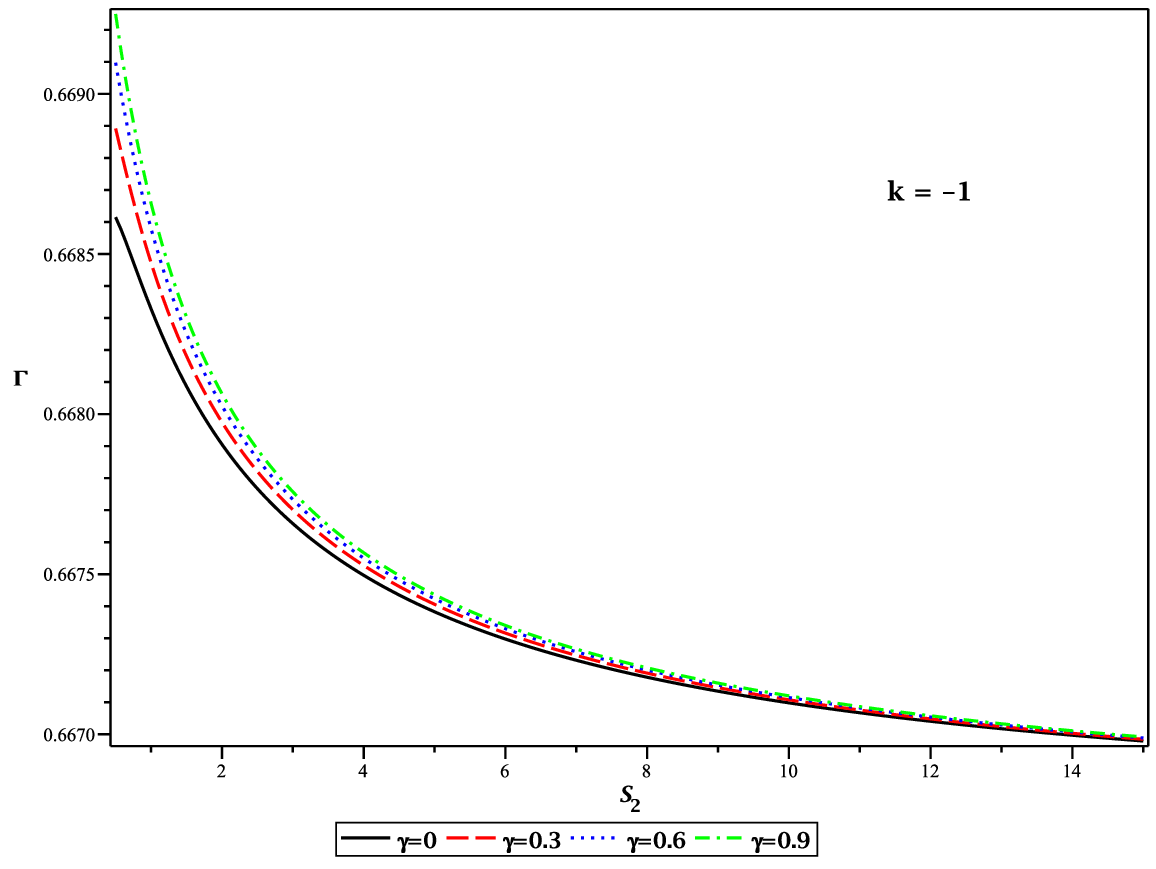} \newline
\caption{Plots of the heat engine efficiency $\Gamma$ as a function of
entropy $S_{2}$ for the ModAMax AdS black hole are presented for various
values of $\protect\gamma$. In this analysis, we have set $P_{1} = 3$, $%
P_{4} = 1$, $S_{1} = 1$, and $q = 1$.}
\label{FigE2}
\end{figure}

In Figs. \ref{FigE1} and \ref{FigE2}, we illustrate the efficiency of
topological Mod(A)Max AdS black holes for various values of $\gamma$.

\begin{description}
\item[\textbf{ModMax ($\protect\eta =+1$):}] In the spherical case ($k=+1$),
the efficiency increases monotonically with entropy and approaches a
constant value as $S_{2}$ becomes large. The influence of $\gamma$ is most
pronounced at low entropies, where a larger $\gamma$ slightly reduces
efficiency. However, all curves converge in the high-entropy regime. In
contrast, for $k=0$ and $k=-1$, efficiency decreases as $S_{2}$ increases.
The dependence on $\gamma$ is again only noticeable at low entropies and
becomes negligible at high $S_{2}$. These findings suggest that NED
corrections tend to suppress the efficiency of the ModMax black hole heat
engine, particularly for small black holes, while for large black holes, the
efficiency becomes nearly independent of both topology and $\gamma$.

\item[\textbf{ModAMax ($\protect\eta =-1$):}] For $k=+1$ and $k=0$, $\Gamma$
increases monotonically with $S_{2}$, approaching an asymptotic value at
high entropy. In contrast, for $k=-1$, $\Gamma$ decreases. In all scenarios,
the nonlinear corrections primarily enhance the efficiency of the ModAMax
black hole at low $S_{2}$, while the dependence on $\gamma$ becomes
negligible at high entropy.
\end{description}


\begin{figure}[tbp]
\centering
\includegraphics[width=0.325\linewidth]{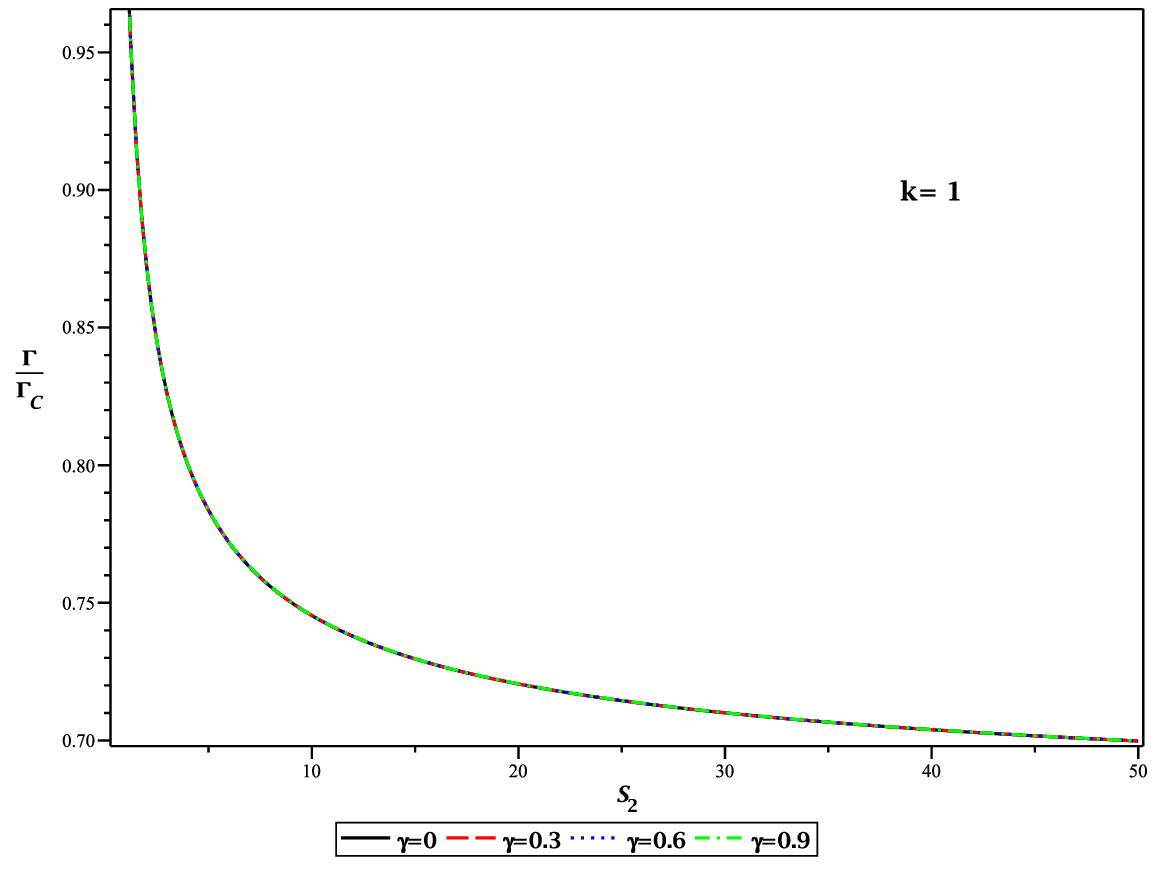} %
\includegraphics[width=0.325\linewidth]{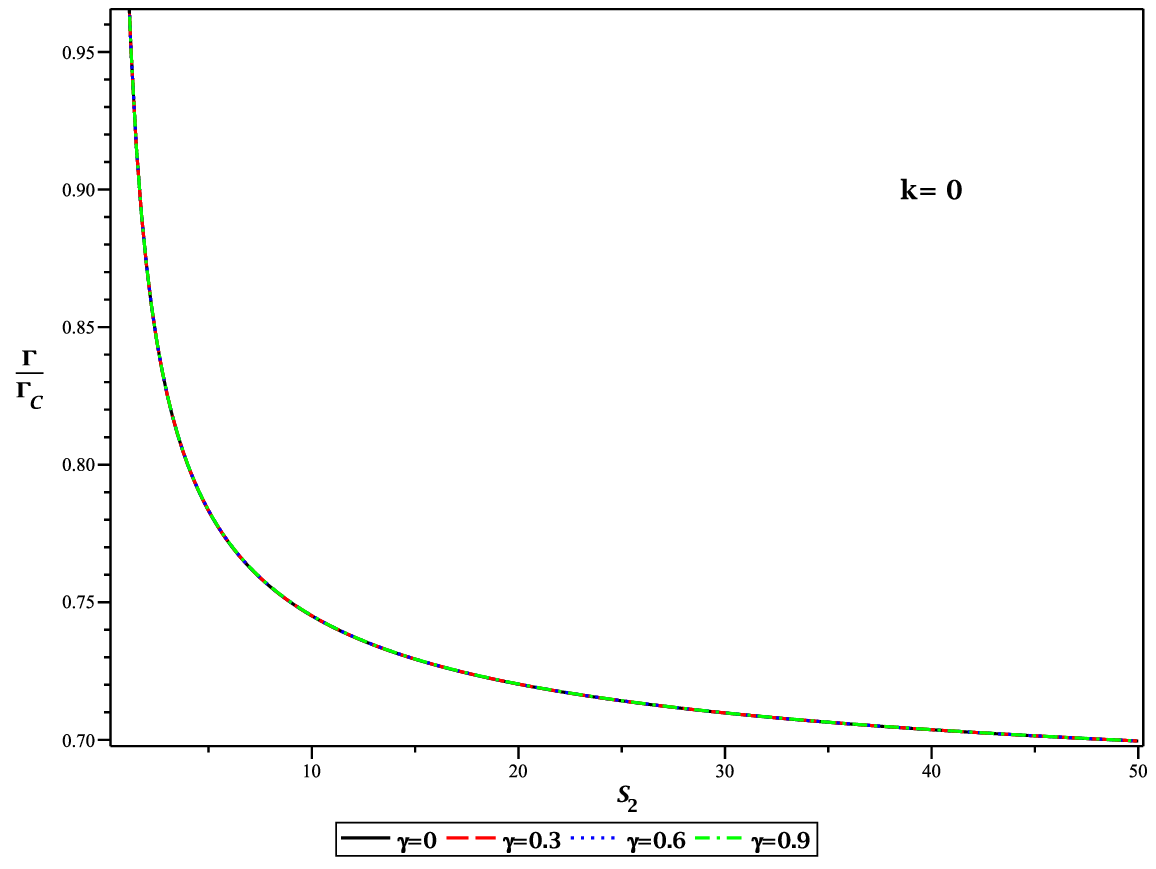} %
\includegraphics[width=0.325\linewidth]{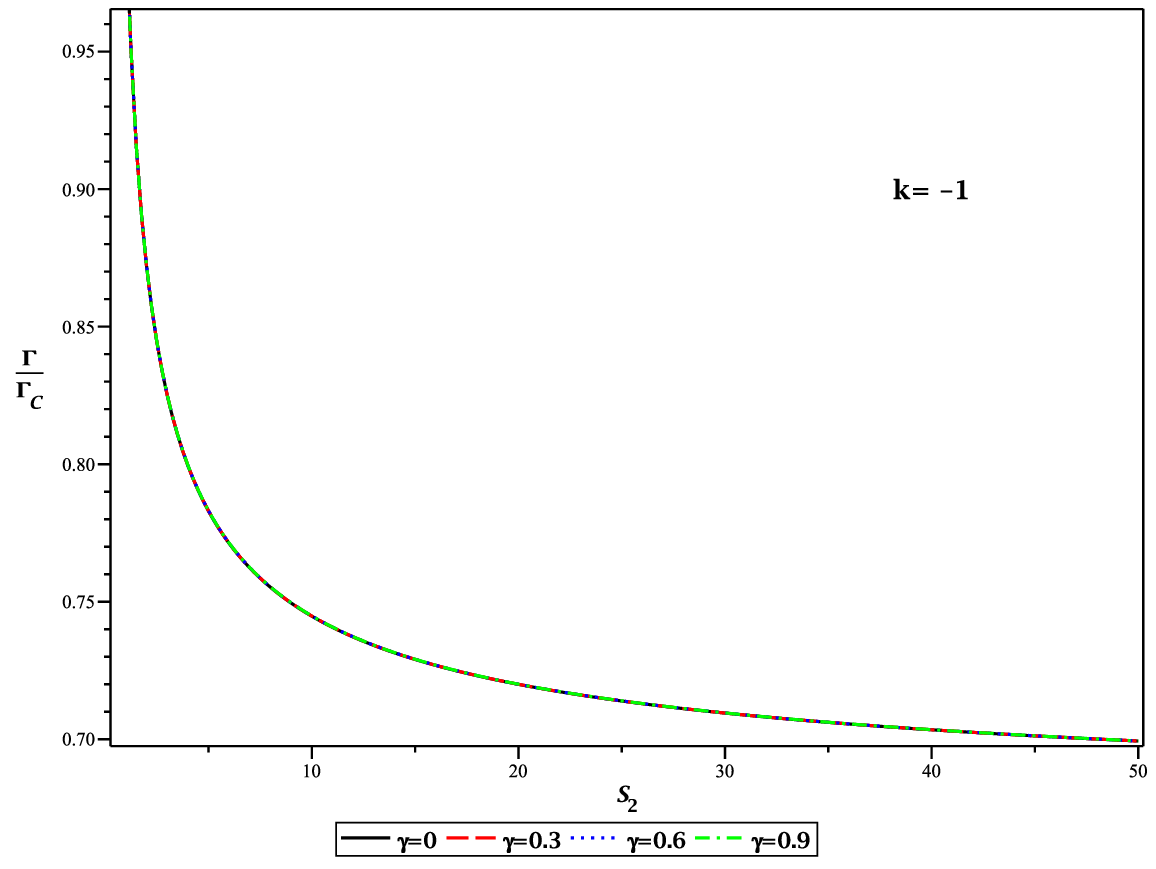} \newline
\caption{Plots of the efficiency ratio $\frac{\Gamma }{\Gamma _{C}}$ as a
function of entropy $S_{2}$ for the ModAMax AdS black hole are shown for
different values of $\protect\gamma $. Here, we set $P_{1}=3$, $P_{4}=1$, $%
S_{1}=1$, and $q=1$.}
\label{FigE3}
\end{figure}

\begin{figure}[tbp]
\centering
\includegraphics[width=0.325\linewidth]{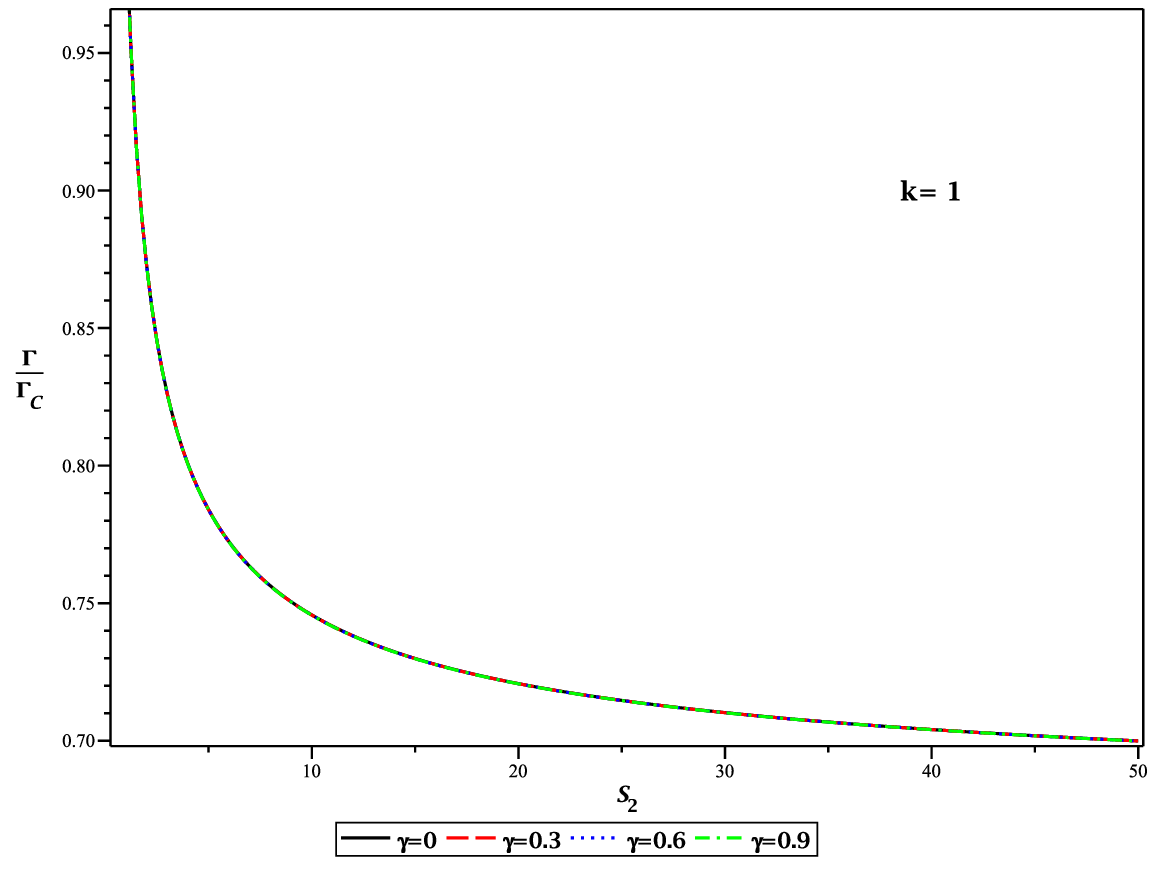} %
\includegraphics[width=0.325\linewidth]{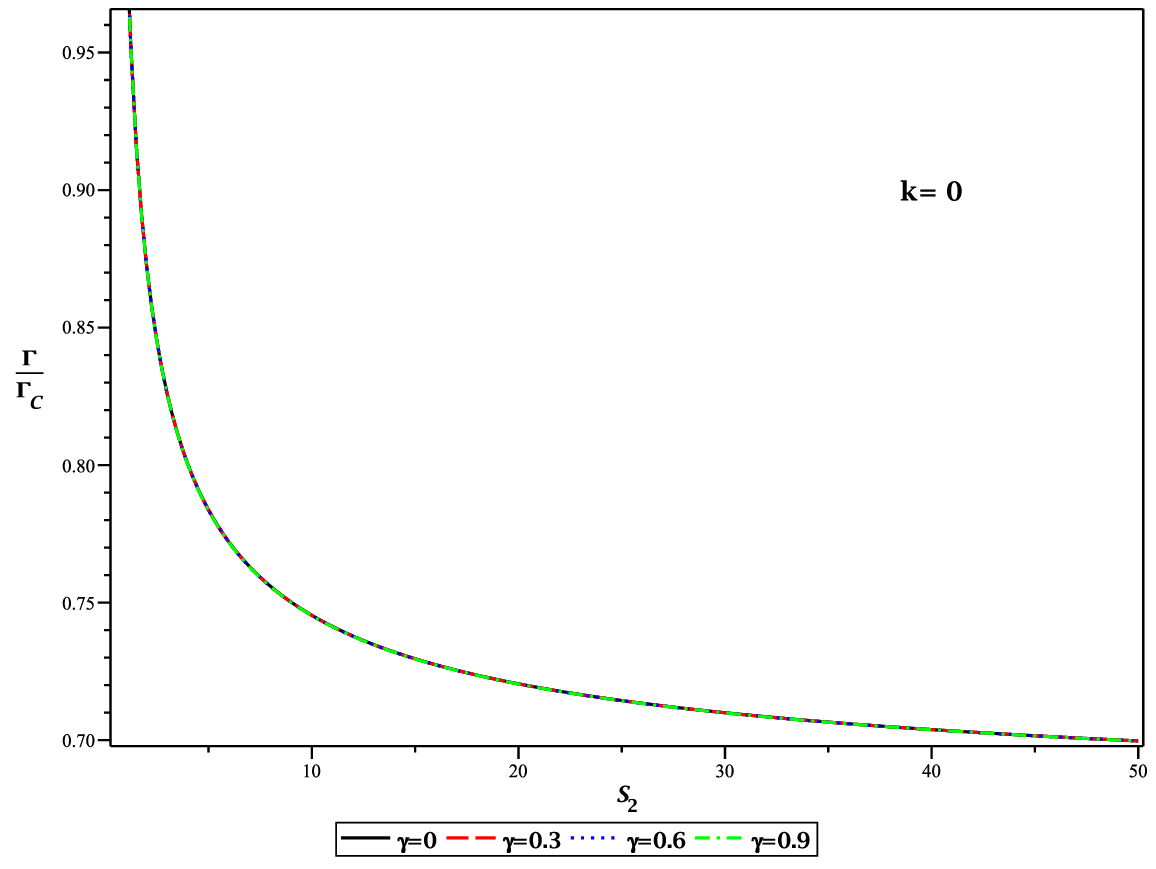} %
\includegraphics[width=0.325\linewidth]{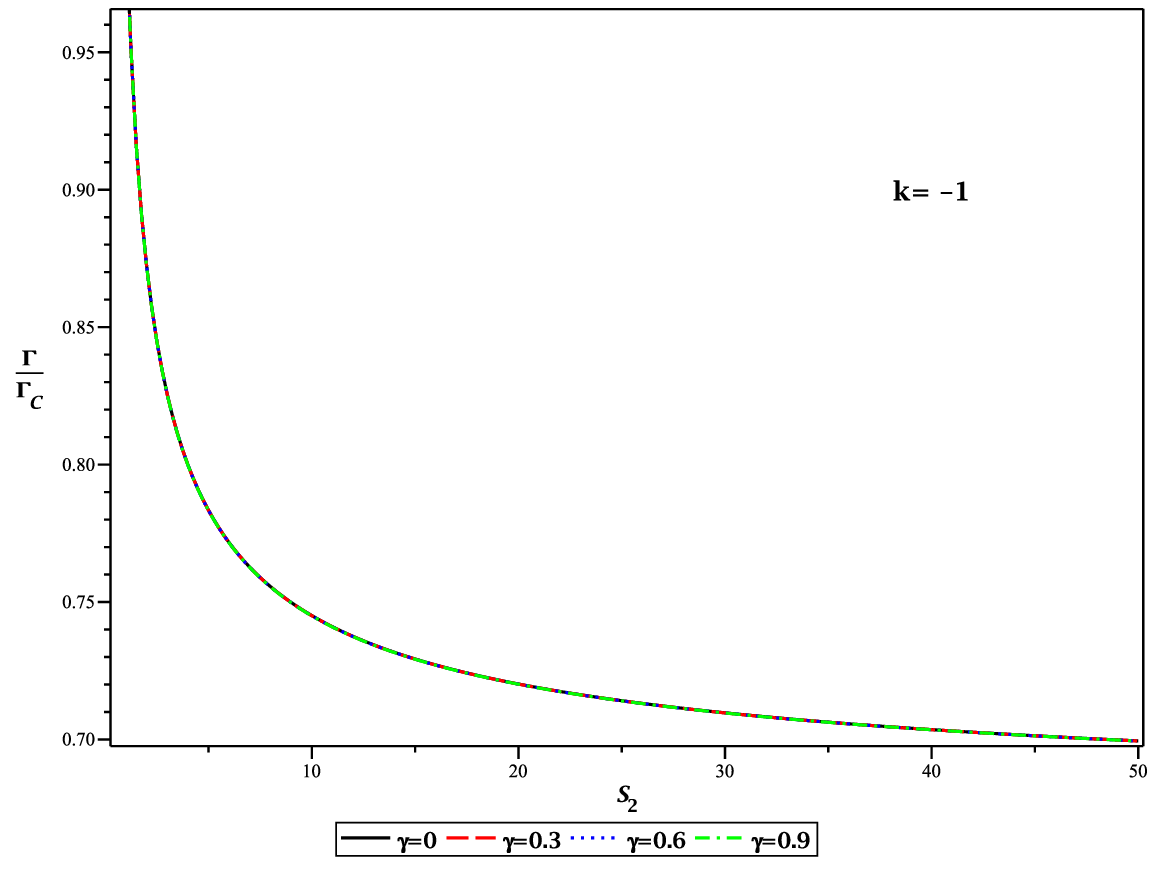} \newline
\caption{Plots of the efficiency ratio $\frac{\Gamma }{\Gamma _{C}}$ as a
function of entropy $S_{2}$ for the ModMax AdS black hole are shown for
different values of $\protect\gamma $. Here, we set $P_{1}=3$, $P_{4}=1$, $%
S_{1}=1$, and $q=1$.}
\label{FigE4}
\end{figure}

The plots in Figures \ref{FigE3}, and \ref{FigE4} display the variation of
the efficiency ratio $\frac{\Gamma}{\Gamma_{C}}$ with respect to the entropy 
$S_{2}$ for Mod(A)Max AdS black holes. In both cases, the ratio $\frac{\Gamma%
}{\Gamma_{C}}$ decreases monotonically as $S_{2}$ increases, indicating that
higher entropies-associated with larger horizon areas or volumes-lead to
lower thermodynamic efficiency compared to the Carnot limit. The ModMax
parameter $\gamma$ has no noticeable effect on the efficiency ratio.


The rectangular (isobar-isochoric) cycle is adopted here for
its analytical tractability, allowing us to isolate the influence of
parameters such as the ModMax coupling $\gamma $ and the topology parameter $%
k$ on work output and efficiency. Its physical interpretation within
extended black-hole thermodynamics, however, indeed warrants explicit
clarification. 

For the static AdS black holes studied here, both the thermodynamic volume
and the Bekenstein--Hawking entropy are monotonic functions of the horizon
radius, 
\begin{equation}
S=\frac{r_{+}^{2}}{4},\qquad V=\frac{r_{+}^{3}}{3}\quad \Rightarrow \quad V=%
\frac{8}{3}\,S^{3/2}.
\end{equation}

Hence, an isochoric process ($V=\mathrm{const}$) necessarily fixes $r_{+}$
and therefore keeps the entropy constant ($dS=0$). From the first law (\ref%
{dh}), it follows that along an isochore with fixed charge one has $T\,dS=0$%
. Consequently, all heat transfer occurs exclusively along the isobaric
segments, which explains why the mechanical work reduces to the geometric
area between the two isobars in the $P$--$V$ plane.

It should also be emphasized that this cycle is an idealization. A
quasi-static variation of the hermodynamic pressure $P$ corresponds to a
slow change of the effective cosmological constant, describing a continuous
path through a family of equilibrium AdS spacetimes. Such cycles therefore
serve as theoretical probes of the thermodynamic structure rather than
operational protocols.

In addition, the fundamental Carnot bound, $\eta _{C}=1-\frac{T_{C}}{T_{H}}$%
, provides the universal upper limit. For the rectangular cycle we identify $%
T_{H}=T(P_{1},S_{2})$ and $T_{C}=T(P_{4},S_{1})$. As required by the second
law, all efficiencies computed in the manuscript satisfy $\Gamma <\Gamma_{C} 
$. This has been verified numerically for representative parameter, across a
range of $S_{2}$ values, as shown in Figures \ref{FigE3} and \ref{FigE4}.

To better contextualize the performance of the rectangular cycle and to
anticipate how its efficiency compares with theoretically optimal processes,
we can consider a Stirling-like cycle composed of two isotherms connected by
isochores. In the present framework the isochores coincide with isentropes,
so an ideal Stirling cycle would operate reversibly between $T_{H}$ and $%
T_{C}$, thereby attaining the Carnot efficiency $\Gamma _{\mathrm{Stirling}%
}\sim \Gamma _{C}$. This comparison therefore provides a natural upper
benchmark against which the rectangular cycle can be evaluated. As expected,
the rectangular cycle is intrinsically less efficient. Nevertheless, its
analytical simplicity allows the influence of physical parameters, most
notably the ModMax coupling $\gamma $ and the horizon topology $k$, to be
identified with clarity. This trade-off between conceptual transparency and
optimal performance is typical of model cycles in exploratory black hole
thermodynamics and motivates our choice of the rectangular cycle as a
diagnostic tool.

\section{Conclusions}

In this paper, we introduced a coupling between a new modification of
Maxwell theory (ModMax) and the anti-Maxwell (phantom) field within the
Einstein-Hilbert action, including the cosmological constant. We then
derived the corresponding field equations. By examining topological
spacetime, we obtained AdS black hole solutions that incorporate both ModMax
and ModAMax (modified anti-Maxwell) fields. We refer to these solutions as
topological Mod(A)Max black holes. This was the first attempt to combine the
phantom field into a NED theory.

We investigated the effects of ModMax's parameter $\gamma $, phantom field $%
\eta $, and topological constant $k$ on AdS black holes. Notably, $\eta$ can
take values of $\pm 1$: $\eta = +1$ corresponds to the Maxwell field, while $%
\eta = -1$ pertains to the anti-Maxwell (phantom) field. Our analysis
revealed that ModMax AdS black holes can exhibit two roots, where the
smaller root is associated with the inner horizon, and the larger root
corresponds to the event horizon. Importantly, as $\gamma$ increases, the
number of roots for ModMax AdS black holes decreases from two to one.
Furthermore, smaller black holes are related to a positive topological
constant ($k = +1$), while larger black holes correspond to a negative
topological constant ($k = -1$). In contrast, ModAMax AdS black holes have
only one root, which is linked to the event horizon. Additionally, the event
horizon of ModAMax AdS black holes decreases as $\gamma$ increases. Similar
to ModMax black holes, smaller to larger ModAMax black holes are associated
with $k = +1$, $k = 0$, and $k = -1$, respectively. Our analysis of the
behavior of Mod(A)Max black holes highlighted two distinct behaviors between
ModMax and ModAMax black holes: i) ModMax AdS black holes can have two
roots, while ModAMax AdS black holes have only one root; ii) Increasing $%
\gamma$ results in an increase in the event horizon for ModMax AdS black
holes, whereas the event horizon of ModAMax AdS black holes decreases as $%
\gamma$ increases.

We calculated the conserved and thermodynamic quantities for Mod(A)Max AdS
black holes, including the Hawking temperature. Our analysis showed that
small ModMax AdS black holes can have a negative temperature when $k=0$ or $%
k=-1$. In contrast, for $k = +1$, the temperature is positive if $%
q^2e^{-\gamma} < r_{+}^2$. Furthermore, for small ModAMax black holes, the
Hawking temperature is always positive when $k = +1$ or $k = 0$. However,
for $k=-1$, it is positive only if $q^2 e^{-\gamma} > r_{+}^2$. For large
Mod(A)Max black holes, the Hawking temperature depends solely on the
cosmological constant and is always positive.

We calculated the total mass of the topological Mod(A)Max black holes. In
the high-energy limit, the total mass depends solely on the electrodynamic
field. Specifically, we found that the mass of small ModMax AdS black holes
can be positive, whereas the total mass of small ModAMax black holes cannot
be positive. This distinction is the primary difference between the masses
of ModMax and ModAMax black holes. In the asymptotic limit, the total mass
relies only on the cosmological constant, indicating that the total mass of
large Mod(A)Max AdS black holes is always positive. Indeed, the mass of
large Mod(A)Max black holes is positive, which is consistent with the
positive Hawking temperature of these black holes.

A simultaneous evaluation of temperature and heat capacity
behavior allowed us to analyze the local stability of Mod(A)Max AdS black
holes across various topological constants ($k$) and ModMax parameter ($%
\gamma$) values. The results demonstrated that: 

i) \textbf{ModMax case}: For $k=+1$, small $\gamma$ yielded an instability
region ($S<S_{{root}_{2}}$) alongside a stable region for larger entropy,
where increasing $\gamma$ enhanced stability. Critically, for a large $%
\gamma<\gamma_{critical}$, two divergence points resulted in three distinct
thermal regimes, while for $\gamma>\gamma_{critical}$, only one divergence
point remained, causing large black holes to satisfy stability. For $k=0$,
stability was maintained when $S>S_{{root}_{2}}$, but this root disappeared
entirely when $\gamma>\gamma_{critical}$, rendering all black holes stable.
Finally, for $k=-1$, a single positive heat capacity root existed across all 
$\gamma$, where stability was always achieved only for sufficiently large
entropy values ($S>S_{{root}_{2}}$). Overall, increasing $\gamma$
consistently led to an expansion of the locally stable region for the black
hole system.

ii) \textbf{ModAMax case}: The investigation revealed that a critical value
of the ModMax parameter ($\gamma _{critical}$) influenced the heat capacity (%
$C_{Q}$) behavior. When $\gamma <\gamma _{critical}$, a divergence point $S_{%
{div}_{2}}$ existed, separating unstable (small) and stable (large) ModAMax
AdS black holes, with increasing $\gamma $ leading to greater stability
overall. Conversely, for $\gamma >\gamma _{critical}$, the thermal
characteristics depended significantly on the topological constant ($k$).
Specifically, for $k=+1$, $S_{{div}_{2}}$ persisted; for $k=0$, $C_{Q}$
remained consistently positive, ensuring stability regardless of entropy;
and for $k=-1$, the presence of $S_{{root}_{2}}$ dictated stability only for
sufficiently large black holes ($S>S_{{root}_{2}}$). These observations
collectively demonstrated that black hole local stability for $\eta =-1$
relies on interplay between the topological constant, the ModMax parameter,
and, crucially, the black hole's size relative to critical entropy values.

The global stability analysis of Mod(A)Max AdS black holes, investigated via
the Helmholtz free energy in the canonical ensemble. Our analysis indicated
that:

i) \textbf{ModMax case}: For $\eta =+1$, a critical parameter $\gamma
_{critical}$ separated the behaviors: below this value, stability ($F<0$)
was achieved only for larger black holes ($S>S_{F=0}$), a domain expanded by
increasing $\gamma $. Above $\gamma _{critical}$ with $k=+1$, one root
existed, yielding stability above $S_{F=0}$. Crucially, for $k=0$ and $k=-1$
under $\eta =+1$, the Helmholtz free energy possessed no real root and
remained strictly negative, inherently ensuring stability.

ii) \textbf{ModAMax case}: For $\eta =-1$, $F$ consistently exhibited no
real roots across all $k$, leading to universal negativity ($F<0$). This
demonstrated that the existence or non-existence of real roots for the
Helmholtz free energy dictated the extent of the thermodynamically stable
phase space. Overall, the findings clarified the precise thermodynamic
stability boundaries dependent on both the ModMax parameter and the
topological constant.

In the grand-canonical ensemble, the global thermodynamic stability for
these black holes was rigorously established by the condition $G<0$.
Investigation yielded the conclusion that:

i) \textbf{ModMax case}: For $\eta =+1$, a critical parameter $\gamma
_{critical}$ governed the stability landscape. When $\gamma <\gamma
_{critical}$, $G$ remained negative, intrinsically satisfying stability,
while $\gamma >\gamma _{critical}$ introduced a transition root where
stability held only for $S>S_{G=0}$ when $k=+1$. Crucially, for $k=0$, and $%
k=-1$ under $\eta =+1$, $G $ was invariably negative, ensuring global
stability across all parameters.

ii) \textbf{ModAMax case}: For $\eta =-1$, no real root for $G=0$ existed,
and $G$ remained universally negative, thus inherently guaranteeing
stability.

We also investigated the Joule--Thomson expansion of topological Mod(A)Max
AdS black holes in the extended phase space. This analysis yielded explicit
expressions for both the Joule--Thomson coefficient and the inversion
temperature. Our findings revealed that the inversion curves and the minimum
inversion temperature are significantly affected by the NED parameter $%
\gamma $ and the topology constant $k$.

By modeling these black holes as heat engines operating through a
rectangular $P-V$ cycle, we assessed their efficiency and compared it to the
Carnot limit. Our findings indicated that the ModMax and ModAMax parameters,
along with the horizon topology, significantly influence the thermodynamic
performance of the engine. A notable result is the distinct topological
dependence of efficiency: for ModMax black holes, efficiency increases with
entropy for spherical topology ($k=+1$) but decreases for both flat ($k=0$)
and hyperbolic ($k=-1$) topologies. In contrast, for ModAMax black holes,
efficiency rises with entropy for spherical ($k=+1$) and flat ($k=0$)
topologies, while it decreases in the hyperbolic case ($k=-1$).
Additionally, we found that NED corrections reduce the efficiency of small
black holes across all configurations.

We emphasise that although our ModAMax solution is obtained through exotic matter, with negative energy density, this does not invalidate it as a possible realistic astrophysical black hole solution. As mentioned in the Introduction, there exist numerous solutions with this exotic–matter feature in GR and in modified gravity theories. Some of them require quantum effects to account for such behaviour. We have shown constraints that render our solutions stable and thermodynamically consistent. In Appendix A, we show that the ModAMax solution with $k=1$ may be linearly stable under small geometric perturbations, through the analysis of QNMs using the third–order WKB approximation. Therefore, even the introduction of exotic matter does not necessarily result in instabilities in general. Naturally, we shall need to establish physical constraints on the parameters so that measurements related to astrophysical black holes, such as QPOs, GWs, and EHT observations, may arise from topological ModAMax solutions.


\begin{acknowledgements}
B. Eslam Panah thanks the University of Mazandaran. This work is based upon research funded by Iran National Science Foundation (INSF) under project No.4035285. Manuel E. Rodrigues thanks Conselho Nacional de Desenvolvimento Cient\'{i}fico e Tecnol\'{o}gico-CNPq, Brazil, for partial financial support. We extend our gratitude to Professor Vitor Cardoso for the confirmation and valuable references provided regarding the WKB approximation for the Quasi-Normal Modes (QNMs) of the Reissner–Nordström Anti-de Sitter (RN–AdS) solution.

\end{acknowledgements}

\appendix

\section{First perturbative check:\newline
scalar modes of the ModAMax solution for $k=+1$}

The linear stability analysis of a black hole solution begins
by introducing small perturbations to the background metric, $g_{\mu \nu
}\rightarrow g_{\mu \nu }+h_{\mu \nu }$, where $h_{\mu \nu }$ denotes an
infinitesimal fluctuation. After decomposing these perturbations into
spherical harmonics on $S^{2}$, they separate into three main classes
according to their transformation properties: scalar modes (spin$-0$),
vector modes (spin$-1$), and tensor modes (spin$-2$) \cite%
{ReggeWheeler1957,Zerilli1970,Moncrief1974,KI2003,KI2004}. Each sector
satisfies an independent master equation, typically cast into a radial
Schrodinger-type form, 

\begin{equation}  \label{eqA1}
\frac{d^2\Psi}{dr_\ast^2} + \left[\omega^2 - V_{\text{eff}}(r)\right]\Psi =
0,
\end{equation}
where $r_\ast$ is the tortoise coordinate and $V_{\text{eff}}$ is the
associated effective potential \cite{Chandrasekhar1983,KokkotasSchmidt1999}.
Among these sectors, scalar modes (spin$-0$) are especially significant in
extended or modified theories of gravity, where additional scalar degrees of
freedom can produce richer effective potentials, featuring multiple peaks,
deep wells, or regions in which the potential becomes negative \cite%
{DottiGleiser2005}, potentially indicating the presence of unstable or
bound-state-type modes.

To investigate dynamical stability, one computes the spectrum of
quasi-normal modes (QNMs). With the convention $\omega = \omega_R +
i\,\omega_I$, stability requires $\omega_I < 0$, ensuring that perturbations
decay exponentially in time. When the effective potential exhibits a single
smooth barrier, an efficient method for estimating the quasi-normal
frequencies is the WKB approximation, originally introduced by Schutz and
Will \cite{SchutzWill1985} and later extended to higher orders by Iyer and
Will \cite{IyerWill1987} and by Konoplya \cite{Konoplya2003}. The $n$%
th-order WKB quantisation condition may be written generically as 
\begin{equation}  \label{WKB}
\frac{i(\omega^2 - V_0)}{\sqrt{-2V_0^{\prime \prime }}} - \Lambda_n = n +
\frac12,
\end{equation}
where $V_0 = V_{\text{eff}}(r_{\text{max}})$, $V_0^{\prime \prime }$ is the
second derivative of the potential at its maximum, and $\Lambda_n$ encodes
the higher-derivative corrections \cite{KonoplyaZhidenko2011}. Owing to its
accuracy for large angular quantum numbers and single-barrier potentials,
the WKB method provides a robust criterion for dynamical stability,
permitting one to verify that the imaginary part of the quasi-normal
spectrum satisfies $\omega_I < 0$ for all modes.

The analysis carried out in asymptotically AdS space-times was initially
performed in \cite{Wang, Berti2} for the Reissner--Nordstrom--AdS solution.
It was concluded that AdS-type solutions indicate stability under linear
perturbations. Subsequently, the linear stability of AdS-type solutions was
analysed, again suggesting stability \cite{Festuccia, Berti1}. The review 
\cite{Berti3} shows that these solutions may be analysed using the
perturbative WKB method (see Sec.~6), in the limit where the angular
momentum number satisfies $l \gg 1$, the so-called eikonal limit. Here we
shall carry out this analysis only for the scalar modes (spin $s=0$) of the
ModAMax solution with $k=1$, employing the third-order WKB approximation.
This is merely a preliminary analysis indicating the stability of our
solution, which may be further improved by enhancing the accuracy
(sixth-order WKB and other high-precision methods) and extending the study
to higher spins ($s=1,2$).

We define the effective potential of the Schrodinger-type equation %
\eqref{eqA1} for our ModAMax solution following \cite{Berti2} 
\begin{eqnarray}
V_{eff}(r) &=&(-g_{tt})\left( \frac{l(l+1)}{g_{\theta \theta }}+\frac{1}{r}%
\frac{d(-g_{tt})}{dr}\right)   \notag \\
&&  \notag \\
&=&\frac{\left( r\left( -3kr+3m+\Lambda r^{3}\right) -3\eta q^{2}e^{-\gamma
}\right) \left( r\left( 2\Lambda r^{3}-3(l(l+1)r+m)\right) +6\eta
q^{2}e^{-\gamma }\right) }{9r^{6}}\,,
\end{eqnarray}%
where $g_{tt}$ and $g_{\theta \theta }$ are the components of the metric %
\eqref{Metric}, and $l$ is the angular momentum number. We now define $r_{%
\mathrm{max}}$ as the value of the radial coordinate at which the effective
potential attains its maximum, that is, $V_{\mathrm{eff}}^{\prime }(r_{%
\mathrm{max}})=0$ and $V_{\mathrm{eff}}^{\prime \prime }(r_{\mathrm{max}})<0$%
. This value depends on $l$, hence $r_{\mathrm{max}}\equiv r_{\mathrm{max}%
}(l)$. We then define $V_{n}(l)=V_{\mathrm{eff}}^{(n)}[r_{\mathrm{max}}(l)]$%
, which is the $n$-th derivative of the effective potential evaluated at $r_{%
\mathrm{max}}$. We also define 
\begin{eqnarray}
&&\alpha (l)=\sqrt{2}\sqrt{-V_{2}(l)}, \\
&&  \notag \\
&&\Gamma _{1}(l)=\frac{1}{\alpha (l)}\left[ \frac{V_{4}(l)}{8V_{2}(l)}-\frac{%
5V_{3}^{2}(l)}{24V_{2}^{2}(l)}\right] , \\
&&  \notag \\
&&\Gamma _{2}(l)=\frac{1}{\alpha ^{3}(l)}\left[ \frac{V_{3}(l)\,V_{5}(l)}{%
48V_{2}^{2}(l)}-\frac{V_{4}^{2}(l)}{288V_{2}^{2}(l)}-\frac{%
5V_{3}^{2}(l)V_{4}(l)}{576V_{2}^{3}(l)}+\frac{5V_{3}^{4}(l)}{3456V_{2}^{4}(l)%
}\right] , \\
&&  \notag \\
&&\Gamma (l)=\Gamma _{1}(l)+\Gamma _{2}(l), \\
&&  \notag \\
&&\omega =\sqrt{\Gamma (l)-i\left( n+\frac{1}{2}\right) \alpha (l)+V_{0}(l)},
\\
&&  \notag \\
&&\omega _{R}=Re[\omega ],~~~\&\;~~\omega _{I}=Im[\omega ]\;.
\end{eqnarray}

The Taylor series for the effective potential are 
\begin{eqnarray}
&&V_{eff}\left(r>>1\right)\sim\frac{2 \Lambda ^2 r^2}{9}-\frac{1}{3} \Lambda
\left(2 k+l^2+l\right)+\frac{\Lambda m}{3 r}+O\left(\left(\frac{1}{r}%
\right)^2\right) \;; \\
&&V_{eff}\left(r<<1\right)\sim -\frac{2 \left(e^{-2 \gamma } \eta ^2
q^4\right)}{r^6}+\frac{3 e^{-\gamma } \eta m q^2}{r^5}+O\left(\frac{1}{r^4}%
\right)\;.
\end{eqnarray}

We can see that the potential tends to infinity in the limit $r\rightarrow
\infty $, and tends to minus infinity in the limit $r\rightarrow 0$. Here we
use values of $l\in \lbrack 30000,50000]$, $M=\{45,55,89.96857502,120\}$ and 
$q=1$, $\gamma =0.2$, $\Lambda =-1$, $k=1$, and $\eta =-1$. For these
values, the effective potential vanishes at the event horizon. In Fig.~\ref%
{Veff1}, we show the effective potential for these parameter values, with $%
l=30000$. The green ($m=45$) and blue ($m=55$) curves possess a maximum and
a minimum, whereas the red ($m=m_{\mathrm{crit}}=89.96857502$) and purple ($%
m=120$) curves have no extrema. This also occurs for AdS solutions in
general, as shown in Fig.~1 of Ref.~\cite{Berti1}. However, the difference
here is that $r_{H(\mathrm{crit})}/L=3.64367$ instead of $r_{H(\mathrm{crit}%
)}/L=1$ ($r_{H(\mathrm{crit})}$ is the critical value at which $V_{\mathrm{%
eff}}$ ceases to possess extrema, and $L=\sqrt{3/(-\Lambda )}$). 
\begin{figure}[tbh]
\centering \includegraphics[width=0.5\textwidth]{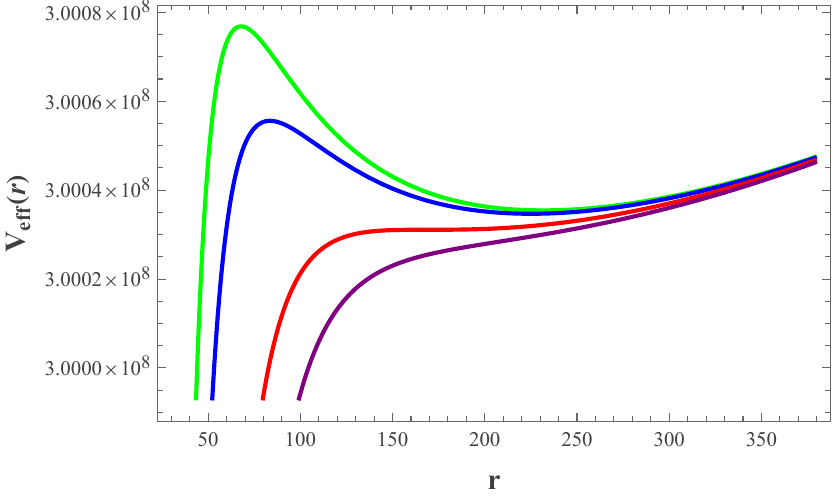}
\caption{Effective potential for the ModAMax solution with $k=+1$.}
\label{Veff1}
\end{figure}

\begin{figure}[htb]
\centering
\par
\includegraphics[width=0.45 \textwidth]{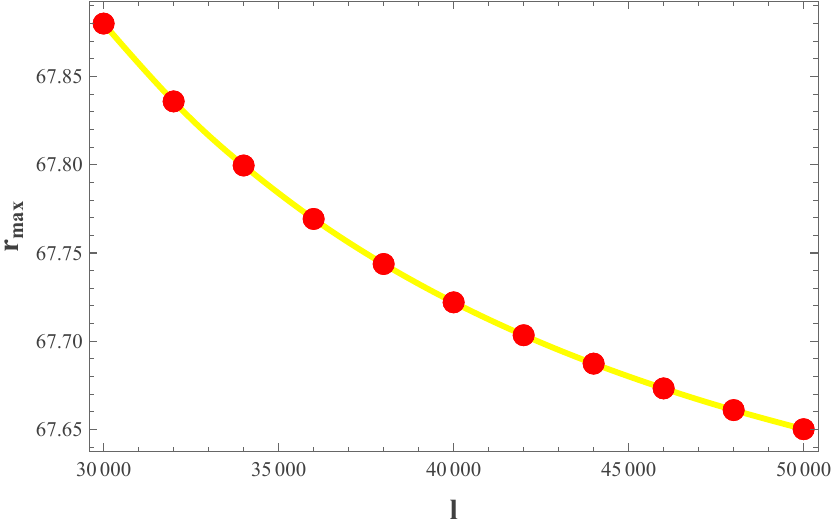}\hfill %
\includegraphics[width=0.45\textwidth]{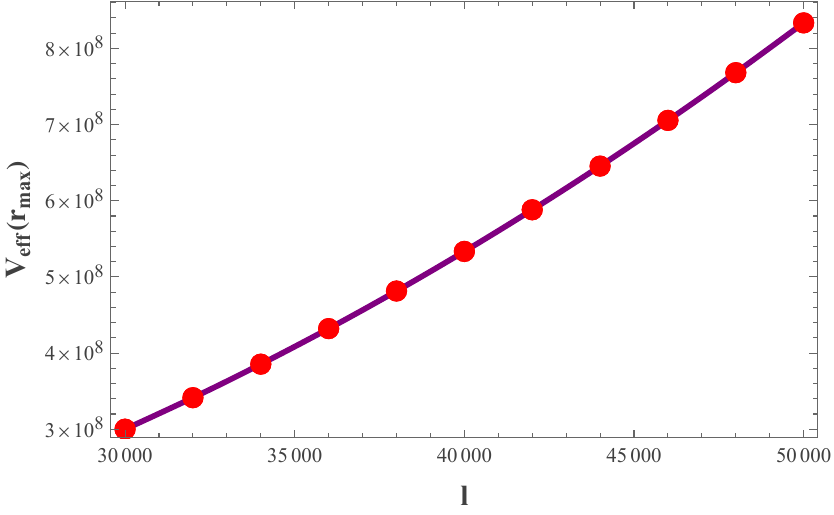}\newline
\par
\includegraphics[width=0.45\textwidth]{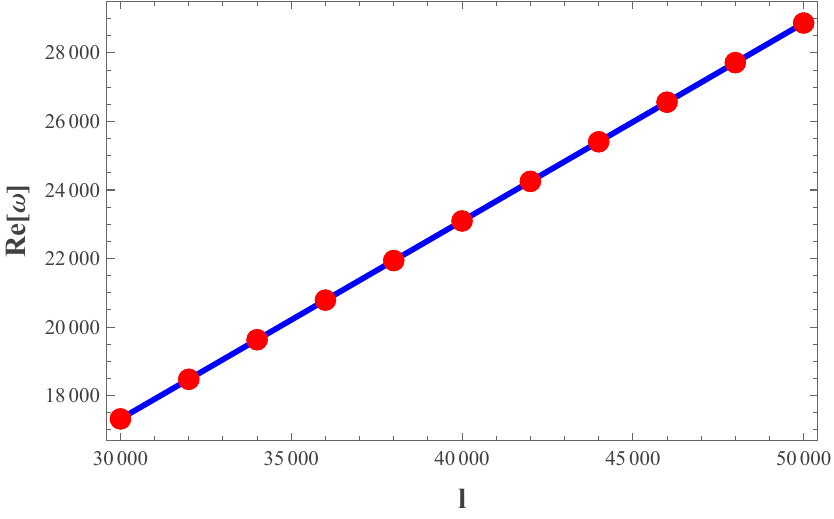}\hfill %
\includegraphics[width=0.45\textwidth]{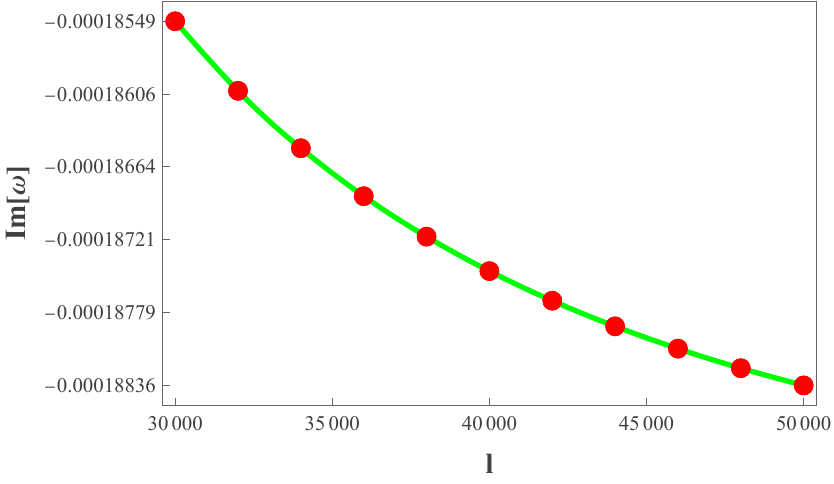}
\caption{Graphical representation of $\{\, r_{\mathrm{max}}(l),\, V_{\mathrm{%
eff}}[r_{\mathrm{max}}(l)],\, \mathrm{Re}[\protect\omega],\, \mathrm{Im}[%
\protect\omega] \,\}$ as functions of $l$, for the ModAMax solution with $k=+1$.}
\label{QNM}
\end{figure}

We have performed the numerical computation of $\{\,r_{\mathrm{max}}(l),\,V_{%
\mathrm{eff}}[r_{\mathrm{max}}(l)],\,\mathrm{Re}[\omega ],\,\mathrm{Im}%
[\omega ]\,\}$ as functions of $l$, and we present the results in Table \ref%
{tab1} and graphically in Fig. \ref{QNM}. We obtain only negative and
decreasing values for the imaginary part of the quasinormal frequency, and
positive and increasing values for the real part. This indicates the linear
stability of the scalar QNMs under a small perturbation of the geometry of
the ModAMax solution for $k=+1$.

\begin{table}[h!]
\centering
\begin{tabular}{|c|c|c|c|c|}
\hline
$l$ & $r_{max}$ & $V_{eff}(r_{max})$ & $Re[\omega]$ & $Im[\omega]$ \\ \hline
30000 & 67.8799886113 & 300076829.3184200000 & 17322.7258050896 & 
-0.0001854857 \\ \hline
32000 & 67.8358974080 & 341419895.7126960000 & 18477.5511286696 & 
-0.0001860350 \\ \hline
34000 & 67.7995735533 & 385430213.8491210000 & 19632.3766734706 & 
-0.0001864884 \\ \hline
36000 & 67.7692841279 & 432107783.7024550000 & 20787.2024020158 & 
-0.0001868671 \\ \hline
38000 & 67.7437562486 & 481452605.2540950000 & 21942.0282848695 & 
-0.0001871866 \\ \hline
40000 & 67.7220372830 & 533464678.4900750000 & 23096.8542985836 & 
-0.0001874588 \\ \hline
42000 & 67.7034024181 & 588144003.3997450000 & 24251.6804242443 & 
-0.0001876926 \\ \hline
44000 & 67.6872920356 & 645490579.9748650000 & 25406.5066464242 & 
-0.0001878949 \\ \hline
46000 & 67.6732683530 & 705504408.2089760000 & 26561.3329524125 & 
-0.0001880711 \\ \hline
48000 & 67.6609848278 & 768185488.0969610000 & 27716.1593316408 & 
-0.0001882256 \\ \hline
50000 & 67.6501641878 & 833533819.6347150000 & 28870.9857752494 & 
-0.0001883617 \\ \hline
\end{tabular}%
\caption{Table of scalar quasinormal modes for the ModAMax solution with $k=+1$, $l=30000$, and $n=0$.}
\label{tab1}
\end{table}

\end{document}